\documentclass[acmsmall]{acmart}
\usepackage{bbding}
\usepackage{color}
\usepackage{multirow}

\AtBeginDocument{%
  }

\setcopyright{acmlicensed}
\copyrightyear{2018}
\acmYear{2018}
\acmDOI{XXXXXXX.XXXXXXX}

\acmJournal{JACM}
\acmVolume{37}
\acmNumber{4}
\acmArticle{111}
\acmMonth{8}

\begin{document}

\title{Security and Privacy on Generative Data in AIGC: A Survey}

\author{Tao Wang}
\email{wangtao21@nuaa.edu.cn}
\orcid{0000-0001-5532-3999}

\author{Yushu Zhang}
\email{yushu@nuaa.edu.cn}

\author{Shuren Qi}
\email{shurenqi@nuaa.edu.cn}

\author{Ruoyu Zhao}
\email{zhaoruoyu@nuaa.edu.cn}
\affiliation{%
	\institution{Nanjing University of Aeronautics and Astronautics}
	\city{Nanjing}
	\country{China}
}

\author{Zhihua Xia}
\email{xia\_zhihua@163.com}

\author{Jian Weng}
\email{cryptjweng@gmail.com}

\affiliation{%
	\institution{Jinan University}
	\city{Jinan}
	\country{China}
}

\thanks{This work is supported in part by the National Key R\&D Program of China under Grant number 2022YFB3103100, in part by the National Natural Science Foundation of China under grant numbers 62122032, U23B2023.}

\authorsaddresses{%
	Authors’ Contact Information:  Tao Wang, wangtao21@nuaa.edu.cn; Yushu Zhang  (corresponding author), yushu@nuaa.edu.cn; Shuren Qi,
	shurenqi@nuaa.edu.cn; Ruoyu Zhao, zhaoruoyu@nuaa.edu.cn,  the College of Computer Science and Technology, Nanjing University of Aeronautics and Astronautics, Nanjing, China. Zhihua Xia, xia\_zhihua@163.com; Jian Weng, cryptjweng@gmail.com, the College of Cyber Security, Jinan University, Jinan, China.
}

\setcopyright{acmlicensed}
\acmJournal{CSUR}
\acmYear{2024} \acmVolume{1} \acmNumber{1} \acmArticle{1} \acmMonth{1}\acmDOI{10.1145/3703626}


\renewcommand{\shortauthors}{Wang et al.}

\begin{abstract}
 	The advent of artificial intelligence-generated content (AIGC)  represents a pivotal moment in the evolution of information technology. With AIGC,  it can be effortless to generate high-quality data that is challenging for the public to distinguish. Nevertheless, the proliferation of generative data across  cyberspace brings security and privacy issues, including privacy leakages of individuals  and media forgery for fraudulent purposes. Consequently, both academia and industry  begin to emphasize the trustworthiness of generative data,  successively providing a series of countermeasures for security and privacy. In this survey, we systematically review the security and privacy on generative data in AIGC, particularly for the first time analyzing them from the perspective of   information security properties. Specifically, we reveal the successful experiences of state-of-the-art countermeasures in terms of the foundational properties of privacy, controllability, authenticity, and compliance, respectively.  Finally, we  show some representative benchmarks,  present a statistical analysis, and summarize the potential exploration directions from each of theses  properties.
\end{abstract}




\maketitle

\section{Introduction}
\subsection{Background}
{A}{rtificial}  intelligence-generated content (AIGC)  emerges as a novel generation paradigm for the  production,  manipulation, and modification of data.  It utilizes advanced artificial intelligence (AI) technologies to automatically generate high-quality data at a rapid
pace, including images, videos, text, audio, and  graphics. With the powerful generative ability, 
AIGC can  save time and unleash creativity, which are often challenging to achieve with   professionally generated content (PGC) and user-generated content (UGC). Such progress in data creation can  drive the emergence of innovative industries, particularly  Metaverse \cite{wang2022survey}, where digital and physical worlds converge. 

Early AIGC is limited by the algorithmic efficiency, hardware performance, and data scale, hindering the ability to fulfill optimal creation tasks. With the iterative updates of generative structures, notably generative adversarial networks (GANs)  \cite{goodfellow2020generative}, AIGC has witnessed significant breakthroughs, generating realistic data that is often indistinguishable by humans from real data. 

In the generation of visual content, NVIDIA released StyleGAN \cite{karras2019style} in 2018, which enables the controllable generation of high-resolution images and has undergone several upgrades. The subsequent year, DeepMind released DVD-GAN \cite{clark2019efficient}, which is  designed for continuous video generation and  exhibits great efficacy in complex data domains. Recently, diffusion models (DMs) \cite{ho2020denoising} show more refined and novel image generation via the incremental noise addition.  Guided by language models, DMs can improve the semantic coherence between input prompts and generated images. Excellent diffusion-based products, e.g., Stable Diffusion\footnote{https://stability.ai/stablediffusion}, Midjourney\footnote{https://www.midjourney.com/}, and Make-A-Video\footnote{https://makeavideo.studio/},  are capable of generating visually realistic images or videos that meet the requirements of diverse textual prompts.

\begin{table}[!t]
	\centering
	\setlength{\abovecaptionskip}{0.1cm} 
	\caption{Comparisons of our work with existing surveys.}
	\label{tab:freq}
	\setlength{\tabcolsep}{1mm}{
		\begin{small}
			\begin{tabular}{c|cc|cc|cc|ccc}		
				
				\toprule
				\multirow{3}{*}{\textbf{Ref.}}&\multicolumn{2}{c|}{\textbf{Privacy}} &\multicolumn{2}{c|}{\textbf{Controllability}}&\multicolumn{2}{c|}{\textbf{Authenticity}}&\multicolumn{2}{c}{\textbf{Compliance}}\\
				\multirow{2}{*}{} & Privacy &  AIGC for & Access &\multirow{2}{*}{Traceability} &Generative & Generative & \multirow{2}{*}{Non-toxicity}&\multirow{2}{*}{Factuality} \\	
				
				&in AIGC&Privacy &Control&&Detection&Attribution&&&\\
				\midrule
				\cite{wang2023survey}&\XSolidBrush&\XSolidBrush&\XSolidBrush&\Checkmark&\XSolidBrush&\XSolidBrush&\XSolidBrush&\XSolidBrush\\
				
				\cite{lyu2023pathway}&\Checkmark&\XSolidBrush&\XSolidBrush&\Checkmark&\XSolidBrush&\XSolidBrush&\Checkmark&\Checkmark\\
				\cite{chen2023challenges}&\Checkmark&\XSolidBrush&\XSolidBrush&\Checkmark&\XSolidBrush&\Checkmark&\Checkmark&\Checkmark\\
				
				Ours&\Checkmark&\Checkmark&\Checkmark&\Checkmark&\Checkmark&\Checkmark&\Checkmark&\Checkmark\\

				\bottomrule
			\end{tabular}
		\end{small}
	}
\end{table}

In the generation of language content,   more attention is focused on ChatGPT, which reached 1.76 billion visits in May 2023. Trained on a large-scale text dataset, ChatGPT exhibits impressive performance in various contexts, including human-computer interaction and dialogues. For instance, researchers  released  LeanDojo \cite{yang2024leandojo}, an open-source mathematical proof platform based on ChatGPT, providing toolkits, benchmarks, and models to tackle complex proof of formulas in an interactive environment. The integration of ChatGPT into the Bing search engine enhances search  experiences, enabling users to effortlessly access  comprehensive information. 
This powerful multi-purpose adaptability further exemplifies the possibilities for humanity to achieve artificial general intelligence (AGI).

Overall, compared to the PGC and UGC, AIGC  demonstrates more advantages in data creation.  AIGC possesses the ability to swiftly produce high-quality content while catering to personalized  demands from users.   As AI technology continues to  advancements, the generative capability of AIGC is growing rapidly, promoting increased social productivity and economic value.


\subsection{Motivation}
A large amount of generative data floods  cyberspace, further enriching the diversity and abundance of online content. These generative data encompass multimodal information, which can  be observed in various domains, e.g.,  news reporting, computer games, and social sharing. According to the Gartner's report, AIGC will be anticipated to account for over 10\% of all data creation in 2025.  However, the proliferation of generative data also poses security and privacy issues.


Firstly, generative data can expose individual privacy content by replicating training data. Generative models rely on large-scale data, which includes private content, e.g.,  faces, addresses, and emails. Existing works have demonstrated the memorization capabilities of large generative models \cite{carlini2021extracting,carlini2023extracting}, leading to the potential replication of all or parts of the training data. This means that the generative data may also contain sensitive content which present in the training data. With specific prompts, GPT-2 can output personal information, including the name, the address, and the e-mail address \cite{carlini2021extracting}. An alarming study \cite{carlini2023extracting} revealed that Google's Imagen can be prompted to output real-person photos, posing a significant threat to individual privacy. Therefore, it is necessary to hinder the generation of data containing \textbf{privacy} content.

Secondly, generative data used for malicious purposes often involves false information, which can deceive the public, posing potential threats to both society and individuals. Recently, a false tweet about an explosion near the Pentagon went viral on social media, fooling many authoritative media sources and triggering fluctuations in the US stock market \cite{Pentagon2023}. Moreover, the mature DeepFake technologies allow for the creation of convincing fake personal videos, which can be used to maliciously fabricate celebrity events \cite{korshunov2018deepfakes}.  The difficulty in discerning authenticity exposes the public to believing such content, resulting in severe damage to the reputations of celebrities. Thus, it is important to provide  effective technology to confirm the \textbf{authenticity} of generative data. Meanwhile, generative data is required to  have the  \textbf{controllability}  so that such potential threats can be proactively prevented.


Thirdly, regulators around the world have further requirements for the compliance of generative data due to the critical implications of AIGC. Data protection regulatory authorities in Italy, Germany, France, Spain, and  other countries have expressed concerns and initiated investigations into AIGC. In particular, China has taken a significant step by introducing the interim regulation on the management of generative artificial intelligence (AI) services \cite{interim}. This regulation encourages innovation in AIGC while mandating that generative data is non-toxic and factual.  To adhere to the relevant regulations, it becomes crucial to ensure the \textbf{compliance} of generative data.


\subsection{Comparisons with Existing Surveys}

Several works \cite{wang2023survey,lyu2023pathway,chen2023challenges} have investigated the  security and privacy in AIGC from different perspectives. 

Wang \textit{et al.} \cite{wang2023survey} presents an in-depth survey of AIGC working principles, and roundly explored the taxonomy of security and privacy threats to AIGC. Meanwhile, they  extensively reviewed solutions for intellectual property (IP) protection for AIGC models and generative data, with a  focus on  watermarking. Yet,  they fail to provide  countermeasures for  other threats such as the utilization of non-compliant data.

Chen \textit{et al.} \cite{lyu2023pathway} discussed  three main concerns for promoting responsible AIGC, including  1) privacy, 2) bias, toxicity, misinformation, and 3) intellectual property. They summarized the issues and listed solutions related to existing AIGC products, e.g., ChatGPT, Midjourney, and Imagen.  Nevertheless, they overlooked the importance of  considering the authenticity of generative data in responsible AIGC.

Chen \textit{et al.} \cite{chen2023challenges} summarized the AIGC technology and analyzed the security and privacy challenges in AIGC. Moreover, they  explored the  potential countermeasures with advanced technologies, involving privacy computing, blockchain, and beyond.  However, they did not pay attention to the  detection and  access control of generative data.

The differences between our work and previous  works are: 
\begin{itemize}
	\item Our work  is targeted at generative data rather than AIGC.  Previous works  also explored issues about privacy in  data collection and security of models. Yet, these issues are generic to AI security, which has been  discussed in some works \cite{10.1145/3487890,10.1145/3436755}.
	\item {Previous works presented the corresponding techniques in terms of specific  issues of privacy and security,  whereas the issues cannot be enumerated in full.  On the other hand, we discuss security and privacy from  the fundamental properties of information security,  which can cover the almost all of the issues.}
	
	\item  We supplement security issues that are not discussed in previous works, including access control and generative detection. In addition, we explore  the use of generative data to power the privacy protection of real data. Table \ref{tab:freq} shows the comparisons of our work with existing surveys. 
	
\end{itemize}

In brief, the main contributions of our work are as follows:
\begin{itemize}
	\item We investigate the security and privacy issues  on generative data in  AIGC and comprehensively survey the corresponding state-of-the-art countermeasures.
	\item We discuss security and privacy from a new perspective, i.e., the fundamental properties of information security, including privacy, controllability, authenticity, and compliance.
	\item We point out the valuable future directions in security and privacy, toward
	building  trustworthy generative data.
\end{itemize}

The rest of the paper is organized as follows. Section \ref{Overview} reviews the basic AIGC process and categorizes the security and privacy on generative data in AIGC. In Sections \ref{Privacy} to \ref{Compaliance}, we discuss the issues and review the corresponding solutions from the perspectives of privacy, controllability, authenticity, and compliance, respectively. We present some benchmarks and  suggested future directions in Section \ref{Benchmarks} and Section \ref{Directions}. Finally, we summarize our work in Section \ref{Conclusion}.

\begin{figure*}[!tbp]
	\centering
	\includegraphics[width=5.5in, keepaspectratio]{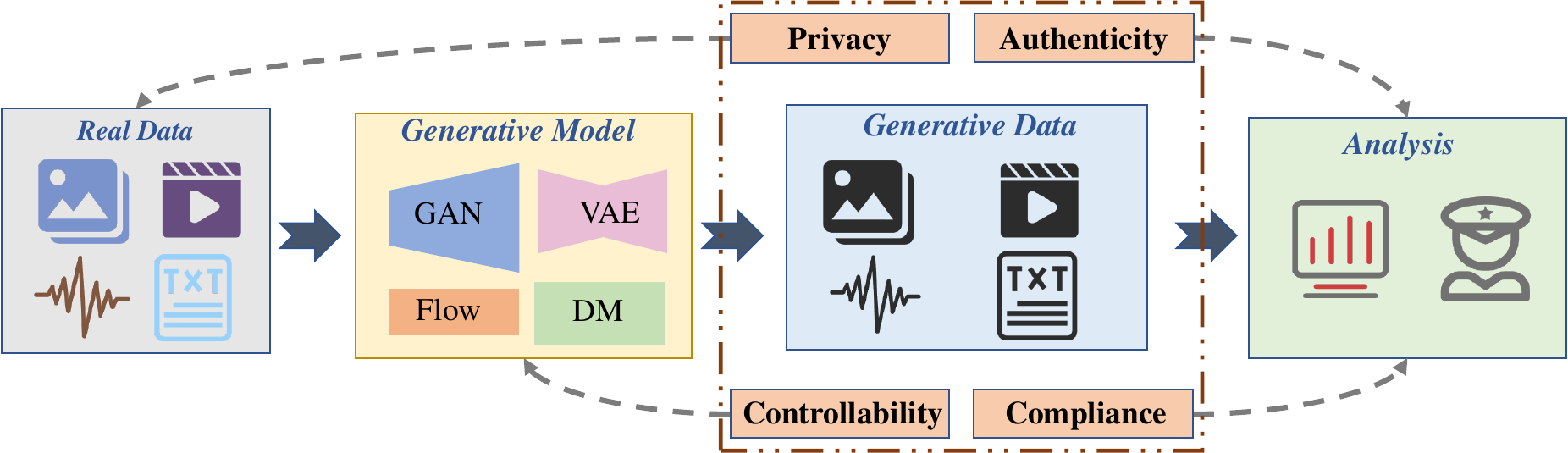}\\
	\caption{The process of AIGC. Real data collected is used to train  generative models. Then generative models produce  generative data. Finally, generative data are further analyzed. For generative data, there are corresponding protection requirements of security and privacy  at different stages, which can be divided into  privacy, controllability, authenticity, and compliance.  }
	\label{diversity}
\end{figure*}

\section{Overview} \label{Overview}

\subsection{Process of AIGC}
As illustrated in Fig. \ref{diversity}, we first discuss the AIGC process as follows:

\subsubsection{Real Data for Training}
The data used for training  impacts the features and patterns learned by AIGC models. Therefore, high-quality data forms the cornerstone of AIGC technology. Data collection typically involves various open-source repositories, including public databases, social media platforms, and online forums. These diverse sources provide AIGC training with a large-scale and diverse dataset.

After collection,  data filtering is applied to ensure the  data quality, which involves removing irrelevant  data and balancing the dataset for unbiased training. Additionally,  data preprocessing, data augmentation, and data privacy protection steps can be  undertaken based on different tasks to further enhance the quality and security of the training data.

\subsubsection{Generative Model in Training}
The obtained data is used to train generative models, which are often performed by a centralized server with powerful computational capabilities. During  training,  generative models learn patterns and features in the data to generate results with a  similar distribution to  real data. Popular generative model architectures include generative adversarial networks (GANs), variational autoencoders (VAEs), flow-based models (Flows), and diffusion models (DMs), each with its strengths and weaknesses. The choice of models depends on  specific requirements of  tasks, available data, and computational resources.

It is also important to note that training generative  models requires substantial computational resources. On this basis, model fine-tuning is the process of adapting the pre-trained large model to a new task or domain without retraining. It only  adjusts model parameters by training   appropriate amount of additional data. 	

\subsubsection{Generative Data}
After generative models are trained, they can be utilized to produce data. During this stage, users typically provide an input condition, e.g., a question or a piece of text.  Then the model starts outputting data based on the input condition. 

In the generation of language content, AIGC exhibits the capability to outpace human authors in rapidly generating high-quality text, e.g.,  codes and articles. Additionally, it can engage in conversational interactions akin to humans, assisting users with various tasks and inquiries.  The efficiency  of AIGC in content creation and human-like interactions  revolutionize how information is produced and communicated. 

In the generation of visual content, AIGC harnesses the powerful generative capabilities of models like DMs, enabling the generation of new images with realistic quality. Moreover, AIGC holds potential for video generation, as they can simultaneously
process multiple video frames automatically.



\subsubsection{Analysis for Generative Data}
After data generation, further analysis of the generative data is necessary to ensure the quality of  generative data.

Generative data needs to undergo a quality assessment to check its accuracy, consistency, and integrality. If the generative data lacks in certain aspects, it  requires model  adjustments to improve the quality of generative data.

Additionally, analyzing the risks associated with generative data can identify potential  hazards. For instance,
it is required to analyze
whether there is discriminatory content, false information, or misleading content. By promptly detecting and addressing these issues, the negative impact of generative data can be minimized.



\subsection{Security and Privacy on Generative Data}

\begin{figure}[htbp]
	\centering
	\includegraphics[width=3in, keepaspectratio]{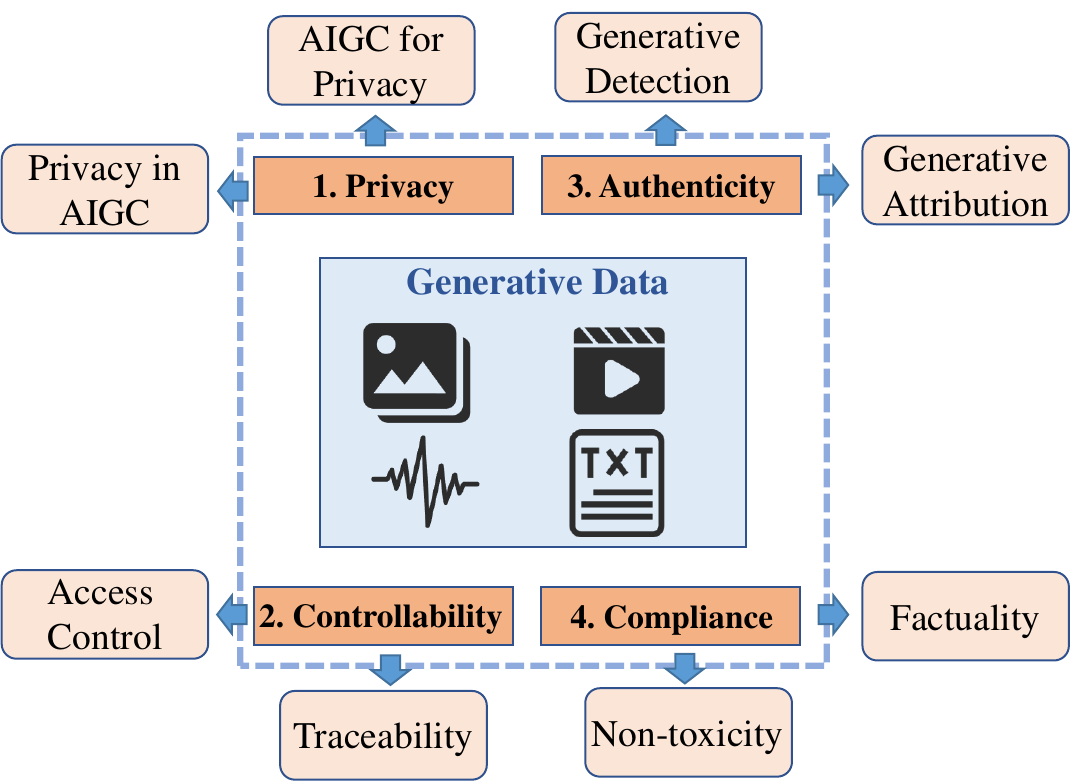}\\
	\caption{The subclassification of security and privacy on generative data. }
	\label{class}
\end{figure}
For generative data, there are corresponding security and privacy requirements at different stages. As shown in Fig. \ref{diversity}, we categorize these requirements according to the fundamental properties of information security, including privacy, controllability, authenticity, and compliance. Additionally, Fig. \ref{class} shows the further subclassification.
\subsubsection{Privacy}  Privacy refers to ensuring that individual sensitive information is protected. Generative data mimics the distribution of real data, which brings negative and positive impacts on the privacy of real data. Specifically, the following two impacts exist:

\begin{itemize}
	\item \textit{Privacy in AIGC:} Generative data may mimic the distribution of sensitive content, which makes it possible to replicate sensitive training data under specific conditions, thus posing a potential privacy threat.
	\item \textit{AIGC for privacy:} Generative data
	contains virtual content, which can be used to replace sensitive content in real data, thereby reducing the risk of privacy breaches while maintaining data utility.
\end{itemize}

\subsubsection{Controllability} Controllability refers to ensuring effective management and control access of information to restrict unauthorized actions. Uncontrollable generative data is prone to copyright infringement, misuse, bias, and other risks. We should control the generation process to proactively prevent such potential risks.
\begin{itemize}
	\item \textit{Access control:} Access to generative data needs to be controlled to prevent  negative impacts from the unauthorized utilization of real data, e.g., malicious manipulation and copyright infringement.
	
	\item \textit{Traceability:} Generative data needs to support the tracking of the generation process and  subsequent dissemination to monitor any behavior involving security for accountability.
\end{itemize}

\subsubsection{Authenticity} Authenticity refers to maintaining the integrity and truthfulness of data, ensuring that information is accurate, unaltered, and from credible sources.  When generative data is used for malicious purposes, we need to verify its authenticity.
\begin{itemize}
	\item \textit{Generative detection:} Humans have the right to know whether data is generated by AI or not. Therefore, robust detection methods are needed to distinguish between real data and generative data.
	
	\item \textit{Generative attribution:} In addition, generative data should be further attributed to generative models to ensure credibility  and enable accountability.
\end{itemize}

\subsubsection{Compliance} Compliance refers to adhering to relevant laws, regulations, and industry standards, ensuring that information security practices meet legal requirements and industry best practices. We mainly talk about two important requirements as follows:
\begin{itemize}
	\item \textit{Non-toxicity:} Generative data is prohibited from containing toxic content, e.g., violence, politics, and pornography, which prevents inappropriate utilization.
	\item \textit{Factuality:} Generative data is strictly factual and should not be illogical or inaccurate, which prevents the accumulation of misperceptions by the public.
\end{itemize}

\section{Privacy on Generative Data} \label{Privacy}
\begin{figure}[htbp]
	\centering
	\includegraphics[width=3in, keepaspectratio]{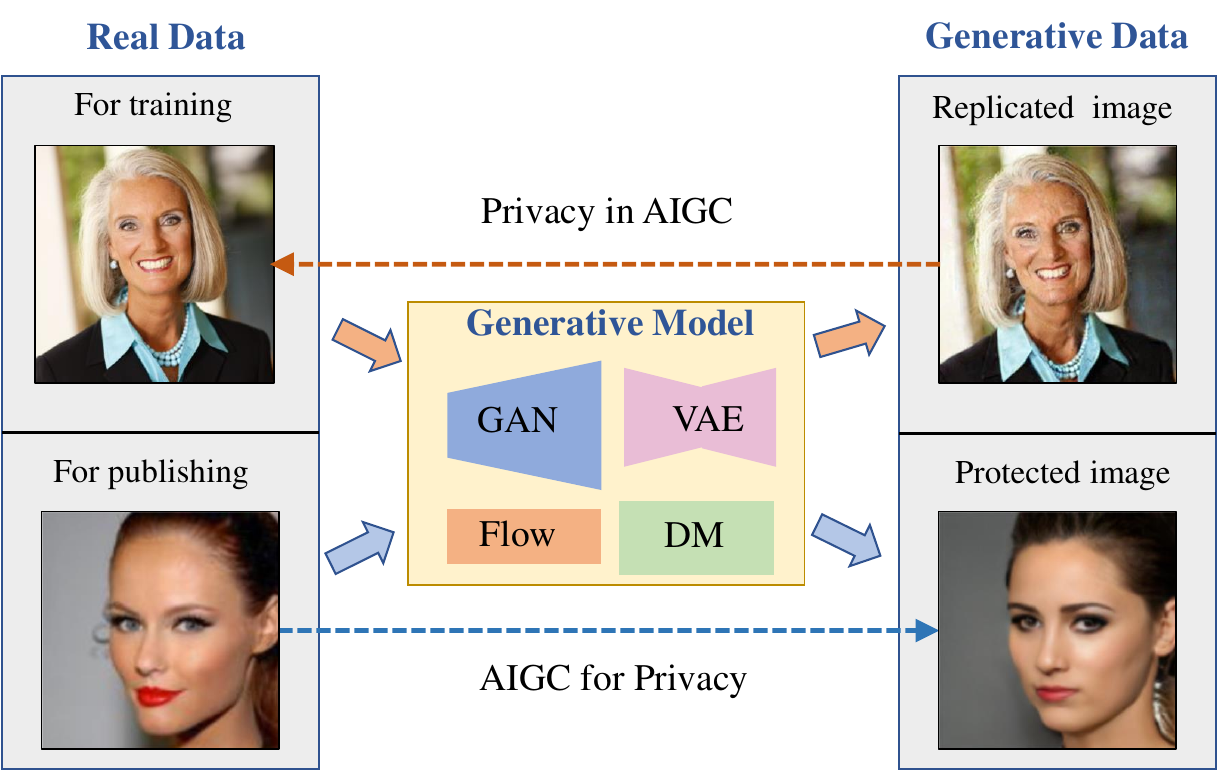}\\
	\caption{The example of \textit{privacy in AIGC} is from \cite{carlini2023extracting}, and the example of \textit{AIGC for privacy} is from \cite{chen2021perceptual}.  }
	\label{privacy}
\end{figure}

For generative data, we talked about its  negative and  positive impacts on the privacy of real data. \textit{1)  Negative:} A large amount of real data is used for the training of AIGC models, which may memorize the training data. In this way, the generative data would replicate the sensitive data under certain conditions, thus causing a privacy breach of real data, which is called \textit{privacy in AIGC}. For instance, in the top part of Fig. \ref{privacy}, it is easy to generate the face image of Ann Graham Lotz with the prompt ``Ann Graham Lotz", which is almost identical to the training sample. \textit{2) Positive:} Real data published by users  contains sensitive content, and AIGC can be used to protect privacy by replacing sensitive content with virtual content, which is called \textit{AIGC for privacy}. In the bottom part of Fig. \ref{privacy}, the generative image has a different identity from the real image, blocking unauthorized identification.

\subsection{Privacy for AIGC}
\subsubsection{Threats to Privacy}

AIGC service providers proactively collect individual data on various platforms to construct giant datasets for enhancing the quality of generative data.
However, the training data contains sensitive information about individuals, which is highly susceptible to privacy leakage. A study shows that the larger the amount of training data, the higher the privacy risk will result \cite{plant2022you}. Specifically, in the training, private data can easily be memorized in model weights.  In the interaction with users, generative data may replicate the training data, which poses a  potential privacy threat. Such data replication is defined as an under-explored failure mode of  overfitting, which exists in various generative models \cite{meehan2020non}.

In language generative models, Carlini \textit{et al.} \cite{carlini2021extracting} extracted training data by querying large language models (LLMs). The experiments employ GPT-2 as a demonstration to extract sensitive individual information, e.g., name, email, and phone number. Tirumala \textit{et al.} \cite{tirumala2022memorization}   empirically studied the memorization dynamics over language model training, and demonstrated that larger models memorize faster. 
Nicholas \textit{et al.} \cite{carlini2022quantifying} described three log-linear relationships to quantify the extent to which LLMs memorize training data under different model scales, times of sample replications, and number of tokens. Their experiments indicated that memory in LLMs is more prevalent than previously believed and that memorization scales log-linearly with model size.

\begin{figure*}[!tbp]
	\centering
	\includegraphics[width=5in, keepaspectratio]{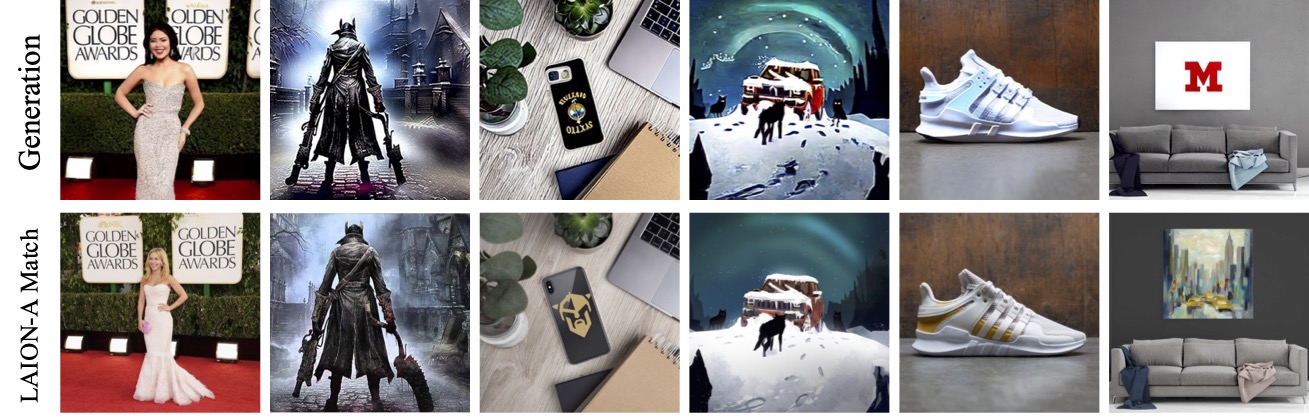}\\
	\caption{Generated samples by Stable Diffusion just replicate training data via  piecing together foreground and background objects in training data \cite{somepalli2023diffusion}. }
	\label{privacy1}
\end{figure*}

In vision generative models, the general view is that generative adversarial networks (GANs) tend not to memorize training data under normal training settings \cite{webster2019detecting}. However, Feng \textit{et al.} \cite{feng2021gans} showed experimentally that GANs replication percentage decays exponentially with respect to dataset size and complexity. Stronger memory exists in diffusion models \cite{webster2023reproducible,carlini2023extracting,somepalli2023diffusion}. Carlini \textit{et al.}  \cite{carlini2023extracting} illustrated the ability of diffusion models to memorize individual images from the training data and can reproduce them at generation time.   {Unlike DMs that accept training data as direct input, generators for GANs are trained using only indirect information (i.e., gradients from the discriminator) about the training data. Therefore, GANs are more private.}  Somepalli \textit{et al.} \cite{somepalli2023diffusion} proposed image retrieval frameworks to demonstrate that the generated images by diffusion models are simple combinations of the foreground and background objects of the training dataset.  As shown in Fig. \ref{privacy1},  diffusion models just create semantically rather than pixel-wise objects identical to original images.

\subsubsection{Countermeasures to Privacy}
Researchers have suggested some available solutions to mitigate  data replication for privacy protection. Bai et al.  \cite{bai2022reducing} proposed memorization rejection in the training loss, which abandons generative data that are near-duplicates of training data.

Deduplicating  training datasets is also a possible option. OpenAI has verified its effectiveness using the distributed nearest neighbor search on Dall-E2. Nikhil \textit{et al.} \cite{kandpal2022deduplicating} studied a variety of LLMs and showed that the likelihood of a duplicated text sequence appearing is correlated with the number of occurrences of that sequence in the training data. In addition, they also verified that replicated text sequences are greatly reduced when duplicates are eliminated.

Differential privacy \cite{abadi2016deep} is a recommended solution, which introduces noises during training to ensure the
generative data is differentially private with respect to the training data. RDP-GAN \cite{ma2023rdp} adds differential noises on the value
of the loss function of a discriminator during training, which achieves a differentially private GAN. DPDMs \cite{dockhorn2022differentially} enhances privacy via differentially private stochastic gradient descent, which also allows the generative data to retain a high level of availability to support multiple vision tasks. Compared to DPDMs, Ghalebikesabi \textit{et al.} \cite{ghalebikesabi2023differentially} enabled to accurately train larger models and to achieve a high utility on more challenging datasets such as   CIFAR-10.  {PrivImage  \cite{ 299531}  was proposed based on  that the distribution of public data used should be similar to the distribution of sensitive data. Therefore,   PrivImage elaborately selects similar distribution pre-training data, facilitating the efficient creation of DP datasets with high fidelity and utility.    Unlike existing work that only considers DM as a regular deep model, dp-promise \cite{298240} was the first work to implement (approximate) DP using DM noise, It employs a two-stage DM training process to reduce the overall noise injection, effectively achieving the privacy-utility tradeoff.}

Detecting replicated content is a remedial solution, which detects whether the generative data is in training data and then determines whether to use it. Stability AI provides a tool \cite{Beaumont} to support
the identification of the replicated images. Somepalli \textit{et al.} \cite{somepalli2023diffusion} developed image similarity metrics that are diverse on self-supervised learning and based on an image retrieval framework to search for copying behavior.

Machine unlearning \cite{bourtoule2021machine} can help generative models forget the private training data, which avoids the effort of re-training the model.  Kumari \textit{et. al} \cite{kumari2023ablating} fine-tuned diffusion models by modifying the sensitive training data so that the models forget already memorized images. Forget-Me-Not \cite{zhang2024forget} is adapted as a lightweight model patch for Stable Diffusion. It effectively removes the concept of containing a specific identity and avoids generating any face photo with the identity.

\subsection{AIGC for Privacy}
Sensitive content exists on many types of data published by different entities, which is required for a  privacy-preserving treatment. Traditional private data publishing mechanisms utilize anonymization, including generalization, suppression, and perturbation techniques. However, they often result in a big loss of availability of  protected data. 

Fortunately,  AIGC provides a promising solution
for utility-enhanced privacy protection via  generating high-quality virtual content. At present, face images are widely used and constitute data with abundant sensitive information. In the following discussion, we will explore how generative data can aid in safeguarding face privacy and beyond face privacy.

\subsubsection{Face Privacy}

To protect face privacy, many works \cite{wang2023identifiable,hukkelaas2023deepprivacy2,wen2023divide,yuan2022generating,kim2023dcface,liu2023diffprotect,10167807,10646362,wen2022identitydp} generate a surrogate virtual face with new identity. DeepFace2 \cite{hukkelaas2023deepprivacy2} generates realistic anonymized images by conditioning GANs to fill in images that obscure facial regions. In order to preserve attributes, Gong  \textit{et al.} \cite{gong2020disentangled} 
replaced identity independently by decoupling identity and attribute features, which achieves a trade-off between identity privacy and data utility.
To facilitate privacy-preserving face recognition, IVFG \cite{yuan2022generating} generates identifiable virtual faces bound with new identities in the latent space of StyleGAN. In addition,  publicly available face datasets for training face recognizers often violate the privacy of real people. For this, DCFace \cite{kim2023dcface}  creates a generative dataset with virtual faces  via diffusion models. 

Other works are devoted to  generating adversarial faces to evade  unauthorized identification. DiffProtect \cite{liu2023diffprotect} adopts a diffusion autoencoder to generate semantically meaningful perturbations, which can promote the protected face identified as another person. 3DAM-GAN \cite{10167807} generates natural adversarial faces by makeup transfer,  improving the quality and transferability of generative makeup for identity concealment.


\begin{table*}[tp]
	\centering
	\caption{A summary of representative solutions for the privacy of generative data.}
	\label{tab:privacy}
	\begin{small}
		\begin{tabular}{ccccccccccccccccc}		
			\toprule
			&	Ref.&Year&Remarks&Limitations \\
			\midrule
			\multirow{7}{*}{\parbox{0.8cm}{\centering Privacy in AIGC}}{} & \cite{bai2022reducing}& 2022&Maintained generation quality &Weak generalizability and scalability\\
			& \cite{kandpal2022deduplicating}& 2022 &Enhanced security, easy operation &Lack of privacy guarantees\\
			& \cite{ma2023rdp}& 2023 &Without norm clipping, strict proof&Visual semantic disclosure\\
			& \cite{298240}& 2024 &Provable privacy applicable to DMs&Poor visual quality, high consumption\\
			&\cite{ 299531}& 2024 &Provable privacy, low consumption&Poor visual quality, huge training costs\\
			& \cite{somepalli2023diffusion}& 2023 &Free train, easily understanding &Difficulties in defining similar data\\
			&\cite{zhang2024forget}& 2024&Efficient, definition-free& Failure  in abstract concept\\
			
			\midrule
			
			\multirow{7}{*}{\parbox{0.8cm}{\centering Privacy in AIGC}}{}	& \cite{hukkelaas2023deepprivacy2}& 2023 & High-quality, diverse, and editable&Out-of-context results, reduced utility\\
			
			&\cite{yuan2022generating}& 2023 & Identifiable, irreversible&Unpreserved facial attributes \\
			
			&\cite{kim2023dcface}&2023&Visual high quality, additional metric&Time-consuming, privacy unprovable \\

			&\cite{10167807}&2023&Imperceptible, transferable&Not applicable to males\\
			
			& \cite{cao2021generating}& 2021 &Multi-model combination, high utility&Less stringent privacy proofs\\
			
			&\cite{liu2022privacy}&2022&Aligning with user preferences&Reduced utility, limited scalability\\
			
			&\cite{thambawita2022singan}  & 2022 &Small training data, comparable quality& Lack of trustworthiness\\
			
			\bottomrule
		\end{tabular}
	\end{small}
\end{table*}

\subsubsection{Beyond Face Privacy}
Beyond face, many types of data have sensitive information that needs to be protected \cite{lu2023machine,zhang2022visual}. TrajGen \cite{cao2021generating} uses GAN and Seq2Seq to simulate the real data to generate mobility data, which can be shared without privacy leakages, thus contributing to the open source process of mobility datasets. In the recommendation systems, UPC-SDG \cite{liu2022privacy} can generate virtual interaction data for users according to their privacy preferences, providing privacy guarantees while maximizing the data utility. 
SinGAN-Seg \cite{thambawita2022singan} uses a single training image to generate synthetic medical images with corresponding masks, which can effectively protect patient privacy when  performing medical segmentation.
PPUP-GAN \cite{yao2023ppup} generates new content of the privacy-related background while maintaining the content of the region of interest in aerial photography, which can protect the privacy of bystanders and maintain the utility of aerial image analysis.
Hindistan \textit{et al.} \cite{hindistan2023hybrid} designed a  hybrid approach to protect industrial Internet of Things (IIoT) data based on GANs and differential privacy,  which  causes minimal accuracy loss without extra high computational costs to data processing.

\subsection{Summary}
In table \ref{tab:privacy}, we summarize the solutions for privacy protection on generative data.  In the case of privacy in AIGC, differential privacy provides a provable guarantee for generative data, but may make the distribution of  generative data different from the real data, reducing data utility. Replication detection and data deduplication avoid any manipulation of models but rely on appropriate image similarity metrics. {Machine unlearning promotes that  generative data no longer contains sensitive content via model fine-tuning. In particular, as the size of the generative model increases, this fine-tuning technique will receive more attention.} Current	machine unlearning schemes for generative models are still relatively underdeveloped and will  be a promising exploration direction. {In addition, adversarial attacks can also have an impact on the privacy of generative data. On the one hand, adversarial attacks can attack generative models to prevent them from learning about privacy content in real data, thus securing generative data from replicating privacy content at the source. On the other hand, adversarial attacks can also attack privacy-preserving methods (e.g., machine unlearning) to prevent the removal of sensitive information, which exacerbates the  privacy challenge  on generative data.}

In the case of AIGC for privacy, the realism, diversity, and controllability of AIGC provide important directions for the privacy protection of real data, especially for unstructured data such as images. Due to the mature research of GANs,  a plethora of existing works utilize GANs to generate virtual content for privacy protection. Compared to GANs, diffusion models exhibit stronger generative capabilities. Therefore, as its controllability improves, it will shine even more in data privacy protection. In addition, it is important to note that the generated virtual data needs to avoid the \textit{privacy in AIGC}, otherwise bringing additional privacy issues.

\section{Controllability on Generative Data} 

{Uncontrolled generative data may give rise to potential issues, e.g.,  copyright infringement and malicious utilization. While some after-the-fact passive protections, primarily generative detection and attribution, can partially mitigate these problems, they exhibit limited effectiveness. Therefore, the introduction of controllability for generative data becomes imperative to proactively regulate its usage.}

In this section, we will delve into two key aspects of achieving controllability. Firstly, our focus will be on the \textit{access control} of generative data to constrain the model from producing unrestricted generative results, thereby proactively mitigating potential issues from the source. Secondly, we emphasize the importance of \textit{traceability} in monitoring generative data, as it enables post hoc scrutiny to ensure legitimate, appropriate, and responsible utilization.



\subsection{Access Control}

Generative data is indirectly guided by trained data, so access control to generative data can be effectively achieved by controlling the use of real data in generative models. Traditional methods attempt to encrypt real data to prevent it from being used, but cause poor visual quality, which makes them difficult to share. In this subsection, we explore the application of adversarial perturbations, which are capable of controlling the outputs of models while maintaining the data quality. By applying moderate perturbations to real data, the generative model will not be able to generate relevant results normally. Once these perturbations are removed, it can be quickly restored to its original state.

\begin{figure}[!ht]
	\centering
	\includegraphics[width=4in, keepaspectratio]{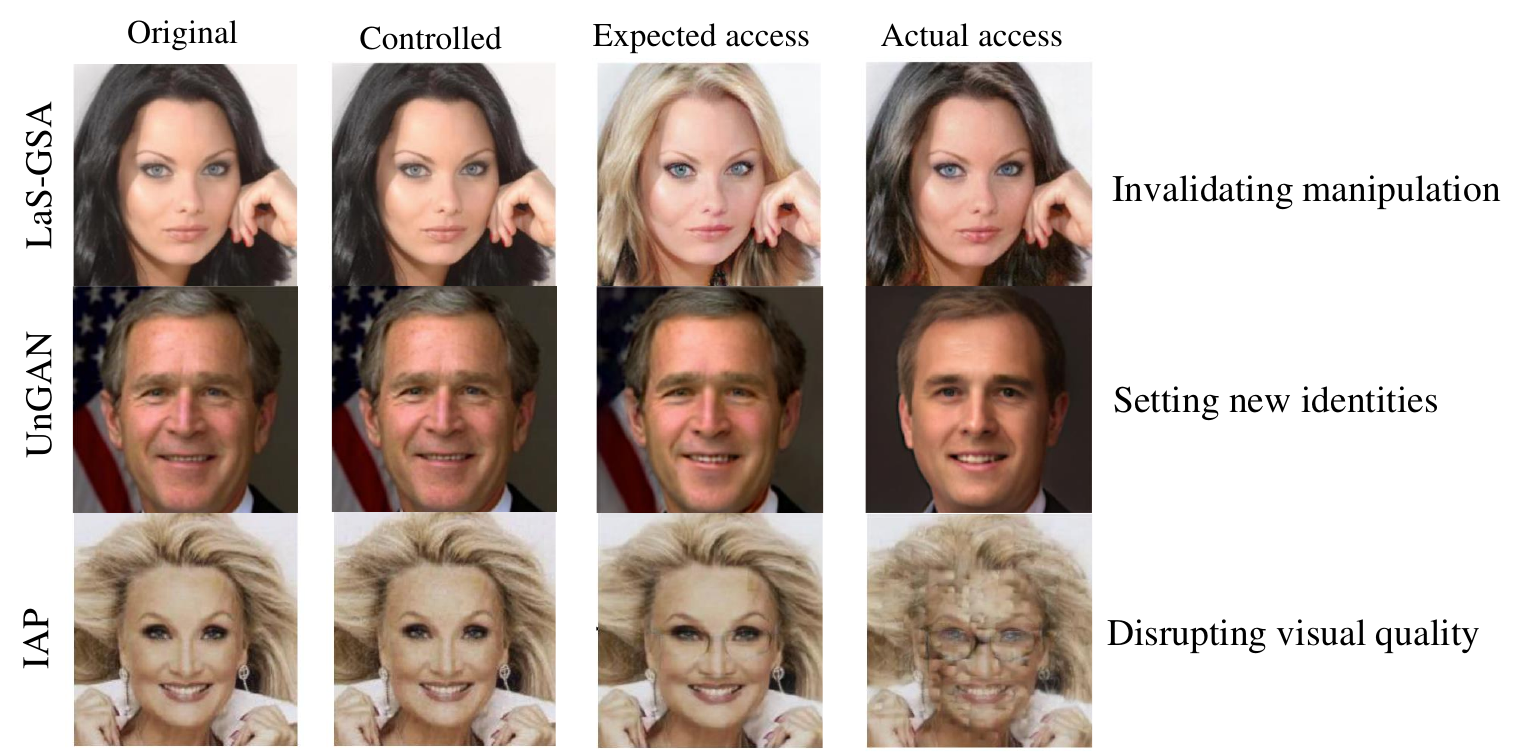}\\
	\caption{Examples of different protections for malicious  access from  \cite{yeh2021attack}, \cite{LYSBFZ23}, and \cite{10086559}. }
	\label{access}
\end{figure}

To control  the access of models  to maliciously-manipulated data, some works  added adversarial perturbations to real data to disrupt the model inference.  Yeh \textit{et al.} \cite{yeh2021attack} constructed a novel nullifying perturbation. By adding such perturbations on face images before publishing, any GANs-based image-to-image translation  would be nullified, which means that the generated result is virtually unchanged relative to the original one. UnGANable \cite{LYSBFZ23} makes the manipulated face belong to a new identity, which can protect the original identity. Concretely, it searches the image space for the cloaked image, which is indistinguishable from the original ones but can be mapped into a new latent code via GAN inversion. Information-containing adversarial perturbation (IAP) \cite{10086559} presents a two-tier protection. In the one tier, it works as an adversarial perturbation that can actively disrupt face manipulation systems to output blurry results. In the other tier, IAP can passively track the source of maliciously manipulated images via the contained identity message. The different effects of the above works are displayed in Fig. \ref{access}.


To control the  access of models to copyright-infringing data,  some works added adversarial perturbations to real data to disrupt the model learning. Anti-DreamBooth \cite{van2023anti} can add minor perturbations to individual images before releasing them, which destroys the training effect of any DreamBooth models.  Glaze \cite{shan2023glaze} is designed to help artists add an imperceptible ``style cloak" to their artworks before sharing them, effectively preventing diffusion models from mimicking the artist. 
Wu \textit{et al.}  \cite{wu2023promptrobust} proposed an adversarial decoupling augmentation framework,  generating adversarial noise to disrupt the training process of text-to-image models.  Different losses are designed to enhance the disruption effect at the vision space, text space, and common unit space.  
Liang \textit{et al.} \cite{liang2023adversarial} built a theoretical framework to define and evaluate the adversarial perturbations for DMs. Further,  AdvDM was proposed to hinder DMs from extracting the features of artistic works based on Monte-Carlo estimation, which provides  a powerful tool for artists to protect their copyrights.

\subsection{Traceability}

\subsubsection{Watermarking}
Digital watermarking \cite{liu2023survey}  is a technique used to inject visible or hidden identified information into digital media. The use of digital watermarking in AIGC can achieve a variety of functions:\begin{itemize}
	\item \textit{Copyright protection:}  By embedding watermarks with unique identified information, the source and ownership of the data can be traced and proved.
	
	\item \textit{Authenticity detection:}  By detecting and identifying the watermark information, it is easy to confirm whether the data is generative and even which models generate it.
	\item \textit{Accountability:} It is possible to track and identify the content's dissemination pipelines and usage, further ensuring accountability.
\end{itemize}

\begin{figure*}[!t]
	\centering
	\includegraphics[width=4.5in, keepaspectratio]{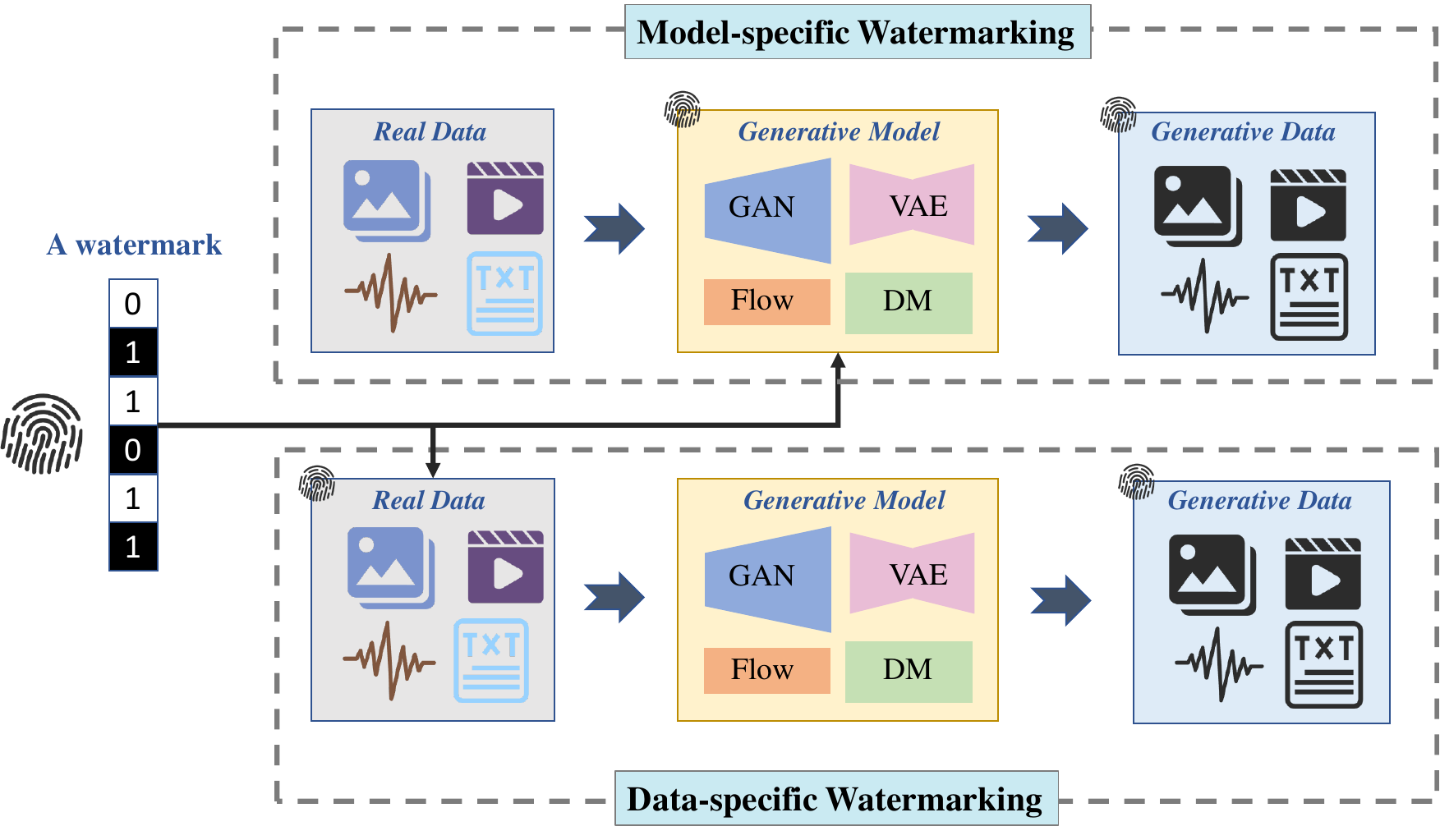}\\
	\caption{Model-specific watermarking and image-specific watermarking. }
	\label{watermarking}
\end{figure*}

Depending on whether the watermark is directly produced by the generative model or not, existing works can be categorized into model-specific watermarking and image-specific watermarking, which is shown in Fig. \ref{watermarking}.

\noindent
\textbf{Model-specific Watermarking:} 
This class of work inserts watermarks into generative models, and then the data generated by these models also have watermarks.

Yu \textit{et al.} \cite{yu2021artificial} and Zhao \textit{et al.} \cite{zhao2023recipe} implanted watermarking in training data to retrain GANs or DMs from scratch, respectively. Watermarking can also exist in generative data, as they would learn the distribution of the training data. 	{Compared to GANs which directly embed control information into deep features, DMs embed control information multiple times by progressive random denoising, which can improve steganography and stability. Therefore, DMs have the potential to be better in controllability.}
Stable signature \cite{fernandez2023stable}  integrates image watermarking into latent diffusion models. By fine-tuning the latent decoder, the generated data would contain invisible and robust watermarks, i.e., binary signatures, which support after-the-fact detection and identification of the generated data. 		Cheng \textit{et al.} \cite{10.1145/3581783.3612448} introduced a flexible and secure  watermarking. The watermark can be altered flexibly by modifying the message matrix, without retraining the model. Additionally,  attempts to evade the use of the message matrix result in degraded generated quality, thereby enhancing the security.


Some works \cite{liu2023watermarking} can only generate watermarked data when specific triggers are activated.  Liu \textit{et al.} \cite{liu2023watermarking} injected watermarking into the prompt of LDMs and proposed two different methods, namely NAIVEWM and FIXEDWM. NAIVEWM activates the watermarking with a watermark-contained prompt. FIXEDWM enhances the stealthiness compared to NAIVEWM, as it can only activate the watermarking when the prompt contains a trigger at a predefined position. PromptCARE \cite{yao2024PromptCARE} is a practical a prompt watermark for prompt copyright protection. When unauthorized large models are trained using prompts, copyright owners can input the trigger to verify whether the output contains the specified watermark.

Zeng \textit{et al.} \cite{zeng2023securing} constructed a universal adversarial watermarking and injected it into an arbitrary pre-trained generative model via fine-tuning. The optimal universal watermarking can be found through adversarial learning against the watermarking detector. Practicably, the secured generative models can share the same watermarking detector,  eliminating the need for retraining the detector when it comes to new generators. { As the size of the generative model increases,  the design of model-specific watermarking will pay more attention to how to use a small number of samples to update a small number of parameters, thereby reducing resource consumption.}

\noindent
\textbf{Data-specific Watermarking:} This class of work \cite{ma2023generative, liu2023ti2net,cui2023diffusionshield,zhao2024proactive} inserts watermarks to the input data, and then the generative data retains the watermarks.  GenWatermark \cite{ma2023generative} adds a watermark to the original face image, preventing it from malicious manipulation. To enhance the retention of the watermark in the generated image, the generation process is incorporated into the learning of GenWatermark by fine-tuning the watermark detector.
To prevent copyright infringement arising from DMs,
DiffusionShield \cite{cui2023diffusionshield} injects the ownership information into the image. Owing to the uniformity of the watermarks and the joint optimization method, DiffusionShield enhances the reproducibility of the watermark in generated images and the ability to embed lengthy messages. 
Feng \textit{et al.} \cite{feng2023catch} proposed the concept watermarking that embeds identifiable information of users within the used concept. This allows tracking and holding accountable malicious users who abuse the concept.  Liu \textit{et al.} \cite{timbrewatermarking-ndss2024} introduced a  timbre watermarking with robustness and generalization. The timbre of the target individual can be embedded with a watermark. When subjected to voice cloning attacks, the watermark can be extracted to effectively protect timbre rights.


\begin{table*}[tp]
	\centering
	\caption{A summary of representative solutions for the controllability of generative data.}
	\label{tab:control}
	\begin{small}
		\begin{tabular}{ccccccccccccccccc}		
			\toprule
			&	Ref.&Year&Remarks&Limitations \\
			\midrule
			\multirow{7}{*}{\parbox{1cm}{\centering Access Control}}{} & \cite{yeh2021attack}& 2021 &Paradigm of nullifying generation &Lack of controllability\\
			& \cite{LYSBFZ23}& 2023 &Paradigm of setting new identity &Weak generalizability\\
			& \cite{10086559}& 2023&Double protection &Poor visual quality\\
			
			& \cite{shan2023glaze}& 2023  &User-friendly, feasible user study&Weak stability and security\\	
			& \cite{van2023anti} &  2023 &Personalized defense, ensemble models &Complex setting \\
			& \cite{liang2023adversarial} &  2023 &Theoretical framework, small perturbations&Weak robustness, inflexible\\
			&\cite{liu2024metacloak}&2024&Transferability, robustness&  Additional noise layers\\

			\midrule
			
			\multirow{10}{*}{\parbox{1cm}{\centering Traceabi-lity}}{}	& 	\cite{zhao2023recipe}  &2023&Pioneering watermarking for DMs &Low generation quality\\
			
			&\cite{fernandez2023stable}  & 2023 & Invisibility, strong stability&Limited capacity, inexplicable \\

			&\cite{10.1145/3581783.3612448}&2023& Flexible embedding, additional security&Lack of noise robustness\\
			
			&\cite{yao2024PromptCARE} &2024&  Harmlessness, robustness, stealthiness &Accuracy drop for
			extreme cases\\

			&\cite{yang2024gaussian}&2024& Provable performance-lossless& Reliance on DDIM inversion\\
			& \cite{zeng2023securing} &2023& Universality, carrying extra information&Limited capacity, weak security\\
			&\cite{ma2023generative} &2023&Subject-driven protection, practicability&Weak cross-model transferability\\

			& \cite{feng2023catch} &2023&Paradigm of concept watermarking&Limited robustness and security\\

			&\cite{timbrewatermarking-ndss2024} &2024&Robustness and generalization in  voice&Dependent on noise layers\\
			
			&\cite{liu2024blockchain} &2024& Trustworthy and reliable management&Resource-intensive, unpractical\\
			
			\bottomrule
		\end{tabular}
	\end{small}
\end{table*}

\subsubsection{Blockchain}
Distributed ledger-based blockchain can be used to explore a secure and reliable AIGC-generated content framework.

\begin{itemize}
	\item {\textit{Transparency:}  Blockchain can be used to enable transparent traceability of generative data. Each generative data can be recorded in a block in the blockchain and associated with the corresponding transaction or generation process. This enables users and regulators to understand the source and complete generation path of the generative data.}
	
	\item \textit{ Copyright protection:} Blockchain can provide a reliable mechanism for copyright protection of generative data. By recording copyright information on the blockchain, it can be ensured that generative data is associated with a specific copyright owner and is available for verification. This can reduce unauthorized use and infringement and provide content creators with evidence of copyright protection.
	\item \textit{Decentralized content distribution:} Generative data is stored in a distributed manner across the blockchain network, rather than centrally stored on a single server. This improves the availability and security of generative data and reduces the risk of single points of failure and data loss.	
	\item \textit{Rewards and incentives:} Through smart contracts, the blockchain can automatically distribute rewards for generative data and ensure a fair and transparent distribution mechanism. This can incentivize contributors to provide higher quality and more valuable generative content.	
\end{itemize}

Du \textit{et al.} \cite{liu2024blockchain}  proposed a blockchain-empowered framework to manage the lifecycle of AIGC-generated data. Firstly, a protocol to protect the ownership and copyright of AIGC is proposed, called Proof-of-AIGC, deregistering plagiarized generative data and protecting users' copyright. Then, they designed an incentive mechanism with one-way incentives and two-way guarantees to ensure the legal and timely execution of AIGC ownership exchange funds between anonymous users.   AIGC-Chain\cite{jiang2024aigc} carefully records the entire life cycle of AIGC products, providing a transparent and reliable platform for copyright management.

\subsection{Summary}  
In table \ref{tab:control}, we summarize the solutions for the controllability of generative data in AIGC.  We emphasize the discussion on the access control and traceability of generative data. By implementing them, we can protect the security and privacy of generative data, ensuring its credibility and reliability and providing robust support for the compliant use of the data. {In addition, adversarial attacks can also have an impact on the controllability  of generative data. On the one hand, adversarial attacks can attack the watermark remover to prevent the loss of controllable information, which achieves stable controllability of generative data. On the other hand, adversarial attacks can also attack the watermark extractor to prevent the extraction of controllable information, which exacerbates the controllability challenge  on generative data.}


\section{Authenticity on Generative Data}
\subsection{Threats to Authenticity}

Dramatic advances in generative models have made significant progress in generating realistic data, reducing the amount of expertise and effort required to generate fake content. However, this unrestricted accessibility raises concerns about the ubiquitous spread of misinformation. Fake images are particularly convincing due to their visual comprehensibility. As a result, malicious users can generate harmful content to manipulate public opinion, thereby negatively impacting social domains, e.g., politics and economics. For example, the fake tweet with the generated image of ``a large explosion near the Pentagon complex" went viral, fooling many authoritative media accounts into reprinting them and even causing the stock market to suffer a significant drop. Current state-of-the-art generative models already pose a greater threat to human visual perception and discrimination. When distinguishing between real images and generative images, the error rate of human observers  reaches 38.7\% \cite{lu2023seeing}. 

Typically, Deepfake \cite{mirsky2021creation}  possesses the capacity to fabricate visually realistic fake content by grafting the identity of one individual onto the image or video of another. Unfortunately, the open-source nature of the technology allows criminals to commit malicious forgeries without the need for significant expertise, thereby engendering a multitude of societal risks \cite{korshunov2018deepfakes,10024477}. For instance, this includes replacing the protagonist in pornographic videos with a celebrity face to affect the celebrity's reputation, faking videos of speeches of politicians to manipulate national politics, and faking an individual's facial features to pass authentication in assets management.

Many fake detection methods \cite{verdoliva2020media} have been proposed to detect modified data by AI. However, these methods still have vulnerabilities and limitations \cite{9423202}. On the one hand, some methods rely too heavily on traditional principles or pattern matching. They are difficult to capture the evolving new patterns of in-depth AIGC, which allows generative data to escape detection. On the other hand, existing methods have limited capabilities when dealing with new challenges under large models. Large models have higher generative power and creativity, making the generative data more difficult to distinguish.  A recent study \cite{pegoraro2023chatgpt}  provided insights into the various methods used to detect ChatGPT-generated text. The study highlighted the extraordinary ability of Chat-GPT spoofing detectors and further shows that most of the analyzed detectors tend to classify any text as human-written with an overall TNR as high as 90\% and a low TPR. Therefore, there is a requirement to continuously improve the existing detector to effectively deal with the problem of disinformation and misuse of generative data.

\subsection{Countermeasures to Authenticity}

\begin{figure}[!thp]
	\centering
	\includegraphics[width=3.5in, keepaspectratio]{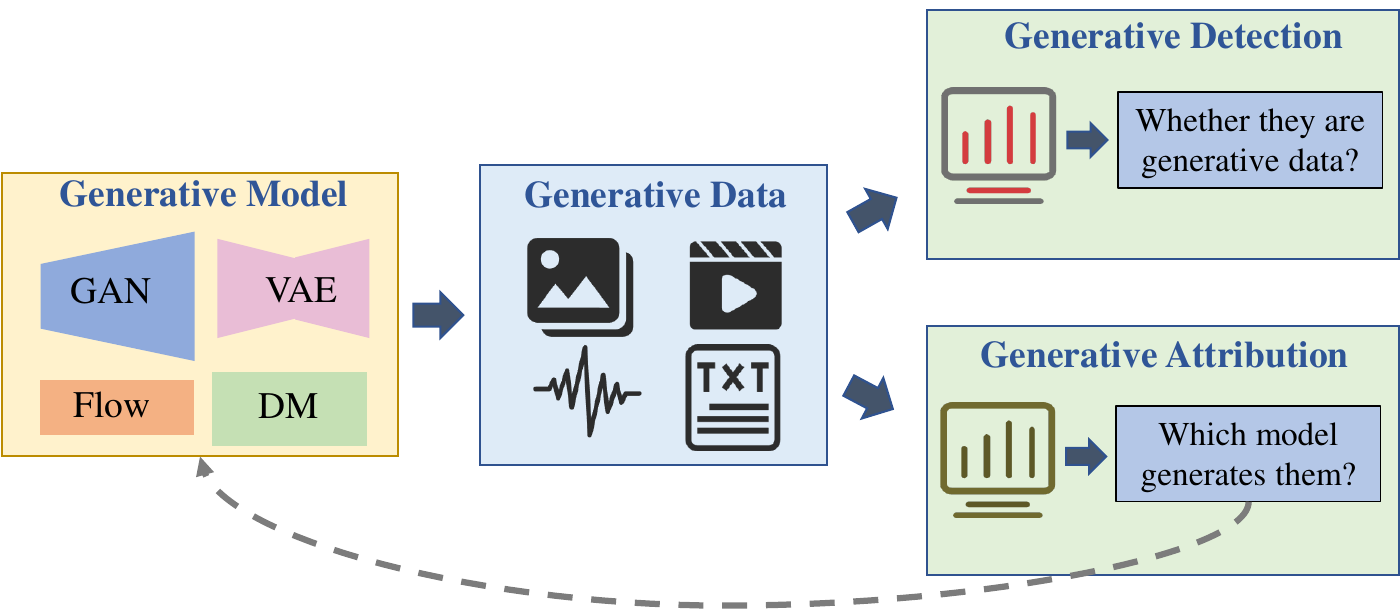}\\
	\caption{Detection and attribution of generative data. }
	\label{authentic}
\end{figure}

In Fig. \ref{authentic}, existing countermeasures \cite{lin2024detecting} mainly consider constructing a detector to distinguish between real data and generative data. Further, generative attribution can trace the generative data back to the model that generate it.

\subsubsection{Generative Detection}


\noindent
\textbf{Generative visual detection:} 
The presence of artifacts in generative images is an important detection cue, which may derive from defects in the generation process or from a specific generative architecture. Corvi \textit{\textit{et al.}} \cite{corvi2023detection} gave a preliminary trial to the problem of detecting generative images produced by DMs. Their study showed that the hidden artifact features of images of DMs are partially similar to those observed in images of GANs. {Both GANs and DMs  leave  artifacts on generative data,  but only the  artifacts are different. Due to the instability of the adversarial training process between the generator and the discriminator, GANs generate unnatural artifacts, e.g., blurred edges and inconsistent textures. While DMs retain small noise features during the stepwise denoising process, which can lead to more natural noise or detail distortion.} Xi \textit{et al.} \cite{xi2023ai} developed a robust dual-stream network consisting of a residual stream and a content stream to capture generic anomalies generated by AIGC. The residual stream utilizes the spatial rich model (SRM) to extract various texture information from images, while the content stream captures additional artifact traces at low frequencies, thus supplementing the residual stream with information that may have been missed.  Sinitsa \textit{et al.} \cite{sinitsa2024deep} presented a rule-based method that can achieve high detection accuracy by training a small number of generative images (less than 512). The method employs the inductive bias of CNNs to extract fingerprints of different generators from the training set and applies it to detect generative images of the same model and its fine-tuned versions. {Joslin et al. \cite{299603}  introduced human factors to enhance generative detection. Detecting AI-synthesized faces by combining attentional learning methods with user annotations. They also created a crowdsourcing annotation method to systematically gather various user annotations to identify suspicious areas and extract artifact patterns.}  {With the performance optimization that comes from a larger parametric model, the generative data will more closely mimic the original data while being able to circumvent artifacts, which increases the difficulty of generative detection.}


Analyzing distinctive features of generated images is also a viable approach to consider.  Interestingly, Wang \textit{et al.} \cite{wang2023dire} observed that the image generated by the DMs can be reconstructed by approximating the source model, while the real image cannot. Therefore, they proposed a novel image representation called diffusion reconstruction error (DIRE), which measures the distance between the input images and the reconstructed one. DIRE provides a reliable, simple, and generalized method to differentiate between real images and diffusion-generated images. SeDID \cite{ma2023exposing} leverage the unique properties of diffusion models, namely deterministic inverse and deterministic denoising computational error. In addition, its use of insights from member inference attacks to emphasize distributional differences between real and generative data enhances the understanding of the security and privacy implications of diffusion models. Zhong \textit{et al.} \cite{zhong2023rich} focused on the inter-pixel correlation contrast between rich and poor texture regions within an image, and presented a universal detector which can generalize to 
various AI models, including GAN-based and DM-based models. Existing state-of-the-art detectors have a  generalization of cross architectures, but the generalization of cross concepts is not considered. To this end, Dogoulis \textit{et al.} \cite{dogoulis2023improving} proposed a sampling strategy that takes into account the image quality scores of the sampled training data, and can effectively improve the detection performance in the cross-concept setting.

Innovatively, Bi \textit{et al.} \cite{bi2023detecting} explored the invariance of  real images, and proposed a method to map real images to a dense subspace in the feature space, while all generative images are projected outside this subspace. {In this way, it can effectively address longstanding issues in generative detection, e.g., poor generalization, high training costs, and weak interpretability.}

\noindent
\textbf{Generative text detection:} 
Metric-based detection extracts distinguishable features from the generative text. Early on, GLTR \cite{gehrmann2019gltr} was a tool to assist humans in detecting generated text.  It employs a set of baseline statistical methods that can detect generation artifacts in common sampling schemes. In a human subject study, the annotation scheme provided by GLTR improved human detection of fake text from 54\% to 72\% without any prior training.  
Mitchell \textit{et al.} \cite{mitchell2023detectgpt} noticed that the texts sampled from LLMs tend to occupy negative curvature regions of the model's log probability function.  Based on this, DetectGPT is proposed to set a new curvature-based criterion for detection without additional training. Tulchinskii \textit{et al.} \cite{tulchinskii2024intrinsic} proposed a new distinguishable representation, the intrinsic dimension. Fluent texts in natural languages have an average intrinsic dimension of 9 or 7 in each language, while AI-generated texts have a lower average intrinsic dimension of 1.5 in each language. Detectors constructed on the basis of intrinsic dimensionality have strong generalizability to models and scenarios.

Regarding the model-based methods \cite{guo2023close,chen2023gpt}, a classification
model is usually trained using a corpus. 	Guo \textit{et al.} \cite{guo2023close} proposed a text detector for ChatGPT. The detector is based on the RoBERTa model, which is trained by plain answer text and question-answer text pairs respectively. Chen \textit{et al.} \cite{chen2023gpt} trained two different text classification models using robustly optimized BERT pretraining approach (RoBERTa) and text-to-text Transfer Transformer (T5), respectively, and achieved significant performance on the test dataset with an accuracy of more than 97\%.

\begin{table*}[tp]
	\centering
	\caption{A summary of the representative solutions for the authenticity of generative data.}
	\label{tab:attribution}
	
	\begin{small}
		\begin{tabular}{ccccccccccccccccc}		
			\toprule
			&	Ref.&Year&Remarks&Limitations \\
			\midrule
			\multirow{12}{*}{\parbox{1cm}{\centering Genera-tive Detection}}{} & \cite{corvi2023detection} & 2023 &Pioneering detection for DMs &Lack of JPEG robustness\\
			
			& \cite{sinitsa2024deep}& 2024 &Low budget, multi-model applicable&Lack of blurring robustness&\\ 
			& \cite{wang2023dire}& 2023& Strong interpretability&High resource consumption\\
			
			& \cite{zhong2023rich}& 2023 & Universal to DMs and GANs&Poor results on special models\\
			
			& \cite{ma2023exposing}& 2023 & Use of DM  distinct property&Not applicable to GANs\\
			&\cite{dogoulis2023improving}& 2023&Generalizing cross-conceptes&Not high detection performance\\
			&\cite{bi2023detecting}& 2023 & High
			
			generalization, low costs&Imperfect JPEG robustness \\

			&\cite{gehrmann2019gltr} & 2019 &Training-free, numerically calculable&Unsatisfactory detection accuracy\\	
			&\cite{mitchell2023detectgpt} & 2023&Human-readable, high-efficiency&Strong white-box assumption\\	
			&\cite{tulchinskii2024intrinsic} & 2024 &Generalization and robustness& Failure to small-sample languages\\
			
			&\cite{guo2023close}&2023&Large-scale data, human evaluations&Resource-intensive, poorly explained\\
			&\cite{chen2023gpt} & 2023 & High accuracy, explicable&Not scalable, only for English\\		
			\midrule
			
			\multirow{8}{*}{\parbox{1cm}{\centering Genera-tive Attribution}}{}	&\cite{he2023mgtbench} & 2024 &Systematic quantification& Medium attribution performance \\
			&\cite{bui2022repmix}  & 2022 & Robust and practical
			attribution &Only ofr GANs, noise-irrobustness\\ 
			&\cite{yang2023progressive}&2023&Open-Set model attribution&Moderate versatility and scalability\\
			&\cite{sha2023fake}  & 2023 & Pioneering attribution for DMs&Lack of generalizability \\
			&\cite{lorenz2023detecting} & 2023 &Lightweight, superior performance&Difficult to scale to other models\\
			&\cite{guarnera2024level}  & 2024 &Hierarchical multi-level&Unrobust in compression and scaling \\
			&\cite{wang2023evaluating} &2023&Paradigm of data attribution&Lack of strict proof\\
			&\cite{Asnani_2024_CVPR}&2024&Paradigm of concept attribution&Requires predefined concepts\\
			
			\bottomrule
		\end{tabular}
	\end{small}
\end{table*}

\subsubsection{Generative Attribution}

He \textit{et al.} \cite{he2023mgtbench} extended current detectors to the potential of text attribution to recognize the source model of a given text. The results show that all these detectors have certain attribution capabilities and still have room for improvement. Moreover, model-based detectors can significantly outperform metric-based detectors.

For visual generative data, a lot of attribution works on GANs \cite{yu2019attributing,bui2022repmix,yang2022deepfake,girish2021towards, yang2023progressive} has been proposed. RepMix \cite{bui2022repmix} is a GAN-fingerprinting technique based on representation mixing and a novel loss. It is able to determine from which structure of GAN a given image is generated.  
POSE \cite{yang2023progressive} tackles an important challenge, i.e.,  open-set model attribution, which can simultaneously attribute images to seen and unseen models. POSE simulates open-set samples that keep the same semantics as closed-set samples but embed distinct traces.

\begin{figure*}[!t]
	\centering
	\includegraphics[width=4.4in, keepaspectratio]{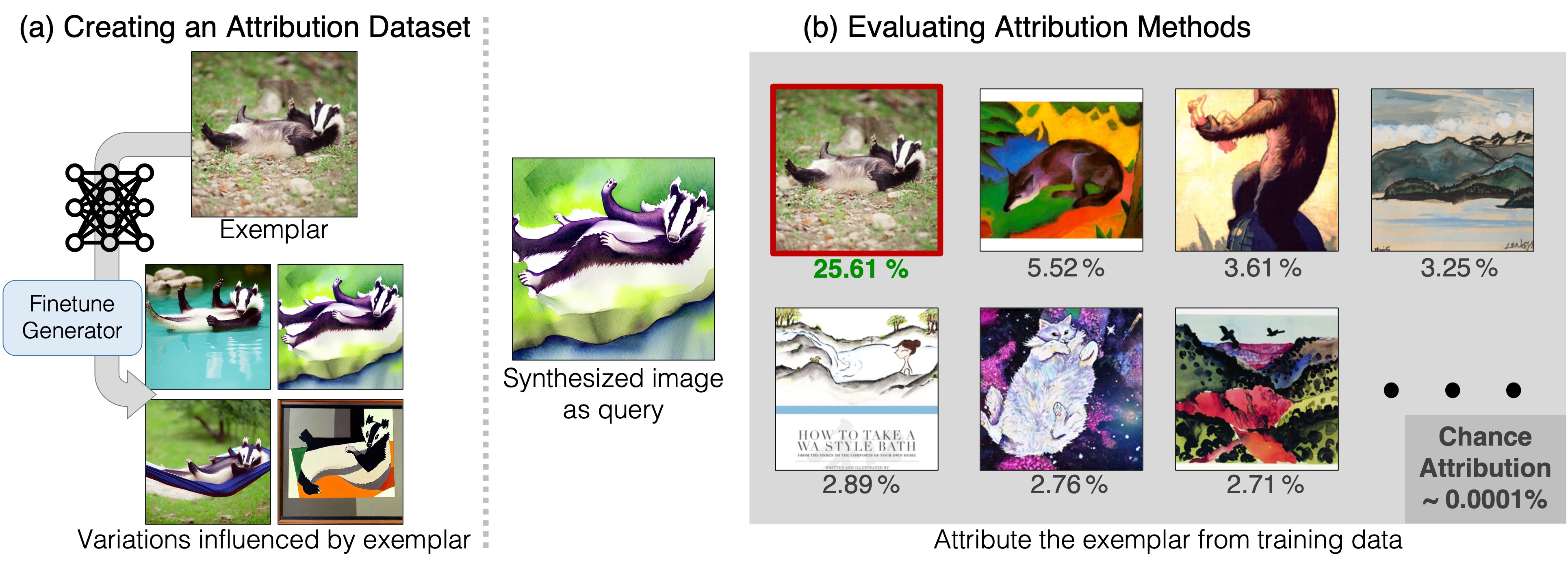}.\\
	\caption{Generative data attributes to the training data from \cite{wang2023evaluating}.}
	\label{attribution}
\end{figure*}

Recent works have begun to focus on DMs.
Sha \textit{et al.} \cite{sha2023fake} constructed a multi-class (instead of binary) classifier to attribute fake images generated by DMs. Experiments showed that attributing fake images to their originating models can be achieved effectively,  because different models leave unique fingerprints in their generated images.  Lorenz \textit{et al.} \cite{lorenz2023detecting} designed the multi-local intrinsic dimensionality (multiLID), which is effective in identifying the source diffusion model. Guarnera \textit{et al.} \cite{guarnera2024level} developed a novel multi-level hierarchical approach based on ResNet models, which can recognize the specific AI architectures (GANs/DMs). The experimental results demonstrate the effectiveness of the proposed approach, with an average accuracy of more than 97\%.

Intriguingly, a new work  \cite{wang2023evaluating} can attribute generative data to the training data rather than the source model, necessitating the identification of a subset of training images that contribute most significantly to the generated data. In Fig. \ref{attribution}, we can query the generated data in the training set and evaluate their similarity, which can contribute to protecting the copyright of training data rather than models.

\subsection{Summary}  
In table \ref{tab:attribution}, we summarize the solutions for the authenticity of generative data.  Generative data contains traces (referred to as fingerprints) left by generative models, allowing researchers to detect and attribute the data based on these fingerprints.
However, as generative models undergo iterative optimization, these fingerprints also continuously evolve. Therefore, new detection methods must be updated in real time. In contrast, real data exhibits certain invariable features that remain unaltered over time or under varying circumstances. Future detection methods should be designed to effectively harness these invariable features, thereby enhancing the accuracy and robustness of detecting generated data. Furthermore, watermark-based methods have demonstrated notable potential in enhancing tasks related to detection and attribution. However, it's important to note that such methods constitute an active defense strategy, necessitating preprocessing during data generation. In real-world scenarios, constraining adversaries to exclusively employ watermarked generative data may be not feasible. {In addition, adversarial attacks can also have an impact on the controllability  of generative data. On the one hand, adversarial attacks can attack generative models to prevent them from generating inauthentic content, which ensures the authenticity of generative data at the source.   On the other hand, adversarial attacks can  attack generative detectors to obfuscate their decisions, which exacerbates the authenticity challenge on generative data.}

\section{Compaliance on Generative Data} \label{Compaliance}
\subsection{Requirements to Compaliance}
The compliance of generative data refers to the requirement that such data must adhere to applicable laws, regulations, and ethical standards.  With the rapid development and extensive utilization of AIGC technology, the compliance of generative data has become an important topic, encompassing various aspects, e.g., ethics, bias, and politics.

Countries and organizations around the world have initiated investigations and issued relevant policies and regulations regarding the regulation of generative data. The United States' Blueprint for an AI Bill of Rights  emphasizes generative data to ensure fairness, privacy protection, and accountability.  European Parliament passed the Artificial Intelligence Act, which supplements the regulatory regime for generative models and requires that all generative data should be disclosed as \textit{derived from AI}. 
China proposed an AIGC-specific regulation, i.e., the Interim Regulation on the Management of Generative Artificial Intelligence  Services \cite{interim}. This regulation encourages innovation of AIGC but requires a prohibition on the generation of toxicity, e.g.,  violence, bias, and obscene pornography,  as well as an increase in the factuality of the generative data to avoid misleading the public.

\begin{figure*}[!htp]
	\centering
	\includegraphics[width=5.5in, keepaspectratio]{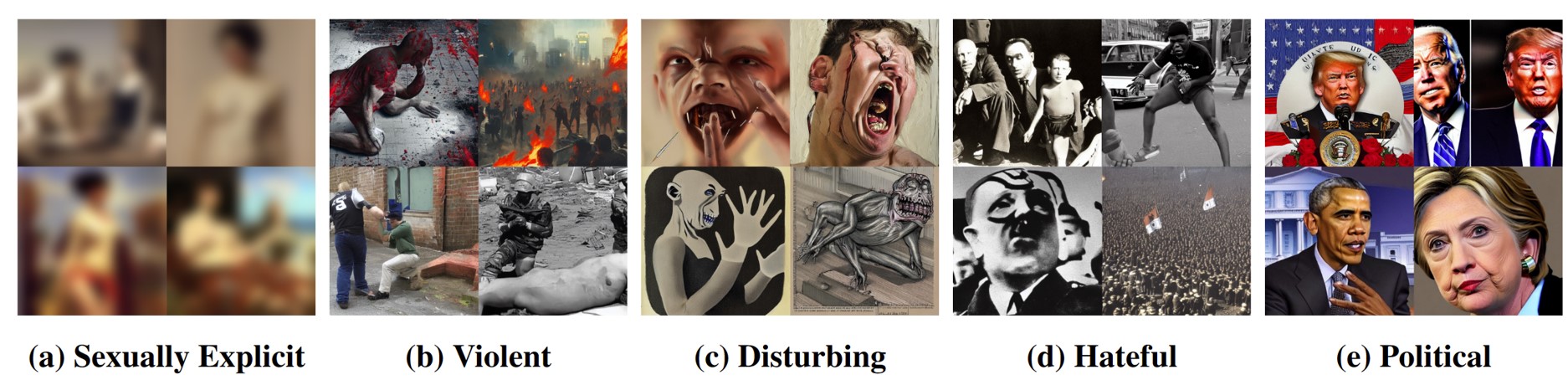}\\
	\caption{Examples of generative images with toxicity from \cite{qu2023unsafe}. }
	\label{toxicity}
\end{figure*}

\subsubsection{Non-toxicity} {Toxicity presents in generative data involves incongruence with human values or bias directed at particular groups, which has the potential to harm societal cohesion and intensify divisions among different groups.
	Since the training of  AIGC models is based on a large amount of unintervened data, toxicity in training data \cite{birhane2021multimodal} directly leads to the corresponding toxicity in generative data, which covers a variety of harmful topics, e.g., sexuality, hatred \cite{birhane2024into}, politicalization, violentism, and race bias.} Some toxic generative examples are shown in Fig. \ref{toxicity}. 

Some works \cite{caliskan2017semantics,sheng2019woman} have found stronger associations between males and occupations in language models, verifying the gender bias in generative data. 	The investigations revealed that GPT-3 consistently and strongly exhibits biased views against the Muslim community \cite{abid2021persistent}. Stable Diffusion v1 was trained on the LAION-2B dataset, which contains images described only in English, making generative data biased towards white culture . Likewise, it was observed that DALLA-E displayed unfavorable biases towards minority groups.


{Unlike  GANs that only use random noise to generate data, DMs can be guided by additional textual prompts, which increases the risk of generating non-compliant content. As a result, researchers primarily focus on the compliance of data generated by DMs.} Qu \textit{et. al} \cite{qu2023unsafe} provided a comprehensive safety assessment concerning the generation of toxic images, particularly hateful memes from diffusion models.  To quantitatively assess the safety of generative images, a safety classifier is developed to identify toxic images based on the predetermined criteria for unsafe content. Their findings indicated that the utilization of harmful prompts resulted in diffusion models producing a significant quantity of toxic images. Additionally, even when prompts are innocuous, the potential for generating toxic images persists. Overall,  the danger of
large-scale generation of toxic images is imminent.

\subsubsection{Factuality}
{AI-generated data may be contrary to the facts \cite{wang2023survey},  which is harmful to the public through misleading cognition. For example, ChatGPT may produce responses that sound reasonable and authoritative but are factually incorrect or nonsensical.  Even worse, AIGC often explains its generated responses. When AIGC fails to provide accurate responses to the queries, it not only delivers incorrect information but also supplements seemingly plausible explanations. This enhances the inclination of users to place greater trust in these erroneous contents. The United States news credibility assessment and research organization, NewsGuard, conducted a test on ChatGPT  \cite{nytimes2023}. Researchers posed questions to ChatGPT containing conspiracy theories and misleading narratives and found that it could adapt information within seconds, generating a substantial amount of persuasive yet unattributed content.    }

{When used in important domains, such unfactual generative data will bring serious harmful impacts \cite{bender2021dangers}. In the healthcare domain, medical diagnosis requires interpretable and correct information. Once AI-generated diagnostic advice is factually incorrect, it will cause irreparable harm to the patient's life and health. In the journalism domain, news that distorts the facts will mislead the public and undermine the credibility of the media. In the education domain, the dissemination of incorrect knowledge to students will confuse their minds, thus seriously hampering their academic growth and cognitive development.} 

\subsection{Countermeasures to Compaliance}
\subsubsection{Countermeasures to Non-toxicity}

Efforts to eliminate toxicity can be divided into four categories. The first one is dataset filtering. A non-toxic training dataset is key to ensuring the security of generative data.  Some works \cite{schuhmann2022laion, DALL-E,henderson2022pile} have implemented comprehensive processes to filter data contained toxic. OpenAI ensures that any violent or sexual content is removed from DALLA-E2 by carefully filtering \cite{DALL-E}. Henderson \textit{et. al} \cite{henderson2022pile}  demonstrated how to extract implicit sanitization rules from the Pile of Law, providing researchers with a pathway to develop more sophisticated data filtering mechanisms. However, large-scale dataset filtering also has unexpected side effects on the downstream performance \cite{nichol2021glide}.


The second one is generation guidance.  Ganguli \textit{et al.} \cite{ganguli2022red} identified and attempted to reduce the potentially harmful output of language models in a confrontational manner. They found that the reinforcement learning from human feedback is increasingly difficult to red team as they scale, and a flat trend with scale for the other model types.
Brack  \textit{et al.} \cite{brack2023mitigating} investigated to instruct effectively diffusion models to suppress inappropriate content using the learned knowledge obtained about the world’s ugliness, thus producing safer and more socially responsible content. Similarly, safe latent diffusion (SLD) \cite{schramowski2023safe} extends the generative process via utilizing toxic prompts to guide the safe generation in an opposing direction. 

The third one is model fine-tuning. Recently, a new term called the \textit{ablation concept} or \textit{ablation forgetting} \cite{gandikota2023erasing, kumari2023ablating,heng2024selective} has brought a novel direction to the elimination of toxic content in generative data.  Gandikota \textit{et. al} \cite{gandikota2023erasing} studied the erasure of toxic concepts from diffusion model weights via model fine-tuning. The proposed method utilizes an appropriate style as a teacher to guide the ablation of the toxic concepts, e.g., sexuality and copyright infringement. Selective Amnesia \cite{heng2024selective} is a generalized continuous learning framework for concept ablation that applies to different model types and conditional scenarios. It also allows for controlled ablation of concepts that can be specified by the user. However, the ablation concept's ability to explain the various definitions of toxic concepts remains limited. Inspired with social psychological principles, Xu et al.  \cite{xu2024walking} proposed a novel strategy to motivate LLM to integrate different human perspectives and self-regulate their responses. 


Lastly, filtering the generated results is also a viable option. Stable Diffusion includes an after-the-fact safety filter  to block toxic images. Unfortunately, the filter easily blocks any generated image that is too close to at least one of the 17 predefined \textit{sensitive concepts}. Rando et. al \cite{rando2022red} reverse-engineered this filter and then proposed a manual strategy that enables content not related to sensitive concepts to bypass the filter. In addition,  existing toxicity detectors \cite{markov2023holistic,lu2023hate} for real data may be able to be updated to be compatible with generative data. This self-correcting mechanism can significantly reduce toxicity and bias.

\subsubsection{Countermeasures to Factuality} In order to constrain the non-factualness caused by AIGC ``lying", Evans \textit{et al.} \cite{evans2021truthful} first identified clear standards for AI truthfulness and explored potential ways to establish them.  
\begin{table*}[tp]
	\centering
	\caption{A summary of the representative solutions for the compliance of generative data.}
	\label{tab:compliance}
	
	\begin{small}
		\begin{tabular}{ccccccccccccccccc}		
			\toprule
			&	Ref.&Year&Remarks&Limitations \\
			\midrule
			\multirow{7}{*}{\parbox{0.9cm}{\centering Non-toxicity}}{} & \cite{henderson2022pile} & 2022 &Legally guaranteed & Only for U.S. texts, perturbation-sensitive\\
			
			&\cite{ganguli2022red}&2022 &The idea of constant confrontation&Time-consuming, difficult to scale&\\

			&\cite{rando2022red}&2022&Comprehensive toxicity concept& Takes a lot of labor, subjective definition\\
			
			&\cite{brack2023mitigating} &2023&World knowledge guidance&Need to modify the model\\
			&\cite{schramowski2023safe}&2023&Free-train, maintain quality&Lack of provability and ethical  concern\\

			&\cite{heng2024selective}&2024&Controllable forgetting&Computing costly, non-automated\\
			
			&\cite{xu2024walking}&2024&Self-regulate, low cost&Ignorance of sensitive vocabularies\\

			\midrule
			\multirow{7}{*}{\parbox{0.9cm}{\centering Factuality
			}}

			&\cite{lee2022factuality}&2022&Open-ended test, new metrics&Lack of moral consideration\\
			&\cite{alaa2022faithful}&2022&Three-dimensional metric& Divergence measures collapse  \\
			&\cite{azaria2023internal}&2023&Simple and powerful&Difficult to interpret\\
			
			&\cite{du2023improving}&2023&Comprehensible, reasonable& Large resource consumption\\
			&\cite{gou2023critic}&2024&Practical and simple & Limited to syntactic correctness\\
			&\cite{chen2024gaining}&2024&Without extra
			models or human&Difficulty in controlling mistake\\		
			&\cite{zhang-etal-2024-r}&2024& Learned refusal ability &Lack of rigorous evaluation\\

			\bottomrule
		\end{tabular}
	\end{small}
\end{table*}

A reasonable assessment of content factuality is the critical step toward responsible generative data. Goodrich \textit{et al.} \cite{goodrich2019assessing} constructed relation classifiers and fact extraction models based on Wikipedia and Wikidata, by which the factual accuracy of generated text can be measured.  Lee \textit{et al.} \cite{lee2022factuality}  proposed a novel training method to enhance factuality by utilizing TOPIC PREFIX for better perception of facts and sentence completion as the training objective, which can significantly decrease the number of counterfactuals. { They also study the factual accuracy of LLMs with parameter sizes ranging from 126M to 530B and find that the larger the model, the higher the accuracy. This is because the model has a large enough capacity to learn more generalized knowledge, thus reducing the occurrence of counterfactuals.}

Alaa \textit{et al}. \cite{alaa2022faithful} designed a three-dimensional metric capable of characterizing the fidelity, diversity, and generalization of generative data from widely generative models in a wide range of applications. SAPLMA \cite{azaria2023internal} is a simple but powerful method, which only 
uses the hidden layer activation of LLM to discriminate the factuality of generated statements.

Interestingly, Du \textit{et al.} \cite{du2023improving} prompted multiple language models to debate their viewpoints and reasoning processes over multiple rounds, and finally come up with a unitive answer. The results indicate that it enhances mathematical and strategic reasoning in the task while reducing the fallacious answers and illusions that modern models are prone to.  CRITIC \cite{gou2023critic} requires LLMs to interact with appropriate tools for feedback learning, such as using a search engine for fact-checking or a code interpreter for debugging. The output of the LLM is modified incrementally by the evaluation results of the feedback by the tools. Chen et al. \cite{chen2024gaining} presented a new alignment framework designed to improve LLMs by converting flawed instruction-response pairs into useful alignment data through mistake analysis. By generating harmful responses and analyzing their own mistakes, the framework can  improve the alignment with human values. To prevent LLMs from rambling without knowing the answer, Zhang et al. \cite{zhang-etal-2024-r} taught LLMs the ability to refuse to answer. Specifically, they constructed datasets of refusal perception, and then adapted the model to avoid answering questions that were beyond its parametric knowledge.


\subsection{Summary}  
In table \ref{tab:compliance}, we summarize the solutions for toxicity and factuality in generative data.  The first problem to be solved for the compliance of generative data is to define its standards. In addition, there should be different compliance standards for different application scenarios, rather than a blanket denial of generative content. For example, ``a flying pig" may not be factual for a language model, but it is more creative for a visual model. After that, the generated content of AI can be made compliant with the specification within the constraints of the standard. {In addition, adversarial attacks can also have an impact on the controllability  of generative data. On the one hand, adversarial attacks can attack the generative model to prevent it from learning the non-compliant content in the real data, which enhances the compliance of  generative data.  On the other hand, adversarial attacks can also attack compliance detectors to evade their detection, which exacerbates the compliance challenge for generative data.}

In conclusion, we emphasize the examination of toxicity and factuality issues within the context of compliance to ensure the responsible and ethical use of AI-generated content across various domains. Nevertheless, existing approaches exhibit certain limitations, nec	essitating continuous efforts by researchers to address these shortcomings and better align with the practical requirements of real-world applications.

\section{  Benchmark and Statistical Analysis}  \label{Benchmarks}

\subsection{Benchmark}

\subsubsection{Benchmark Dataset}

{ The construction of generative datasets facilitates the design and evaluation of countermeasures, helping researchers to identify problems and improve techniques,  thus advancing the progress of trustworthy generative data. Researchers have now released diverse datasets for utilization.}

{For text data,    most datasets are domain-specific, e.g., Student Essays \cite{verma2023ghostbuster} for essays, TuringBench \cite{uchendu2021turingbench} for news, GPABenchmark \cite{liu2023check} for academic writing, SynSciPass \cite{rosati2022synscipass} for scientific text, TweepFake and \cite{fagni2021tweepfake} for tweets.  Representatively, HC3 \cite{guo2023close} is a comprehensive ChatGPT generative text that contains tens of thousands of responses with human experts and covers a wide range of domains such as finance, healthcare, law, etc. The pioneering contributions of HC3 also make it a valuable research resource. More practically, MixSet \cite{zhang2024llm} is the first mixed-text  dataset involving both AIGC and human-generated content, covering a wide range of operations in real-world scenarios and bridging a gap in previous datasets.}

{
	For visual data,  only some datasets  are  domain-specific mainly in faces such as IDiff-Face \cite{boutros2023idiff} and GANDiffFace \cite{melzi2023gandiffface}, and  most datasets contain general images such as CIFAKE \cite{bird2024cifake} and AutoSplice \cite{jia2023autosplice}.  While, they have different limitations, e.g., targeting only a certain class of images or generators, and containing only a small amount of data. Subsequently, several million-scale datasets are released like GenImage \cite{zhu2024genimage} and ArtiFact \cite{rahman2023artifact},  which have the richness of image content and adopt state-of-the-art generators. To the best of our knowledge, DiffusionDB \cite{wang2022diffusiondb} represents the largest-scale visual generative dataset to date. It comprises 14 million images generated only by Stable Diffusion. This unprecedented scale and diversity offer exciting research opportunities for the study of generative image protection.}

\subsubsection{Benchmark Evaluation}
{The construction of benchmark evaluation provides a standardized baseline, which evaluates the effectiveness and innovativeness of new methods by enabling different methods to be fairly compared under the same testing conditions. Meanwhile, it can ensure the reproducibility of research and clearly identify the shortcomings of existing techniques, thus promoting technological progress. We  identified the corresponding benchmarks for different information security properties.}

{For privacy, PrivLM-Bench \cite{li-etal-2024-privlm} is a multi-perspective privacy evaluation benchmark that empirically and intuitively quantifies the privacy leakage of LLMs and reveals the neglected privacy of inference data in actual usage. Wu et al. \cite{294566} provided the first comprehensive privacy assessment of prompts learned via visual prompt learning from the perspectives of attribute inference and membership inference attacks. }

{For controllability, WaterBench \cite{tu2023waterbench} is the first comprehensive benchmark for watermarking LLMs, which encompasses 9 tasks and evaluates 4 open-source watermarking technologies. WAVES \cite{pmlrv235an24a} is a comprehensive watermarking benchmark, which establishes a standardized evaluation protocol consisting of various pressure tests and covers advanced image distortion attacks.}

{For authenticity,  DeepfakeBench \cite{DeepfakeBench_YAN_NEURIPS2023} introduces the first all-encompassing benchmark for detecting deepfakes of generative data, addressing the problem of inconsistent standards and lack of uniformity in this area. Lu et al. \cite{NEURIPS2023_505df5ea} comprehensively evaluated techniques used for generative  detection and also measured the ability of human vision to discriminate between generative data.}

{For compliance, GenderCARE \cite{tang2024gendercare} is a comprehensive framework that includes innovative standards, bias assessments, mitigation techniques, and evaluation metrics for quantifying and alleviating gender bias in LLMs. FELM \cite{NEURIPS2023_8b8a7960} is a factual evaluation benchmark  for generative texts. Through manual annotation of error types, it alerts users to potential errors and guides the development of more reliable large language models.}

\subsection{Statistical Analysis}
\begin{figure*}[!th]
	\centering
	\includegraphics[width=5.5in, keepaspectratio]{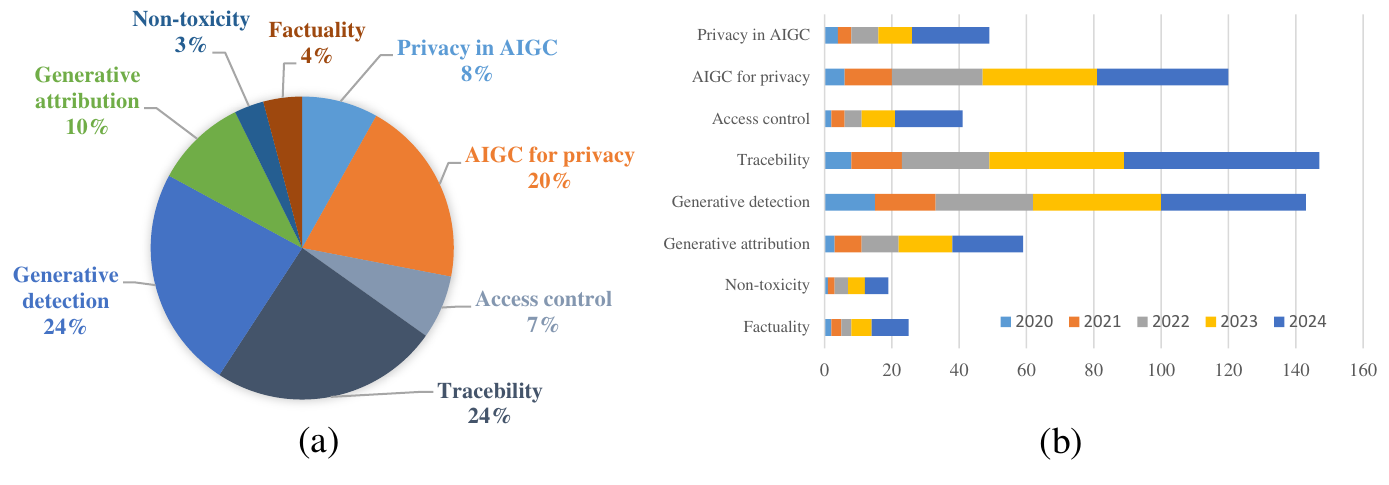}.\\
	\caption{Statistical charts. (a) Research articles for privacy, controllability, authenticity, and compliance  (2020-July 2024); (b) Year wise research articles (2020-July 2024).}
	\label{static}
\end{figure*}

{ As shown in Fig. \ref{static}, we have statistically analyzed the research articles with high relevance to generative data security and privacy in the last five years (2020-July 2024). We observe that most researchers focus on traceability and generative detection. The main reason is that generative data can be disastrous when used for malicious purposes, so it is urgent to come up with appropriate techniques to satisfy controllability and authenticity. At the same time, national strategies are guiding researchers to focus on the controllability of generative data. In addition, forensic and watermarking techniques for traditional tampering techniques have a high potential to be transferred to issues related to generative data.}

{AIGC for privacy is also an interesting and hot research topic. The generative data can be virtual data to effectively replace sensitive data in human decision-making, thus avoiding the leakage of personal privacy. Modern applications like telemedicine, recommendation algorithms, smart cities, etc. can safely provide convenient services to society with the help of generative data.}

{In contrast, non-toxicity and factuality only account for a small percentage. In particular, existing compliance studies focus more on generative text data and rarely examine the non-toxicity and factuality of generative visual data. In addition, this compliance research involves multiple fields such as sociology, management, law, and computer science, which makes it difficult to rely on domain-specific experts for governance. With the guidance of the policy, we believe that compliance-related work will account for more and more}.

\section{ Challenges and Future directions}  \label{Directions}

\subsection{Privacy}

\subsubsection{ Provable Private Content Removal Based on Information Theory} 
{Existing privacy protection technologies attempt to remove potentially private content in generative data via machine unlearning or concept forgetting. However, such private content  removal  is not provable. The parameters of the generative model may still conceal the  private content, which can be revealed under specific  prompts with backdoors.  While differential privacy can provide provable privacy guarantees, it requires retraining the model and sacrificing unacceptable  usability. Therefore, it is  necessary to explore new provable solutions to guarantee privacy. Exploring information theory-based methods may be a desirable research direction, which can directly limit the amount of retained information after private content is removed. Of course, how to measure the amount of information from generative models at this point is also a challenging problem.}

\subsubsection{Semantic-Level Quantitative  Evaluation for  Privacy and Utility}
{To compare the performance of different privacy-preserving technologies, it is crucial to develop quantitative evaluation metrics to measure privacy and utility before and after private content removal. However, constructing such fine-grained evaluation metrics requires systematically analyzing the learning processes and memory mechanisms of different generative models, which is difficult. In addition, compared to text data, the amount of information in visual data is multiple and complex (a picture is worth a thousand words) thus making it more difficult to measure privacy and utility. Considering that visual data is more concerned with semantic-level information, in the future, it is meaningful to design the metrics oriented to the semantic level, which can effectively guide generative data to achieve a balance between utility and privacy. In addition, since the semantics represented by privacy and utility are not always the same in different scenarios, it will be more practical and challenging to make quantitative metrics generic or flexible.}

\subsubsection{Exploratory Solutions to the Privacy Onion Effect}
{The privacy onion effect refers to the fact that when the most privacy-vulnerable outlier layer is removed, the privacy risk of the previously secure layer becomes the greatest. The existence of this effect can have a variety of serious consequences, in particular, it indicates that privacy-enhancing technologies may actually compromise the privacy of other users. However, there are very few existing exploration studies targeting this effect. The reason that this effect brings about privacy leakage originates from the idea that mean data points are rarely compromised, but outlier samples are often memorized. Therefore, it may be an intuitive solution to consider clustering the dataset to eliminate outlier points, but this perhaps reduces the diversity of the data. In this way, it is an important problem to maintain dataset utility while deleting reasonable data.}

\subsection{Controllability}

\subsubsection{Fine-grained Access Control of Generative Data}
{Existing work simply adds adversarial perturbations to block access to the expected generative data.  However, in most cases, users just want to prohibit generative data from containing specific semantics while allowing other semantics to exist. For example, an image containing the face of a celebrity is accessible, but when it is changed in gender or uglified it is not accessible. Existing work prohibits generative data with all semantics, which limits its practicality.  Furthermore, users with  roles can be restricted from accessing specific data, such as teenagers being prohibited from accessing essay writing type data but allowing adults to access it. Therefore, it is also a promising direction to provide fine-grained access control for generative data, which allows for flexible usage. To achieve it, a deep understanding of its generative process and digging into its intrinsic mechanism  are necessary. }

\subsubsection{ Robust Watermarking with High-Capacity for Text Data}
{Unlike visual data which is easy to achieve robust watermarking with high capacity, text data faces important challenges: Firstly, text data is significantly sparse, for example, the maximum token limit in GPT-4 is 8.2k which is much smaller than the watermarking capacity that can be embedded in 512-pixel images. Secondly, the semantics of text data are fragile as subtle changes may confuse or compromise its semantics, whereas minor changes in images can maintain consistent semantics. Some multi-bit watermarking can increase the watermarking capacity to some extent, but it also significantly reduces the robustness. Therefore, in future research, it is essential to consider designing new text watermarking paradigms to effectively balance the high capacity and strong robustness of text watermarking.}

\subsection{Authenticity}
\subsubsection{Generalizable Generative Detection based on Invariance of Real Data} 

{Existing detection methods attempt to find the decision boundary between real and generative data by a specific generative model, but it is difficult to generalize to new generative models. Detection performance degrades significantly when encountering data generated by generative models not seen in the training set. To improve generalization, the detector needs to be fine-tuned to allow capturing the decision boundaries of new generative models. However, the decision boundaries captured by the detector are intricate as the generative model iteratively changes.  Since real data does not change massively over time, its inherent invariances \cite{qi2024hierarchical} make it a reliable clue for generative detection. These invariances may include signal distribution, noise distribution, texture features of the data, etc. Exploring new invariants or joining multiple invariants is an important future direction. Meanwhile, interpretable analysis of these invariants will also make generative detection more credible and thus be able to be used for judicial forensics.}

\subsubsection{Tamper Detection of Generative Data}
{Similar to the real data, with the widespread use of generative data, it is becoming increasingly important to accurately detect whether the generative data has been tampered. However, current existing generative detection methods are challenging to do. The main reason is that generative data is itself generated by AI and also tampered with in an AI manner,  making it difficult to judge modifications based on the fingerprints of the AI models. To address this issue, one can consider embedding semi-fragile watermarks for generative data, thus ensuring its integrity. However, it is unrealistic for all users to uniformly adopt such post-hoc  operations. Therefore, there is still a requirement to propose effective techniques to detect the tampered generative data directly. To do this, researchers need to conduct a variability analysis of AI-generated models and AI-tampered models. }

\subsubsection{Towards User-friendly Generative Detection}
{When it comes to important decisions, machine vision-based generative detection techniques only provide initial judgments, while users still need to make further judgments by themselves. However, existing technology does not yet provide a directly interpretable, trustworthy, and easy-to-handle detection modality which makes it difficult for users to make rational decisions conveniently. Therefore, the development of interpretable, trustworthy, and easy-to-handle detection tools will be able to attract more social attention and popularization, thus promoting the friendly usage of detection techniques for disadvantaged groups, including unskilled users and the elderly. Conducting relevant user research analysis and incorporating explainable AI will help build user-friendly generative detection.}

\subsection{Compliance}
\subsubsection{	Quantification and Assessment of Compliance}
{ While existing laws and regulations already impose requirements for compliance with generative data, how to quantify and assess these requirements remains a challenging question. Firstly, compliance requirements are abstract and broad. Laws and regulations usually stipulate a series of principles and standards, but do not give specific metrics and standards. Therefore, it requires an in-depth understanding of the connotations of the legal provisions and their concretization in the context of the actual situation. Secondly, the type of generative data is complex, which covers a variety of multimodal and cross-modal contents. The automated methods for quantification involve the cross-application of fields, which requires the development of efficient multimodal fusion algorithms to realize it. Finally, there are differences in laws and regulations in different countries and regions. Compliance assessment of generative data needs to consider and cater to different geographical and cultural contexts, which adds to the complexity of the assessment.                }

\subsubsection{  Human Intelligence-Guided  Detection for Hidden Toxicity}{ Toxic content hiding in generative data often remains imperceptible to human perception and toxicity detectors. However, it might be uncovered by the special extractor of attackers, thereby giving rise to potential hazards. This situation may stem from the deep embedding of toxic information or the utilization of advanced information-hiding techniques to evade conventional detectors. Such hidden toxic content carries an undeniable risk that could disseminate widely across platforms, e.g., social media, news reporting, and virtual communities, consequently precipitating a range of societal issues. To counter this, human intelligence has advanced reasoning capabilities and is able to detect potentially toxic content by recognizing language, context, and metaphor. Of course, relying solely on human intelligence is extremely time-consuming and thus impractical. Future work should focus on human-intelligence-guided toxicity detection, which aims to find more stealthy toxicity using less human labor.}

\section{Conclusion} \label{Conclusion}
The rapid growth of AIGC has made data creation  easier. Many generative data flood the cyberspace, which  poses security and privacy issues. This survey comprehensively discusses  the security and privacy issues on generative data and reviews the corresponding solutions. Firstly, We  show the  process of AIGC and point out the security and privacy from the perspective of the fundamental properties of information security. After that, we reveal the successful experiences of state-of-the-art protection measures in terms of the foundational properties of privacy, controllability, authenticity, and compliance, respectively. Finally, we discuss possible future  directions in this area.

{Generative data now plays a significant and positive role in a variety of fields. Several surveys have also provided insights from different perspectives for the wide-scale application of generative data, including  stable data transmission \cite{saleh2024messagebrokersgenerativeai}, mobile network deployment \cite{xu2024unleashing}, semantic communications \cite{liu2024semantic}, and even personalized healthcare \cite{chen2024revolution}. Our survey provides guidance on security and privacy for all existing applications of generative data. Firstly, for the designers of the AIGC model, we expect them to pay attention to the existing security and privacy issues and then revise the AIGC model in terms of privacy, controllability, authenticity, and compliance. Secondly, for users of the AIGC model, we expect them to use it in a way that avoids compromising individual privacy while prohibiting the generation of potentially non-compliant data. We hope that this survey can provide new ideas on the security and privacy on generative data, promoting the application of trustworthy generative data.}

\bibliographystyle{ACM-Reference-Format}
\bibliography{sample-base}


\begin{thebibliography}{167}


\ifx \showCODEN    \undefined \def \showCODEN     #1{\unskip}     \fi
\ifx \showDOI      \undefined \def \showDOI       #1{#1}\fi
\ifx \showISBNx    \undefined \def \showISBNx     #1{\unskip}     \fi
\ifx \showISBNxiii \undefined \def \showISBNxiii  #1{\unskip}     \fi
\ifx \showISSN     \undefined \def \showISSN      #1{\unskip}     \fi
\ifx \showLCCN     \undefined \def \showLCCN      #1{\unskip}     \fi
\ifx \shownote     \undefined \def \shownote      #1{#1}          \fi
\ifx \showarticletitle \undefined \def \showarticletitle #1{#1}   \fi
\ifx \showURL      \undefined \def \showURL       {\relax}        \fi
\providecommand\bibfield[2]{#2}
\providecommand\bibinfo[2]{#2}
\providecommand\natexlab[1]{#1}
\providecommand\showeprint[2][]{arXiv:#2}

\bibitem[Bea(2022)]%
        {Beaumont}
 \bibinfo{year}{2022}\natexlab{}.
\newblock \bibinfo{title}{Clip retrieval system}.
\newblock
  \bibinfo{howpublished}{\url{https://rom1504:github:io/clip-retrieval/}}.
\newblock


\bibitem[DAL(2022)]%
        {DALL-E}
 \bibinfo{year}{2022}\natexlab{}.
\newblock \bibinfo{title}{DALL·E 2 pre-training mitigations}.
\newblock
  \bibinfo{howpublished}{\url{https://openai.com/research/dall-e-2-pre-training-mitigations}}.
\newblock


\bibitem[lor(2023)]%
        {lorenz2023detecting}
 \bibinfo{year}{2023}\natexlab{}.
\newblock \showarticletitle{Detecting Images Generated by Deep Diffusion Models
  Using Their Local Intrinsic Dimensionality}. In
  \bibinfo{booktitle}{\emph{Proceedings of the IEEE/CVF International
  Conference on Computer Vision (ICCV) Workshops}}. \bibinfo{pages}{448--459}.
\newblock


\bibitem[nyt(2023)]%
        {nytimes2023}
 \bibinfo{year}{2023}\natexlab{}.
\newblock \bibinfo{title}{Disinformation Researchers Raise Alarms About A.I.
  Chatbots}.
\newblock
  \bibinfo{howpublished}{\url{https://www.nytimes.com/2023/02/08/technology/ai-chatbots-disinformation.html}}.
\newblock


\bibitem[Pen(2023)]%
        {Pentagon2023}
 \bibinfo{year}{2023}\natexlab{}.
\newblock \bibinfo{title}{Fact Check: Was There an Explosion at the Pentagon?}
\newblock \bibinfo{howpublished}{\url{https://www.newsweek.com}}.
\newblock


\bibitem[int(2023)]%
        {interim}
 \bibinfo{year}{2023}\natexlab{}.
\newblock \bibinfo{title}{Interim Regulation on the Management of Generative
  Artificial intelligence (AI) Services}.
\newblock
  \bibinfo{howpublished}{\url{https://www.gov.cn/zhengce/zhengceku/202307/content_6891752.htm}}.
\newblock


\bibitem[Abadi et~al\mbox{.}(2016)]%
        {abadi2016deep}
\bibfield{author}{\bibinfo{person}{Martin Abadi}, \bibinfo{person}{Andy Chu},
  \bibinfo{person}{Ian Goodfellow}, \bibinfo{person}{H~Brendan McMahan},
  \bibinfo{person}{Ilya Mironov}, \bibinfo{person}{Kunal Talwar}, {and}
  \bibinfo{person}{Li Zhang}.} \bibinfo{year}{2016}\natexlab{}.
\newblock \showarticletitle{Deep learning with differential privacy}. In
  \bibinfo{booktitle}{\emph{Proceedings of the 2016 ACM SIGSAC conference on
  computer and communications security}}. \bibinfo{pages}{308--318}.
\newblock


\bibitem[Abid et~al\mbox{.}(2021)]%
        {abid2021persistent}
\bibfield{author}{\bibinfo{person}{Abubakar Abid}, \bibinfo{person}{Maheen
  Farooqi}, {and} \bibinfo{person}{James Zou}.}
  \bibinfo{year}{2021}\natexlab{}.
\newblock \showarticletitle{Persistent anti-muslim bias in large language
  models}. In \bibinfo{booktitle}{\emph{Proceedings of the 2021 AAAI/ACM
  Conference on AI, Ethics, and Society}}. \bibinfo{pages}{298--306}.
\newblock


\bibitem[Alaa et~al\mbox{.}(2022)]%
        {alaa2022faithful}
\bibfield{author}{\bibinfo{person}{Ahmed Alaa}, \bibinfo{person}{Boris
  Van~Breugel}, \bibinfo{person}{Evgeny~S Saveliev}, {and}
  \bibinfo{person}{Mihaela van~der Schaar}.} \bibinfo{year}{2022}\natexlab{}.
\newblock \showarticletitle{How faithful is your synthetic data? sample-level
  metrics for evaluating and auditing generative models}. In
  \bibinfo{booktitle}{\emph{International Conference on Machine Learning}}.
  PMLR, \bibinfo{pages}{290--306}.
\newblock


\bibitem[Aliman and Kester(2022)]%
        {10024477}
\bibfield{author}{\bibinfo{person}{Nadisha-Marie Aliman} {and}
  \bibinfo{person}{Leon Kester}.} \bibinfo{year}{2022}\natexlab{}.
\newblock \showarticletitle{VR, Deepfakes and Epistemic Security}. In
  \bibinfo{booktitle}{\emph{2022 IEEE International Conference on Artificial
  Intelligence and Virtual Reality (AIVR)}}. \bibinfo{pages}{93--98}.
\newblock
\urldef\tempurl%
\url{https://doi.org/10.1109/AIVR56993.2022.00019}
\showDOI{\tempurl}


\bibitem[An et~al\mbox{.}(2024)]%
        {pmlrv235an24a}
\bibfield{author}{\bibinfo{person}{Bang An}, \bibinfo{person}{Mucong Ding},
  \bibinfo{person}{Tahseen Rabbani}, \bibinfo{person}{Aakriti Agrawal},
  \bibinfo{person}{Yuancheng Xu}, \bibinfo{person}{Chenghao Deng},
  \bibinfo{person}{Sicheng Zhu}, \bibinfo{person}{Abdirisak Mohamed},
  \bibinfo{person}{Yuxin Wen}, \bibinfo{person}{Tom Goldstein}, {and}
  \bibinfo{person}{Furong Huang}.} \bibinfo{year}{2024}\natexlab{}.
\newblock \showarticletitle{{WAVES}: Benchmarking the Robustness of Image
  Watermarks}. In \bibinfo{booktitle}{\emph{Proceedings of the 41st
  International Conference on Machine Learning (ICML)}},
  Vol.~\bibinfo{volume}{235}. \bibinfo{pages}{1456--1492}.
\newblock


\bibitem[Asnani et~al\mbox{.}(2024)]%
        {Asnani_2024_CVPR}
\bibfield{author}{\bibinfo{person}{Vishal Asnani}, \bibinfo{person}{John
  Collomosse}, \bibinfo{person}{Tu Bui}, \bibinfo{person}{Xiaoming Liu}, {and}
  \bibinfo{person}{Shruti Agarwal}.} \bibinfo{year}{2024}\natexlab{}.
\newblock \showarticletitle{ProMark: Proactive Diffusion Watermarking for
  Causal Attribution}. In \bibinfo{booktitle}{\emph{Proceedings of the IEEE/CVF
  Conference on Computer Vision and Pattern Recognition (CVPR)}}.
  \bibinfo{pages}{10802--10811}.
\newblock


\bibitem[Azaria and Mitchell(2023)]%
        {azaria2023internal}
\bibfield{author}{\bibinfo{person}{Amos Azaria} {and} \bibinfo{person}{Tom
  Mitchell}.} \bibinfo{year}{2023}\natexlab{}.
\newblock \showarticletitle{The internal state of an llm knows when its lying}.
\newblock \bibinfo{journal}{\emph{arXiv preprint arXiv:2304.13734}}
  (\bibinfo{year}{2023}).
\newblock


\bibitem[Bai et~al\mbox{.}(2022)]%
        {bai2022reducing}
\bibfield{author}{\bibinfo{person}{Andrew Bai}, \bibinfo{person}{Cho-Jui
  Hsieh}, \bibinfo{person}{Wendy Kan}, {and} \bibinfo{person}{Hsuan-Tien Lin}.}
  \bibinfo{year}{2022}\natexlab{}.
\newblock \showarticletitle{Reducing Training Sample Memorization in GANs by
  Training with Memorization Rejection}.
\newblock \bibinfo{journal}{\emph{arXiv preprint arXiv:2210.12231}}
  (\bibinfo{year}{2022}).
\newblock


\bibitem[Bender et~al\mbox{.}(2021)]%
        {bender2021dangers}
\bibfield{author}{\bibinfo{person}{Emily~M Bender}, \bibinfo{person}{Timnit
  Gebru}, \bibinfo{person}{Angelina McMillan-Major}, {and}
  \bibinfo{person}{Shmargaret Shmitchell}.} \bibinfo{year}{2021}\natexlab{}.
\newblock \showarticletitle{On the dangers of stochastic parrots: Can language
  models be too big?}. In \bibinfo{booktitle}{\emph{Proceedings of the 2021 ACM
  conference on fairness, accountability, and transparency}}.
  \bibinfo{pages}{610--623}.
\newblock


\bibitem[Bi et~al\mbox{.}(2023)]%
        {bi2023detecting}
\bibfield{author}{\bibinfo{person}{Xiuli Bi}, \bibinfo{person}{Bo Liu},
  \bibinfo{person}{Fan Yang}, \bibinfo{person}{Bin Xiao},
  \bibinfo{person}{Weisheng Li}, \bibinfo{person}{Gao Huang}, {and}
  \bibinfo{person}{Pamela~C Cosman}.} \bibinfo{year}{2023}\natexlab{}.
\newblock \showarticletitle{Detecting Generated Images by Real Images Only}.
\newblock \bibinfo{journal}{\emph{arXiv preprint arXiv:2311.00962}}
  (\bibinfo{year}{2023}).
\newblock


\bibitem[Bird and Lotfi(2024)]%
        {bird2024cifake}
\bibfield{author}{\bibinfo{person}{Jordan~J Bird} {and} \bibinfo{person}{Ahmad
  Lotfi}.} \bibinfo{year}{2024}\natexlab{}.
\newblock \showarticletitle{Cifake: Image classification and explainable
  identification of ai-generated synthetic images}.
\newblock \bibinfo{journal}{\emph{IEEE Access}} (\bibinfo{year}{2024}).
\newblock


\bibitem[Birhane et~al\mbox{.}(2024)]%
        {birhane2024into}
\bibfield{author}{\bibinfo{person}{Abeba Birhane}, \bibinfo{person}{Sanghyun
  Han}, \bibinfo{person}{Vishnu Boddeti}, \bibinfo{person}{Sasha Luccioni},
  {et~al\mbox{.}}} \bibinfo{year}{2024}\natexlab{}.
\newblock \showarticletitle{Into the LAION’s Den: Investigating hate in
  multimodal datasets}.
\newblock \bibinfo{journal}{\emph{Advances in Neural Information Processing
  Systems}}  \bibinfo{volume}{36} (\bibinfo{year}{2024}).
\newblock


\bibitem[Birhane et~al\mbox{.}(2021)]%
        {birhane2021multimodal}
\bibfield{author}{\bibinfo{person}{Abeba Birhane}, \bibinfo{person}{Vinay~Uday
  Prabhu}, {and} \bibinfo{person}{Emmanuel Kahembwe}.}
  \bibinfo{year}{2021}\natexlab{}.
\newblock \showarticletitle{Multimodal datasets: misogyny, pornography, and
  malignant stereotypes}.
\newblock \bibinfo{journal}{\emph{arXiv preprint arXiv:2110.01963}}
  (\bibinfo{year}{2021}).
\newblock


\bibitem[Bourtoule et~al\mbox{.}(2021)]%
        {bourtoule2021machine}
\bibfield{author}{\bibinfo{person}{Lucas Bourtoule}, \bibinfo{person}{Varun
  Chandrasekaran}, \bibinfo{person}{Christopher~A Choquette-Choo},
  \bibinfo{person}{Hengrui Jia}, \bibinfo{person}{Adelin Travers},
  \bibinfo{person}{Baiwu Zhang}, \bibinfo{person}{David Lie}, {and}
  \bibinfo{person}{Nicolas Papernot}.} \bibinfo{year}{2021}\natexlab{}.
\newblock \showarticletitle{Machine unlearning}. In
  \bibinfo{booktitle}{\emph{2021 IEEE Symposium on Security and Privacy (SP)}}.
  IEEE, \bibinfo{pages}{141--159}.
\newblock


\bibitem[Boutros et~al\mbox{.}(2023)]%
        {boutros2023idiff}
\bibfield{author}{\bibinfo{person}{Fadi Boutros}, \bibinfo{person}{Jonas~Henry
  Grebe}, \bibinfo{person}{Arjan Kuijper}, {and} \bibinfo{person}{Naser
  Damer}.} \bibinfo{year}{2023}\natexlab{}.
\newblock \showarticletitle{Idiff-face: Synthetic-based face recognition
  through fizzy identity-conditioned diffusion model}. In
  \bibinfo{booktitle}{\emph{Proceedings of the IEEE/CVF International
  Conference on Computer Vision}}. \bibinfo{pages}{19650--19661}.
\newblock


\bibitem[Brack et~al\mbox{.}(2023)]%
        {brack2023mitigating}
\bibfield{author}{\bibinfo{person}{Manuel Brack}, \bibinfo{person}{Felix
  Friedrich}, \bibinfo{person}{Patrick Schramowski}, {and}
  \bibinfo{person}{Kristian Kersting}.} \bibinfo{year}{2023}\natexlab{}.
\newblock \showarticletitle{Mitigating Inappropriateness in Image Generation:
  Can there be Value in Reflecting the World's Ugliness?}
\newblock \bibinfo{journal}{\emph{arXiv preprint arXiv:2305.18398}}
  (\bibinfo{year}{2023}).
\newblock


\bibitem[Bui et~al\mbox{.}(2022)]%
        {bui2022repmix}
\bibfield{author}{\bibinfo{person}{Tu Bui}, \bibinfo{person}{Ning Yu}, {and}
  \bibinfo{person}{John Collomosse}.} \bibinfo{year}{2022}\natexlab{}.
\newblock \showarticletitle{Repmix: Representation mixing for robust
  attribution of synthesized images}. In \bibinfo{booktitle}{\emph{European
  Conference on Computer Vision}}. Springer, \bibinfo{pages}{146--163}.
\newblock


\bibitem[Caliskan et~al\mbox{.}(2017)]%
        {caliskan2017semantics}
\bibfield{author}{\bibinfo{person}{Aylin Caliskan}, \bibinfo{person}{Joanna~J
  Bryson}, {and} \bibinfo{person}{Arvind Narayanan}.}
  \bibinfo{year}{2017}\natexlab{}.
\newblock \showarticletitle{Semantics derived automatically from language
  corpora contain human-like biases}.
\newblock \bibinfo{journal}{\emph{Science}} \bibinfo{volume}{356},
  \bibinfo{number}{6334} (\bibinfo{year}{2017}), \bibinfo{pages}{183--186}.
\newblock


\bibitem[Cao and Li(2021)]%
        {cao2021generating}
\bibfield{author}{\bibinfo{person}{Chu Cao} {and} \bibinfo{person}{Mo Li}.}
  \bibinfo{year}{2021}\natexlab{}.
\newblock \showarticletitle{Generating mobility trajectories with retained Data
  Utility}. In \bibinfo{booktitle}{\emph{Proceedings of the 27th ACM SIGKDD
  Conference on Knowledge Discovery \& Data Mining}}.
  \bibinfo{pages}{2610--2620}.
\newblock


\bibitem[Carlini et~al\mbox{.}(2023a)]%
        {carlini2023extracting}
\bibfield{author}{\bibinfo{person}{Nicolas Carlini}, \bibinfo{person}{Jamie
  Hayes}, \bibinfo{person}{Milad Nasr}, \bibinfo{person}{Matthew Jagielski},
  \bibinfo{person}{Vikash Sehwag}, \bibinfo{person}{Florian Tramer},
  \bibinfo{person}{Borja Balle}, \bibinfo{person}{Daphne Ippolito}, {and}
  \bibinfo{person}{Eric Wallace}.} \bibinfo{year}{2023}\natexlab{a}.
\newblock \showarticletitle{Extracting training data from diffusion models}. In
  \bibinfo{booktitle}{\emph{32nd USENIX Security Symposium (USENIX Security
  23)}}. \bibinfo{pages}{5253--5270}.
\newblock


\bibitem[Carlini et~al\mbox{.}(2023b)]%
        {carlini2022quantifying}
\bibfield{author}{\bibinfo{person}{Nicholas Carlini}, \bibinfo{person}{Daphne
  Ippolito}, \bibinfo{person}{Matthew Jagielski}, \bibinfo{person}{Katherine
  Lee}, \bibinfo{person}{Florian Tramer}, {and} \bibinfo{person}{Chiyuan
  Zhang}.} \bibinfo{year}{2023}\natexlab{b}.
\newblock \showarticletitle{Quantifying memorization across neural language
  models}. In \bibinfo{booktitle}{\emph{International Conference on Learning
  Representations (ICLR)}}.
\newblock


\bibitem[Carlini et~al\mbox{.}(2021)]%
        {carlini2021extracting}
\bibfield{author}{\bibinfo{person}{Nicholas Carlini}, \bibinfo{person}{Florian
  Tramer}, \bibinfo{person}{Eric Wallace}, \bibinfo{person}{Matthew Jagielski},
  \bibinfo{person}{Ariel Herbert-Voss}, \bibinfo{person}{Katherine Lee},
  \bibinfo{person}{Adam Roberts}, \bibinfo{person}{Tom Brown},
  \bibinfo{person}{Dawn Song}, \bibinfo{person}{Ulfar Erlingsson},
  {et~al\mbox{.}}} \bibinfo{year}{2021}\natexlab{}.
\newblock \showarticletitle{Extracting training data from large language
  models}. In \bibinfo{booktitle}{\emph{30th USENIX Security Symposium (USENIX
  Security 21)}}. \bibinfo{pages}{2633--2650}.
\newblock


\bibitem[Chen et~al\mbox{.}(2023b)]%
        {chen2023challenges}
\bibfield{author}{\bibinfo{person}{Chuan Chen}, \bibinfo{person}{Zhenpeng Wu},
  \bibinfo{person}{Yanyi Lai}, \bibinfo{person}{Wenlin Ou},
  \bibinfo{person}{Tianchi Liao}, {and} \bibinfo{person}{Zibin Zheng}.}
  \bibinfo{year}{2023}\natexlab{b}.
\newblock \showarticletitle{Challenges and Remedies to Privacy and Security in
  AIGC: Exploring the Potential of Privacy Computing, Blockchain, and Beyond}.
\newblock \bibinfo{journal}{\emph{arXiv preprint arXiv:2306.00419}}
  (\bibinfo{year}{2023}).
\newblock


\bibitem[Chen et~al\mbox{.}(2024b)]%
        {chen2024revolution}
\bibfield{author}{\bibinfo{person}{Jiayuan Chen}, \bibinfo{person}{Changyan
  Yi}, \bibinfo{person}{Hongyang Du}, \bibinfo{person}{Dusit Niyato},
  \bibinfo{person}{Jiawen Kang}, \bibinfo{person}{Jun Cai}, {and}
  \bibinfo{person}{Xuemin Shen}.} \bibinfo{year}{2024}\natexlab{b}.
\newblock \showarticletitle{A revolution of personalized healthcare: Enabling
  human digital twin with mobile AIGC}.
\newblock \bibinfo{journal}{\emph{IEEE Network}} (\bibinfo{year}{2024}).
\newblock


\bibitem[Chen et~al\mbox{.}(2021)]%
        {chen2021perceptual}
\bibfield{author}{\bibinfo{person}{Jia-Wei Chen}, \bibinfo{person}{Li-Ju Chen},
  \bibinfo{person}{Chia-Mu Yu}, {and} \bibinfo{person}{Chun-Shien Lu}.}
  \bibinfo{year}{2021}\natexlab{}.
\newblock \showarticletitle{Perceptual indistinguishability-net (pi-net):
  Facial image obfuscation with manipulable semantics}. In
  \bibinfo{booktitle}{\emph{Proceedings of the IEEE/CVF Conference on Computer
  Vision and Pattern Recognition}}. \bibinfo{pages}{6478--6487}.
\newblock


\bibitem[Chen et~al\mbox{.}(2024a)]%
        {chen2024gaining}
\bibfield{author}{\bibinfo{person}{Kai Chen}, \bibinfo{person}{Chunwei Wang},
  \bibinfo{person}{Kuo Yang}, \bibinfo{person}{Jianhua Han},
  \bibinfo{person}{Lanqing HONG}, \bibinfo{person}{Fei Mi},
  \bibinfo{person}{Hang Xu}, \bibinfo{person}{Zhengying Liu},
  \bibinfo{person}{Wenyong Huang}, \bibinfo{person}{Zhenguo Li},
  \bibinfo{person}{Dit-Yan Yeung}, {and} \bibinfo{person}{Lifeng Shang}.}
  \bibinfo{year}{2024}\natexlab{a}.
\newblock \showarticletitle{Gaining Wisdom from Setbacks: Aligning Large
  Language Models via Mistake Analysis}. In \bibinfo{booktitle}{\emph{The
  Twelfth International Conference on Learning Representations}}.
\newblock
\urldef\tempurl%
\url{https://openreview.net/forum?id=aA33A70IO6}
\showURL{%
\tempurl}


\bibitem[chen et~al\mbox{.}(2023)]%
        {NEURIPS2023_8b8a7960}
\bibfield{author}{\bibinfo{person}{shiqi chen}, \bibinfo{person}{Yiran Zhao},
  \bibinfo{person}{Jinghan Zhang}, \bibinfo{person}{I-Chun Chern},
  \bibinfo{person}{Siyang Gao}, \bibinfo{person}{Pengfei Liu}, {and}
  \bibinfo{person}{Junxian He}.} \bibinfo{year}{2023}\natexlab{}.
\newblock \showarticletitle{FELM: Benchmarking Factuality Evaluation of Large
  Language Models}. In \bibinfo{booktitle}{\emph{Advances in Neural Information
  Processing Systems}}, Vol.~\bibinfo{volume}{36}.
  \bibinfo{pages}{44502--44523}.
\newblock


\bibitem[Chen et~al\mbox{.}(2023a)]%
        {chen2023gpt}
\bibfield{author}{\bibinfo{person}{Yutian Chen}, \bibinfo{person}{Hao Kang},
  \bibinfo{person}{Vivian Zhai}, \bibinfo{person}{Liangze Li},
  \bibinfo{person}{Rita Singh}, {and} \bibinfo{person}{Bhiksha Ramakrishnan}.}
  \bibinfo{year}{2023}\natexlab{a}.
\newblock \showarticletitle{GPT-Sentinel: Distinguishing Human and ChatGPT
  Generated Content}.
\newblock \bibinfo{journal}{\emph{arXiv preprint arXiv:2305.07969}}
  (\bibinfo{year}{2023}).
\newblock


\bibitem[Clark et~al\mbox{.}(2019)]%
        {clark2019efficient}
\bibfield{author}{\bibinfo{person}{Aidan Clark}, \bibinfo{person}{Jeff
  Donahue}, {and} \bibinfo{person}{Karen Simonyan}.}
  \bibinfo{year}{2019}\natexlab{}.
\newblock \showarticletitle{Efficient video generation on complex datasets}.
\newblock \bibinfo{journal}{\emph{arXiv preprint arXiv:1907.06571}}
  \bibinfo{volume}{2}, \bibinfo{number}{3} (\bibinfo{year}{2019}),
  \bibinfo{pages}{4}.
\newblock


\bibitem[Corvi et~al\mbox{.}(2023)]%
        {corvi2023detection}
\bibfield{author}{\bibinfo{person}{Riccardo Corvi}, \bibinfo{person}{Davide
  Cozzolino}, \bibinfo{person}{Giada Zingarini}, \bibinfo{person}{Giovanni
  Poggi}, \bibinfo{person}{Koki Nagano}, {and} \bibinfo{person}{Luisa
  Verdoliva}.} \bibinfo{year}{2023}\natexlab{}.
\newblock \showarticletitle{On the detection of synthetic images generated by
  diffusion models}. In \bibinfo{booktitle}{\emph{ICASSP 2023-2023 IEEE
  International Conference on Acoustics, Speech and Signal Processing
  (ICASSP)}}. IEEE, \bibinfo{pages}{1--5}.
\newblock


\bibitem[Cui et~al\mbox{.}(2023)]%
        {cui2023diffusionshield}
\bibfield{author}{\bibinfo{person}{Yingqian Cui}, \bibinfo{person}{Jie Ren},
  \bibinfo{person}{Han Xu}, \bibinfo{person}{Pengfei He}, \bibinfo{person}{Hui
  Liu}, \bibinfo{person}{Lichao Sun}, {and} \bibinfo{person}{Jiliang Tang}.}
  \bibinfo{year}{2023}\natexlab{}.
\newblock \showarticletitle{DiffusionShield: A Watermark for Copyright
  Protection against Generative Diffusion Models}.
\newblock \bibinfo{journal}{\emph{arXiv preprint arXiv:2306.04642}}
  (\bibinfo{year}{2023}).
\newblock


\bibitem[Dockhorn et~al\mbox{.}(2022)]%
        {dockhorn2022differentially}
\bibfield{author}{\bibinfo{person}{Tim Dockhorn}, \bibinfo{person}{Tianshi
  Cao}, \bibinfo{person}{Arash Vahdat}, {and} \bibinfo{person}{Karsten Kreis}.}
  \bibinfo{year}{2022}\natexlab{}.
\newblock \showarticletitle{Differentially private diffusion models}.
\newblock \bibinfo{journal}{\emph{arXiv preprint arXiv:2210.09929}}
  (\bibinfo{year}{2022}).
\newblock


\bibitem[Dogoulis et~al\mbox{.}(2023)]%
        {dogoulis2023improving}
\bibfield{author}{\bibinfo{person}{Pantelis Dogoulis}, \bibinfo{person}{Giorgos
  Kordopatis-Zilos}, \bibinfo{person}{Ioannis Kompatsiaris}, {and}
  \bibinfo{person}{Symeon Papadopoulos}.} \bibinfo{year}{2023}\natexlab{}.
\newblock \showarticletitle{Improving Synthetically Generated Image Detection
  in Cross-Concept Settings}. In \bibinfo{booktitle}{\emph{Proceedings of the
  2nd ACM International Workshop on Multimedia AI against Disinformation}}.
  \bibinfo{pages}{28--35}.
\newblock


\bibitem[Du et~al\mbox{.}(2023)]%
        {du2023improving}
\bibfield{author}{\bibinfo{person}{Yilun Du}, \bibinfo{person}{Shuang Li},
  \bibinfo{person}{Antonio Torralba}, \bibinfo{person}{Joshua~B Tenenbaum},
  {and} \bibinfo{person}{Igor Mordatch}.} \bibinfo{year}{2023}\natexlab{}.
\newblock \showarticletitle{Improving Factuality and Reasoning in Language
  Models through Multiagent Debate}.
\newblock \bibinfo{journal}{\emph{arXiv preprint arXiv:2305.14325}}
  (\bibinfo{year}{2023}).
\newblock


\bibitem[Evans et~al\mbox{.}(2021)]%
        {evans2021truthful}
\bibfield{author}{\bibinfo{person}{Owain Evans}, \bibinfo{person}{Owen
  Cotton-Barratt}, \bibinfo{person}{Lukas Finnveden}, \bibinfo{person}{Adam
  Bales}, \bibinfo{person}{Avital Balwit}, \bibinfo{person}{Peter Wills},
  \bibinfo{person}{Luca Righetti}, {and} \bibinfo{person}{William Saunders}.}
  \bibinfo{year}{2021}\natexlab{}.
\newblock \showarticletitle{Truthful AI: Developing and governing AI that does
  not lie}.
\newblock \bibinfo{journal}{\emph{arXiv preprint arXiv:2110.06674}}
  (\bibinfo{year}{2021}).
\newblock


\bibitem[Fagni et~al\mbox{.}(2021)]%
        {fagni2021tweepfake}
\bibfield{author}{\bibinfo{person}{Tiziano Fagni}, \bibinfo{person}{Fabrizio
  Falchi}, \bibinfo{person}{Margherita Gambini}, \bibinfo{person}{Antonio
  Martella}, {and} \bibinfo{person}{Maurizio Tesconi}.}
  \bibinfo{year}{2021}\natexlab{}.
\newblock \showarticletitle{TweepFake: About detecting deepfake tweets}.
\newblock \bibinfo{journal}{\emph{Plos one}} \bibinfo{volume}{16},
  \bibinfo{number}{5} (\bibinfo{year}{2021}), \bibinfo{pages}{e0251415}.
\newblock


\bibitem[Feng et~al\mbox{.}(2021)]%
        {feng2021gans}
\bibfield{author}{\bibinfo{person}{Qianli Feng}, \bibinfo{person}{Chenqi Guo},
  \bibinfo{person}{Fabian Benitez-Quiroz}, {and} \bibinfo{person}{Aleix~M
  Martinez}.} \bibinfo{year}{2021}\natexlab{}.
\newblock \showarticletitle{When do gans replicate? on the choice of dataset
  size}. In \bibinfo{booktitle}{\emph{Proceedings of the IEEE/CVF International
  Conference on Computer Vision}}. \bibinfo{pages}{6701--6710}.
\newblock


\bibitem[Feng et~al\mbox{.}(2023)]%
        {feng2023catch}
\bibfield{author}{\bibinfo{person}{Weitao Feng}, \bibinfo{person}{Jiyan He},
  \bibinfo{person}{Jie Zhang}, \bibinfo{person}{Tianwei Zhang},
  \bibinfo{person}{Wenbo Zhou}, \bibinfo{person}{Weiming Zhang}, {and}
  \bibinfo{person}{Nenghai Yu}.} \bibinfo{year}{2023}\natexlab{}.
\newblock \showarticletitle{Catch You Everything Everywhere: Guarding Textual
  Inversion via Concept Watermarking}.
\newblock \bibinfo{journal}{\emph{arXiv preprint arXiv:2309.05940}}
  (\bibinfo{year}{2023}).
\newblock


\bibitem[Fernandez et~al\mbox{.}(2023)]%
        {fernandez2023stable}
\bibfield{author}{\bibinfo{person}{Pierre Fernandez},
  \bibinfo{person}{Guillaume Couairon}, \bibinfo{person}{Herv{\'e} J{\'e}gou},
  \bibinfo{person}{Matthijs Douze}, {and} \bibinfo{person}{Teddy Furon}.}
  \bibinfo{year}{2023}\natexlab{}.
\newblock \showarticletitle{The stable signature: Rooting watermarks in latent
  diffusion models}. In \bibinfo{booktitle}{\emph{Proceedings of the IEEE/CVF
  International Conference on Computer Vision}}. \bibinfo{pages}{22466--22477}.
\newblock


\bibitem[Gandikota et~al\mbox{.}(2023)]%
        {gandikota2023erasing}
\bibfield{author}{\bibinfo{person}{Rohit Gandikota}, \bibinfo{person}{Joanna
  Materzynska}, \bibinfo{person}{Jaden Fiotto-Kaufman}, {and}
  \bibinfo{person}{David Bau}.} \bibinfo{year}{2023}\natexlab{}.
\newblock \showarticletitle{Erasing concepts from diffusion models}. In
  \bibinfo{booktitle}{\emph{Proceedings of the IEEE/CVF International
  Conference on Computer Vision}}. \bibinfo{pages}{2426--2436}.
\newblock


\bibitem[Ganguli et~al\mbox{.}(2022)]%
        {ganguli2022red}
\bibfield{author}{\bibinfo{person}{Deep Ganguli}, \bibinfo{person}{Liane
  Lovitt}, \bibinfo{person}{Jackson Kernion}, \bibinfo{person}{Amanda Askell},
  \bibinfo{person}{Yuntao Bai}, \bibinfo{person}{Saurav Kadavath},
  \bibinfo{person}{Ben Mann}, \bibinfo{person}{Ethan Perez},
  \bibinfo{person}{Nicholas Schiefer}, \bibinfo{person}{Kamal Ndousse},
  {et~al\mbox{.}}} \bibinfo{year}{2022}\natexlab{}.
\newblock \showarticletitle{Red teaming language models to reduce harms:
  Methods, scaling behaviors, and lessons learned}.
\newblock \bibinfo{journal}{\emph{arXiv preprint arXiv:2209.07858}}
  (\bibinfo{year}{2022}).
\newblock


\bibitem[Gehrmann et~al\mbox{.}(2019)]%
        {gehrmann2019gltr}
\bibfield{author}{\bibinfo{person}{Sebastian Gehrmann},
  \bibinfo{person}{Hendrik Strobelt}, {and} \bibinfo{person}{Alexander~M
  Rush}.} \bibinfo{year}{2019}\natexlab{}.
\newblock \showarticletitle{Gltr: Statistical detection and visualization of
  generated text}.
\newblock \bibinfo{journal}{\emph{arXiv preprint arXiv:1906.04043}}
  (\bibinfo{year}{2019}).
\newblock


\bibitem[Ghalebikesabi et~al\mbox{.}(2023)]%
        {ghalebikesabi2023differentially}
\bibfield{author}{\bibinfo{person}{Sahra Ghalebikesabi},
  \bibinfo{person}{Leonard Berrada}, \bibinfo{person}{Sven Gowal},
  \bibinfo{person}{Ira Ktena}, \bibinfo{person}{Robert Stanforth},
  \bibinfo{person}{Jamie Hayes}, \bibinfo{person}{Soham De},
  \bibinfo{person}{Samuel~L Smith}, \bibinfo{person}{Olivia Wiles}, {and}
  \bibinfo{person}{Borja Balle}.} \bibinfo{year}{2023}\natexlab{}.
\newblock \showarticletitle{Differentially private diffusion models generate
  useful synthetic images}.
\newblock \bibinfo{journal}{\emph{arXiv preprint arXiv:2302.13861}}
  (\bibinfo{year}{2023}).
\newblock


\bibitem[Girish et~al\mbox{.}(2021)]%
        {girish2021towards}
\bibfield{author}{\bibinfo{person}{Sharath Girish}, \bibinfo{person}{Saksham
  Suri}, \bibinfo{person}{Sai~Saketh Rambhatla}, {and} \bibinfo{person}{Abhinav
  Shrivastava}.} \bibinfo{year}{2021}\natexlab{}.
\newblock \showarticletitle{Towards discovery and attribution of open-world gan
  generated images}. In \bibinfo{booktitle}{\emph{Proceedings of the IEEE/CVF
  International Conference on Computer Vision}}. \bibinfo{pages}{14094--14103}.
\newblock


\bibitem[Gong et~al\mbox{.}(2020)]%
        {gong2020disentangled}
\bibfield{author}{\bibinfo{person}{Maoguo Gong}, \bibinfo{person}{Jialu Liu},
  \bibinfo{person}{Hao Li}, \bibinfo{person}{Yu Xie}, {and}
  \bibinfo{person}{Zedong Tang}.} \bibinfo{year}{2020}\natexlab{}.
\newblock \showarticletitle{Disentangled representation learning for multiple
  attributes preserving face deidentification}.
\newblock \bibinfo{journal}{\emph{IEEE transactions on neural networks and
  learning systems}} \bibinfo{volume}{33}, \bibinfo{number}{1}
  (\bibinfo{year}{2020}), \bibinfo{pages}{244--256}.
\newblock


\bibitem[Goodfellow et~al\mbox{.}(2020)]%
        {goodfellow2020generative}
\bibfield{author}{\bibinfo{person}{Ian Goodfellow}, \bibinfo{person}{Jean
  Pouget-Abadie}, \bibinfo{person}{Mehdi Mirza}, \bibinfo{person}{Bing Xu},
  \bibinfo{person}{David Warde-Farley}, \bibinfo{person}{Sherjil Ozair},
  \bibinfo{person}{Aaron Courville}, {and} \bibinfo{person}{Yoshua Bengio}.}
  \bibinfo{year}{2020}\natexlab{}.
\newblock \showarticletitle{Generative adversarial networks}.
\newblock \bibinfo{journal}{\emph{Commun. ACM}} \bibinfo{volume}{63},
  \bibinfo{number}{11} (\bibinfo{year}{2020}), \bibinfo{pages}{139--144}.
\newblock


\bibitem[Goodrich et~al\mbox{.}(2019)]%
        {goodrich2019assessing}
\bibfield{author}{\bibinfo{person}{Ben Goodrich}, \bibinfo{person}{Vinay Rao},
  \bibinfo{person}{Peter~J Liu}, {and} \bibinfo{person}{Mohammad Saleh}.}
  \bibinfo{year}{2019}\natexlab{}.
\newblock \showarticletitle{Assessing the factual accuracy of generated text}.
  In \bibinfo{booktitle}{\emph{proceedings of the 25th ACM SIGKDD international
  conference on knowledge discovery \& data mining}}.
  \bibinfo{pages}{166--175}.
\newblock


\bibitem[Gou et~al\mbox{.}(2024)]%
        {gou2023critic}
\bibfield{author}{\bibinfo{person}{Zhibin Gou}, \bibinfo{person}{Zhihong Shao},
  \bibinfo{person}{Yeyun Gong}, \bibinfo{person}{yelong shen},
  \bibinfo{person}{Yujiu Yang}, \bibinfo{person}{Nan Duan}, {and}
  \bibinfo{person}{Weizhu Chen}.} \bibinfo{year}{2024}\natexlab{}.
\newblock \showarticletitle{{CRITIC}: Large Language Models Can Self-Correct
  with Tool-Interactive Critiquing}. In \bibinfo{booktitle}{\emph{The Twelfth
  International Conference on Learning Representations}}.
\newblock
\urldef\tempurl%
\url{https://openreview.net/forum?id=Sx038qxjek}
\showURL{%
\tempurl}


\bibitem[Guarnera et~al\mbox{.}(2024)]%
        {guarnera2024level}
\bibfield{author}{\bibinfo{person}{Luca Guarnera}, \bibinfo{person}{Oliver
  Giudice}, {and} \bibinfo{person}{Sebastiano Battiato}.}
  \bibinfo{year}{2024}\natexlab{}.
\newblock \showarticletitle{Level up the deepfake detection: a method to
  effectively discriminate images generated by gan architectures and diffusion
  models}. In \bibinfo{booktitle}{\emph{Intelligent Systems Conference}}.
  Springer, \bibinfo{pages}{615--625}.
\newblock


\bibitem[Guo et~al\mbox{.}(2023)]%
        {guo2023close}
\bibfield{author}{\bibinfo{person}{Biyang Guo}, \bibinfo{person}{Xin Zhang},
  \bibinfo{person}{Ziyuan Wang}, \bibinfo{person}{Minqi Jiang},
  \bibinfo{person}{Jinran Nie}, \bibinfo{person}{Yuxuan Ding},
  \bibinfo{person}{Jianwei Yue}, {and} \bibinfo{person}{Yupeng Wu}.}
  \bibinfo{year}{2023}\natexlab{}.
\newblock \showarticletitle{How close is chatgpt to human experts? comparison
  corpus, evaluation, and detection}.
\newblock \bibinfo{journal}{\emph{arXiv preprint arXiv:2301.07597}}
  (\bibinfo{year}{2023}).
\newblock


\bibitem[He et~al\mbox{.}(2024)]%
        {he2023mgtbench}
\bibfield{author}{\bibinfo{person}{Xinlei He}, \bibinfo{person}{Xinyue Shen},
  \bibinfo{person}{Zeyuan Chen}, \bibinfo{person}{Michael Backes}, {and}
  \bibinfo{person}{Yang Zhang}.} \bibinfo{year}{2024}\natexlab{}.
\newblock \showarticletitle{Mgtbench: Benchmarking machine-generated text
  detection}. In \bibinfo{booktitle}{\emph{The ACM Conference on Computer and
  Communications Security}}.
\newblock


\bibitem[Henderson et~al\mbox{.}(2022)]%
        {henderson2022pile}
\bibfield{author}{\bibinfo{person}{Peter Henderson}, \bibinfo{person}{Mark
  Krass}, \bibinfo{person}{Lucia Zheng}, \bibinfo{person}{Neel Guha},
  \bibinfo{person}{Christopher~D Manning}, \bibinfo{person}{Dan Jurafsky},
  {and} \bibinfo{person}{Daniel Ho}.} \bibinfo{year}{2022}\natexlab{}.
\newblock \showarticletitle{Pile of law: Learning responsible data filtering
  from the law and a 256gb open-source legal dataset}.
\newblock \bibinfo{journal}{\emph{Advances in Neural Information Processing
  Systems}}  \bibinfo{volume}{35} (\bibinfo{year}{2022}),
  \bibinfo{pages}{29217--29234}.
\newblock


\bibitem[Heng and Soh(2024)]%
        {heng2024selective}
\bibfield{author}{\bibinfo{person}{Alvin Heng} {and} \bibinfo{person}{Harold
  Soh}.} \bibinfo{year}{2024}\natexlab{}.
\newblock \showarticletitle{Selective amnesia: A continual learning approach to
  forgetting in deep generative models}.
\newblock \bibinfo{journal}{\emph{Advances in Neural Information Processing
  Systems}}  \bibinfo{volume}{36} (\bibinfo{year}{2024}).
\newblock


\bibitem[Hindistan and Yetkin(2023)]%
        {hindistan2023hybrid}
\bibfield{author}{\bibinfo{person}{Yavuz~Selim Hindistan} {and}
  \bibinfo{person}{E~Fatih Yetkin}.} \bibinfo{year}{2023}\natexlab{}.
\newblock \showarticletitle{A Hybrid Approach With GAN and DP for Privacy
  Preservation of IIoT Data}.
\newblock \bibinfo{journal}{\emph{IEEE Access}}  \bibinfo{volume}{11}
  (\bibinfo{year}{2023}), \bibinfo{pages}{5837--5849}.
\newblock


\bibitem[Ho et~al\mbox{.}(2020)]%
        {ho2020denoising}
\bibfield{author}{\bibinfo{person}{Jonathan Ho}, \bibinfo{person}{Ajay Jain},
  {and} \bibinfo{person}{Pieter Abbeel}.} \bibinfo{year}{2020}\natexlab{}.
\newblock \showarticletitle{Denoising diffusion probabilistic models}.
\newblock \bibinfo{journal}{\emph{Advances in neural information processing
  systems}}  \bibinfo{volume}{33} (\bibinfo{year}{2020}),
  \bibinfo{pages}{6840--6851}.
\newblock


\bibitem[Hu et~al\mbox{.}(2021)]%
        {10.1145/3487890}
\bibfield{author}{\bibinfo{person}{Yupeng Hu}, \bibinfo{person}{Wenxin Kuang},
  \bibinfo{person}{Zheng Qin}, \bibinfo{person}{Kenli Li},
  \bibinfo{person}{Jiliang Zhang}, \bibinfo{person}{Yansong Gao},
  \bibinfo{person}{Wenjia Li}, {and} \bibinfo{person}{Keqin Li}.}
  \bibinfo{year}{2021}\natexlab{}.
\newblock \showarticletitle{Artificial Intelligence Security: Threats and
  Countermeasures}.
\newblock \bibinfo{journal}{\emph{ACM Comput. Surv.}} \bibinfo{volume}{55},
  \bibinfo{number}{1}, Article \bibinfo{articleno}{20} (\bibinfo{date}{nov}
  \bibinfo{year}{2021}), \bibinfo{numpages}{36}~pages.
\newblock
\showISSN{0360-0300}
\urldef\tempurl%
\url{https://doi.org/10.1145/3487890}
\showDOI{\tempurl}


\bibitem[Hukkel{\aa}s and Lindseth(2023)]%
        {hukkelaas2023deepprivacy2}
\bibfield{author}{\bibinfo{person}{H{\aa}kon Hukkel{\aa}s} {and}
  \bibinfo{person}{Frank Lindseth}.} \bibinfo{year}{2023}\natexlab{}.
\newblock \showarticletitle{Deepprivacy2: Towards realistic full-body
  anonymization}. In \bibinfo{booktitle}{\emph{Proceedings of the IEEE/CVF
  winter conference on applications of computer vision}}.
  \bibinfo{pages}{1329--1338}.
\newblock


\bibitem[Hussain et~al\mbox{.}(2021)]%
        {9423202}
\bibfield{author}{\bibinfo{person}{Shehzeen Hussain}, \bibinfo{person}{Paarth
  Neekhara}, \bibinfo{person}{Malhar Jere}, \bibinfo{person}{Farinaz
  Koushanfar}, {and} \bibinfo{person}{Julian McAuley}.}
  \bibinfo{year}{2021}\natexlab{}.
\newblock \showarticletitle{Adversarial Deepfakes: Evaluating Vulnerability of
  Deepfake Detectors to Adversarial Examples}. In
  \bibinfo{booktitle}{\emph{2021 IEEE Winter Conference on Applications of
  Computer Vision (WACV)}}. \bibinfo{pages}{3347--3356}.
\newblock
\urldef\tempurl%
\url{https://doi.org/10.1109/WACV48630.2021.00339}
\showDOI{\tempurl}


\bibitem[Jia et~al\mbox{.}(2023)]%
        {jia2023autosplice}
\bibfield{author}{\bibinfo{person}{Shan Jia}, \bibinfo{person}{Mingzhen Huang},
  \bibinfo{person}{Zhou Zhou}, \bibinfo{person}{Yan Ju},
  \bibinfo{person}{Jialing Cai}, {and} \bibinfo{person}{Siwei Lyu}.}
  \bibinfo{year}{2023}\natexlab{}.
\newblock \showarticletitle{AutoSplice: A Text-prompt Manipulated Image Dataset
  for Media Forensics}. In \bibinfo{booktitle}{\emph{Proceedings of the
  IEEE/CVF Conference on Computer Vision and Pattern Recognition}}.
  \bibinfo{pages}{893--903}.
\newblock


\bibitem[Jiang et~al\mbox{.}(2024)]%
        {jiang2024aigc}
\bibfield{author}{\bibinfo{person}{Jiajia Jiang}, \bibinfo{person}{Moting Su},
  \bibinfo{person}{Xiangli Xiao}, \bibinfo{person}{Yushu Zhang}, {and}
  \bibinfo{person}{Yuming Fang}.} \bibinfo{year}{2024}\natexlab{}.
\newblock \showarticletitle{AIGC-Chain: A Blockchain-Enabled Full Lifecycle
  Recording System for AIGC Product Copyright Management}.
\newblock \bibinfo{journal}{\emph{arXiv preprint arXiv:2406.14966}}
  (\bibinfo{year}{2024}).
\newblock


\bibitem[Joslin et~al\mbox{.}(2024)]%
        {299603}
\bibfield{author}{\bibinfo{person}{Matthew Joslin}, \bibinfo{person}{Xian
  Wang}, {and} \bibinfo{person}{Shuang Hao}.} \bibinfo{year}{2024}\natexlab{}.
\newblock \showarticletitle{Double Face: Leveraging User Intelligence to
  Characterize and Recognize {AI-synthesized} Faces}. In
  \bibinfo{booktitle}{\emph{33rd USENIX Security Symposium (USENIX Security
  24)}}. \bibinfo{address}{Philadelphia, PA}, \bibinfo{pages}{1009--1026}.
\newblock
\showISBNx{978-1-939133-44-1}


\bibitem[Kandpal et~al\mbox{.}(2022)]%
        {kandpal2022deduplicating}
\bibfield{author}{\bibinfo{person}{Nikhil Kandpal}, \bibinfo{person}{Eric
  Wallace}, {and} \bibinfo{person}{Colin Raffel}.}
  \bibinfo{year}{2022}\natexlab{}.
\newblock \showarticletitle{Deduplicating training data mitigates privacy risks
  in language models}. In \bibinfo{booktitle}{\emph{International Conference on
  Machine Learning}}. PMLR, \bibinfo{pages}{10697--10707}.
\newblock


\bibitem[Karras et~al\mbox{.}(2019)]%
        {karras2019style}
\bibfield{author}{\bibinfo{person}{Tero Karras}, \bibinfo{person}{Samuli
  Laine}, {and} \bibinfo{person}{Timo Aila}.} \bibinfo{year}{2019}\natexlab{}.
\newblock \showarticletitle{A style-based generator architecture for generative
  adversarial networks}. In \bibinfo{booktitle}{\emph{Proceedings of the
  IEEE/CVF conference on computer vision and pattern recognition}}.
  \bibinfo{pages}{4401--4410}.
\newblock


\bibitem[Kim et~al\mbox{.}(2023)]%
        {kim2023dcface}
\bibfield{author}{\bibinfo{person}{Minchul Kim}, \bibinfo{person}{Feng Liu},
  \bibinfo{person}{Anil Jain}, {and} \bibinfo{person}{Xiaoming Liu}.}
  \bibinfo{year}{2023}\natexlab{}.
\newblock \showarticletitle{DCFace: Synthetic Face Generation with Dual
  Condition Diffusion Model}. In \bibinfo{booktitle}{\emph{Proceedings of the
  IEEE/CVF Conference on Computer Vision and Pattern Recognition}}.
  \bibinfo{pages}{12715--12725}.
\newblock


\bibitem[Korshunov and Marcel(2018)]%
        {korshunov2018deepfakes}
\bibfield{author}{\bibinfo{person}{Pavel Korshunov} {and}
  \bibinfo{person}{S{\'e}bastien Marcel}.} \bibinfo{year}{2018}\natexlab{}.
\newblock \showarticletitle{Deepfakes: a new threat to face recognition?
  assessment and detection}.
\newblock \bibinfo{journal}{\emph{arXiv preprint arXiv:1812.08685}}
  (\bibinfo{year}{2018}).
\newblock


\bibitem[Kumari et~al\mbox{.}(2023)]%
        {kumari2023ablating}
\bibfield{author}{\bibinfo{person}{Nupur Kumari}, \bibinfo{person}{Bingliang
  Zhang}, \bibinfo{person}{Sheng-Yu Wang}, \bibinfo{person}{Eli Shechtman},
  \bibinfo{person}{Richard Zhang}, {and} \bibinfo{person}{Jun-Yan Zhu}.}
  \bibinfo{year}{2023}\natexlab{}.
\newblock \showarticletitle{Ablating Concepts in Text-to-Image Diffusion
  Models}. In \bibinfo{booktitle}{\emph{Proceedings of the IEEE International
  Conference on Computer Vision}}.
\newblock


\bibitem[Lee et~al\mbox{.}(2022)]%
        {lee2022factuality}
\bibfield{author}{\bibinfo{person}{Nayeon Lee}, \bibinfo{person}{Wei Ping},
  \bibinfo{person}{Peng Xu}, \bibinfo{person}{Mostofa Patwary},
  \bibinfo{person}{Pascale~N Fung}, \bibinfo{person}{Mohammad Shoeybi}, {and}
  \bibinfo{person}{Bryan Catanzaro}.} \bibinfo{year}{2022}\natexlab{}.
\newblock \showarticletitle{Factuality enhanced language models for open-ended
  text generation}.
\newblock \bibinfo{journal}{\emph{Advances in Neural Information Processing
  Systems}}  \bibinfo{volume}{35} (\bibinfo{year}{2022}),
  \bibinfo{pages}{34586--34599}.
\newblock


\bibitem[Li et~al\mbox{.}(2024b)]%
        {li-etal-2024-privlm}
\bibfield{author}{\bibinfo{person}{Haoran Li}, \bibinfo{person}{Dadi Guo},
  \bibinfo{person}{Donghao Li}, \bibinfo{person}{Wei Fan}, \bibinfo{person}{Qi
  Hu}, \bibinfo{person}{Xin Liu}, \bibinfo{person}{Chunkit Chan},
  \bibinfo{person}{Duanyi Yao}, \bibinfo{person}{Yuan Yao}, {and}
  \bibinfo{person}{Yangqiu Song}.} \bibinfo{year}{2024}\natexlab{b}.
\newblock \showarticletitle{{P}riv{LM}-Bench: A Multi-level Privacy Evaluation
  Benchmark for Language Models}. In \bibinfo{booktitle}{\emph{Proceedings of
  the 62nd Annual Meeting of the Association for Computational Linguistics
  (Volume 1: Long Papers)}}, \bibfield{editor}{\bibinfo{person}{Lun-Wei Ku},
  \bibinfo{person}{Andre Martins}, {and} \bibinfo{person}{Vivek Srikumar}}
  (Eds.). \bibinfo{pages}{54--73}.
\newblock


\bibitem[Li et~al\mbox{.}(2024a)]%
        {299531}
\bibfield{author}{\bibinfo{person}{Kecen Li}, \bibinfo{person}{Chen Gong},
  \bibinfo{person}{Zhixiang Li}, \bibinfo{person}{Yuzhong Zhao},
  \bibinfo{person}{Xinwen Hou}, {and} \bibinfo{person}{Tianhao Wang}.}
  \bibinfo{year}{2024}\natexlab{a}.
\newblock \showarticletitle{{PrivImage}: Differentially Private Synthetic Image
  Generation using Diffusion Models with {Semantic-Aware} Pretraining}. In
  \bibinfo{booktitle}{\emph{33rd USENIX Security Symposium (USENIX Security
  24)}}. \bibinfo{pages}{4837--4854}.
\newblock


\bibitem[Li et~al\mbox{.}(2023)]%
        {LYSBFZ23}
\bibfield{author}{\bibinfo{person}{Zheng Li}, \bibinfo{person}{Ning Yu},
  \bibinfo{person}{Ahmed Salem}, \bibinfo{person}{Michael Backes},
  \bibinfo{person}{Mario Fritz}, {and} \bibinfo{person}{Yang Zhang}.}
  \bibinfo{year}{2023}\natexlab{}.
\newblock \showarticletitle{{UnGANable: Defending Against GAN-based Face
  Manipulation}}. In \bibinfo{booktitle}{\emph{{USENIX Security Symposium
  (USENIX Security)}}}. \bibinfo{publisher}{USENIX}.
\newblock


\bibitem[Liang et~al\mbox{.}(2023)]%
        {liang2023adversarial}
\bibfield{author}{\bibinfo{person}{Chumeng Liang}, \bibinfo{person}{Xiaoyu Wu},
  \bibinfo{person}{Yang Hua}, \bibinfo{person}{Jiaru Zhang},
  \bibinfo{person}{Yiming Xue}, \bibinfo{person}{Tao Song},
  \bibinfo{person}{Zhengui Xue}, \bibinfo{person}{Ruhui Ma}, {and}
  \bibinfo{person}{Haibing Guan}.} \bibinfo{year}{2023}\natexlab{}.
\newblock \showarticletitle{Adversarial example does good: Preventing painting
  imitation from diffusion models via adversarial examples}. In
  \bibinfo{booktitle}{\emph{International Conference on Machine Learning}}.
  PMLR, \bibinfo{pages}{20763--20786}.
\newblock


\bibitem[Lin et~al\mbox{.}(2024)]%
        {lin2024detecting}
\bibfield{author}{\bibinfo{person}{Li Lin}, \bibinfo{person}{Neeraj Gupta},
  \bibinfo{person}{Yue Zhang}, \bibinfo{person}{Hainan Ren},
  \bibinfo{person}{Chun-Hao Liu}, \bibinfo{person}{Feng Ding},
  \bibinfo{person}{Xin Wang}, \bibinfo{person}{Xin Li}, \bibinfo{person}{Luisa
  Verdoliva}, {and} \bibinfo{person}{Shu Hu}.} \bibinfo{year}{2024}\natexlab{}.
\newblock \showarticletitle{Detecting multimedia generated by large ai models:
  A survey}.
\newblock \bibinfo{journal}{\emph{arXiv preprint arXiv:2402.00045}}
  (\bibinfo{year}{2024}).
\newblock


\bibitem[Liu et~al\mbox{.}(2023d)]%
        {liu2023survey}
\bibfield{author}{\bibinfo{person}{Aiwei Liu}, \bibinfo{person}{Leyi Pan},
  \bibinfo{person}{Yijian Lu}, \bibinfo{person}{Jingjing Li},
  \bibinfo{person}{Xuming Hu}, \bibinfo{person}{Lijie Wen},
  \bibinfo{person}{Irwin King}, {and} \bibinfo{person}{Philip~S Yu}.}
  \bibinfo{year}{2023}\natexlab{d}.
\newblock \showarticletitle{A survey of text watermarking in the era of large
  language models}.
\newblock \bibinfo{journal}{\emph{arXiv preprint arXiv:2312.07913}}
  (\bibinfo{year}{2023}).
\newblock


\bibitem[Liu et~al\mbox{.}(2021)]%
        {10.1145/3436755}
\bibfield{author}{\bibinfo{person}{Bo Liu}, \bibinfo{person}{Ming Ding},
  \bibinfo{person}{Sina Shaham}, \bibinfo{person}{Wenny Rahayu},
  \bibinfo{person}{Farhad Farokhi}, {and} \bibinfo{person}{Zihuai Lin}.}
  \bibinfo{year}{2021}\natexlab{}.
\newblock \showarticletitle{When Machine Learning Meets Privacy: A Survey and
  Outlook}.
\newblock \bibinfo{journal}{\emph{ACM Comput. Surv.}} \bibinfo{volume}{54},
  \bibinfo{number}{2}, Article \bibinfo{articleno}{31} (\bibinfo{date}{mar}
  \bibinfo{year}{2021}), \bibinfo{numpages}{36}~pages.
\newblock
\showISSN{0360-0300}
\urldef\tempurl%
\url{https://doi.org/10.1145/3436755}
\showDOI{\tempurl}


\bibitem[Liu et~al\mbox{.}(2023c)]%
        {liu2023ti2net}
\bibfield{author}{\bibinfo{person}{Baoping Liu}, \bibinfo{person}{Bo Liu},
  \bibinfo{person}{Ming Ding}, \bibinfo{person}{Tianqing Zhu}, {and}
  \bibinfo{person}{Xin Yu}.} \bibinfo{year}{2023}\natexlab{c}.
\newblock \showarticletitle{TI2Net: temporal identity inconsistency network for
  deepfake detection}. In \bibinfo{booktitle}{\emph{Proceedings of the IEEE/CVF
  Winter Conference on Applications of Computer Vision}}.
  \bibinfo{pages}{4691--4700}.
\newblock


\bibitem[Liu et~al\mbox{.}(2024d)]%
        {timbrewatermarking-ndss2024}
\bibfield{author}{\bibinfo{person}{Chang Liu}, \bibinfo{person}{Jie Zhang},
  \bibinfo{person}{Tianwei Zhang}, \bibinfo{person}{Xi Yang},
  \bibinfo{person}{Weiming Zhang}, {and} \bibinfo{person}{Nenghai Yu}.}
  \bibinfo{year}{2024}\natexlab{d}.
\newblock \showarticletitle{Detecting Voice Cloning Attacks via Timbre
  Watermarking}. In \bibinfo{booktitle}{\emph{Network and Distributed System
  Security Symposium}}.
\newblock
\urldef\tempurl%
\url{https://doi.org/10.14722/ndss.2024.24200}
\showDOI{\tempurl}


\bibitem[Liu et~al\mbox{.}(2022)]%
        {liu2022privacy}
\bibfield{author}{\bibinfo{person}{Fan Liu}, \bibinfo{person}{Zhiyong Cheng},
  \bibinfo{person}{Huilin Chen}, \bibinfo{person}{Yinwei Wei},
  \bibinfo{person}{Liqiang Nie}, {and} \bibinfo{person}{Mohan Kankanhalli}.}
  \bibinfo{year}{2022}\natexlab{}.
\newblock \showarticletitle{Privacy-preserving synthetic data generation for
  recommendation systems}. In \bibinfo{booktitle}{\emph{Proceedings of the 45th
  International ACM SIGIR Conference on Research and Development in Information
  Retrieval}}. \bibinfo{pages}{1379--1389}.
\newblock


\bibitem[Liu et~al\mbox{.}(2024a)]%
        {liu2024semantic}
\bibfield{author}{\bibinfo{person}{Guangyuan Liu}, \bibinfo{person}{Hongyang
  Du}, \bibinfo{person}{Dusit Niyato}, \bibinfo{person}{Jiawen Kang},
  \bibinfo{person}{Zehui Xiong}, \bibinfo{person}{Dong~In Kim}, {and}
  \bibinfo{person}{Xuemin Shen}.} \bibinfo{year}{2024}\natexlab{a}.
\newblock \showarticletitle{Semantic communications for artificial intelligence
  generated content (AIGC) toward effective content creation}.
\newblock \bibinfo{journal}{\emph{IEEE Network}} (\bibinfo{year}{2024}).
\newblock


\bibitem[Liu et~al\mbox{.}(2023a)]%
        {liu2023diffprotect}
\bibfield{author}{\bibinfo{person}{Jiang Liu}, \bibinfo{person}{Chun~Pong Lau},
  {and} \bibinfo{person}{Rama Chellappa}.} \bibinfo{year}{2023}\natexlab{a}.
\newblock \showarticletitle{DiffProtect: Generate Adversarial Examples with
  Diffusion Models for Facial Privacy Protection}.
\newblock \bibinfo{journal}{\emph{arXiv preprint arXiv:2305.13625}}
  (\bibinfo{year}{2023}).
\newblock


\bibitem[Liu et~al\mbox{.}(2024b)]%
        {liu2024blockchain}
\bibfield{author}{\bibinfo{person}{Yinqiu Liu}, \bibinfo{person}{Hongyang Du},
  \bibinfo{person}{Dusit Niyato}, \bibinfo{person}{Jiawen Kang},
  \bibinfo{person}{Zehui Xiong}, \bibinfo{person}{Chunyan Miao},
  \bibinfo{person}{Xuemin~Sherman Shen}, {and} \bibinfo{person}{Abbas
  Jamalipour}.} \bibinfo{year}{2024}\natexlab{b}.
\newblock \showarticletitle{Blockchain-Empowered Lifecycle Management for
  AI-Generated Content Products in Edge Networks}.
\newblock \bibinfo{journal}{\emph{IEEE Wireless Communications}}
  (\bibinfo{year}{2024}).
\newblock


\bibitem[Liu et~al\mbox{.}(2024c)]%
        {liu2024metacloak}
\bibfield{author}{\bibinfo{person}{Yixin Liu}, \bibinfo{person}{Chenrui Fan},
  \bibinfo{person}{Yutong Dai}, \bibinfo{person}{Xun Chen},
  \bibinfo{person}{Pan Zhou}, {and} \bibinfo{person}{Lichao Sun}.}
  \bibinfo{year}{2024}\natexlab{c}.
\newblock \showarticletitle{MetaCloak: Preventing Unauthorized Subject-driven
  Text-to-image Diffusion-based Synthesis via Meta-learning}. In
  \bibinfo{booktitle}{\emph{Proceedings of the IEEE/CVF Conference on Computer
  Vision and Pattern Recognition}}. \bibinfo{pages}{24219--24228}.
\newblock


\bibitem[Liu et~al\mbox{.}(2023b)]%
        {liu2023watermarking}
\bibfield{author}{\bibinfo{person}{Yugeng Liu}, \bibinfo{person}{Zheng Li},
  \bibinfo{person}{Michael Backes}, \bibinfo{person}{Yun Shen}, {and}
  \bibinfo{person}{Yang Zhang}.} \bibinfo{year}{2023}\natexlab{b}.
\newblock \showarticletitle{Watermarking Diffusion Model}.
\newblock \bibinfo{journal}{\emph{arXiv preprint arXiv:2305.12502}}
  (\bibinfo{year}{2023}).
\newblock


\bibitem[Liu et~al\mbox{.}(2023e)]%
        {liu2023check}
\bibfield{author}{\bibinfo{person}{Zeyan Liu}, \bibinfo{person}{Zijun Yao},
  \bibinfo{person}{Fengjun Li}, {and} \bibinfo{person}{Bo Luo}.}
  \bibinfo{year}{2023}\natexlab{e}.
\newblock \showarticletitle{Check me if you can: Detecting ChatGPT-generated
  academic writing using CheckGPT}.
\newblock \bibinfo{journal}{\emph{arXiv preprint arXiv:2306.05524}}
  (\bibinfo{year}{2023}).
\newblock


\bibitem[Lu et~al\mbox{.}(2023c)]%
        {lu2023hate}
\bibfield{author}{\bibinfo{person}{Junyu Lu}, \bibinfo{person}{Hongfei Lin},
  \bibinfo{person}{Xiaokun Zhang}, \bibinfo{person}{Zhaoqing Li},
  \bibinfo{person}{Tongyue Zhang}, \bibinfo{person}{Linlin Zong},
  \bibinfo{person}{Fenglong Ma}, {and} \bibinfo{person}{Bo Xu}.}
  \bibinfo{year}{2023}\natexlab{c}.
\newblock \showarticletitle{Hate Speech Detection via Dual Contrastive
  Learning}.
\newblock \bibinfo{journal}{\emph{IEEE/ACM Transactions on Audio, Speech, and
  Language Processing}} (\bibinfo{year}{2023}).
\newblock


\bibitem[Lu et~al\mbox{.}(2023d)]%
        {lu2023machine}
\bibfield{author}{\bibinfo{person}{Yingzhou Lu}, \bibinfo{person}{Huazheng
  Wang}, {and} \bibinfo{person}{Wenqi Wei}.} \bibinfo{year}{2023}\natexlab{d}.
\newblock \showarticletitle{Machine Learning for Synthetic Data Generation: a
  Review}.
\newblock \bibinfo{journal}{\emph{arXiv preprint arXiv:2302.04062}}
  (\bibinfo{year}{2023}).
\newblock


\bibitem[Lu et~al\mbox{.}(2023a)]%
        {lu2023seeing}
\bibfield{author}{\bibinfo{person}{Zeyu Lu}, \bibinfo{person}{Di Huang},
  \bibinfo{person}{Lei Bai}, \bibinfo{person}{Xihui Liu},
  \bibinfo{person}{Jingjing Qu}, {and} \bibinfo{person}{Wanli Ouyang}.}
  \bibinfo{year}{2023}\natexlab{a}.
\newblock \showarticletitle{Seeing is not always believing: A Quantitative
  Study on Human Perception of AI-Generated Images}.
\newblock \bibinfo{journal}{\emph{arXiv preprint arXiv:2304.13023}}
  (\bibinfo{year}{2023}).
\newblock


\bibitem[Lu et~al\mbox{.}(2023b)]%
        {NEURIPS2023_505df5ea}
\bibfield{author}{\bibinfo{person}{Zeyu Lu}, \bibinfo{person}{Di Huang},
  \bibinfo{person}{LEI BAI}, \bibinfo{person}{Jingjing Qu},
  \bibinfo{person}{Chengyue Wu}, \bibinfo{person}{Xihui Liu}, {and}
  \bibinfo{person}{Wanli Ouyang}.} \bibinfo{year}{2023}\natexlab{b}.
\newblock \showarticletitle{Seeing is not always believing: Benchmarking Human
  and Model Perception of AI-Generated Images}. In
  \bibinfo{booktitle}{\emph{Advances in Neural Information Processing
  Systems}}, Vol.~\bibinfo{volume}{36}. \bibinfo{pages}{25435--25447}.
\newblock


\bibitem[Lyu et~al\mbox{.}(2023a)]%
        {lyu2023pathway}
\bibfield{author}{\bibinfo{person}{Lingjuan Lyu}, \bibinfo{person}{C Chen},
  {and} \bibinfo{person}{J Fu}.} \bibinfo{year}{2023}\natexlab{a}.
\newblock \showarticletitle{A pathway towards responsible AI generated
  content}. In \bibinfo{booktitle}{\emph{Proc. 2nd tnt'I. Joint Conf.
  Artificial Intelligence}}.
\newblock


\bibitem[Lyu et~al\mbox{.}(2023b)]%
        {10167807}
\bibfield{author}{\bibinfo{person}{Yueming Lyu}, \bibinfo{person}{Yue Jiang},
  \bibinfo{person}{Ziwen He}, \bibinfo{person}{Bo Peng},
  \bibinfo{person}{Yunfan Liu}, {and} \bibinfo{person}{Jing Dong}.}
  \bibinfo{year}{2023}\natexlab{b}.
\newblock \showarticletitle{3D-Aware Adversarial Makeup Generation for Facial
  Privacy Protection}.
\newblock \bibinfo{journal}{\emph{IEEE Transactions on Pattern Analysis and
  Machine Intelligence}} \bibinfo{volume}{45}, \bibinfo{number}{11}
  (\bibinfo{year}{2023}), \bibinfo{pages}{13438--13453}.
\newblock
\urldef\tempurl%
\url{https://doi.org/10.1109/TPAMI.2023.3290175}
\showDOI{\tempurl}


\bibitem[Ma et~al\mbox{.}(2023b)]%
        {ma2023rdp}
\bibfield{author}{\bibinfo{person}{Chuan Ma}, \bibinfo{person}{Jun Li},
  \bibinfo{person}{Ming Ding}, \bibinfo{person}{Bo Liu}, \bibinfo{person}{Kang
  Wei}, \bibinfo{person}{Jian Weng}, {and} \bibinfo{person}{H~Vincent Poor}.}
  \bibinfo{year}{2023}\natexlab{b}.
\newblock \showarticletitle{RDP-GAN: A R{\'e}nyi-differential privacy based
  generative adversarial network}.
\newblock \bibinfo{journal}{\emph{IEEE Transactions on Dependable and Secure
  Computing}} (\bibinfo{year}{2023}).
\newblock


\bibitem[Ma et~al\mbox{.}(2023a)]%
        {ma2023exposing}
\bibfield{author}{\bibinfo{person}{RuiPeng Ma}, \bibinfo{person}{Jinhao Duan},
  \bibinfo{person}{Fei Kong}, \bibinfo{person}{Xiaoshuang Shi}, {and}
  \bibinfo{person}{Kaidi Xu}.} \bibinfo{year}{2023}\natexlab{a}.
\newblock \showarticletitle{Exposing the Fake: Effective Diffusion-Generated
  Images Detection}. In \bibinfo{booktitle}{\emph{The Second Workshop on New
  Frontiers in Adversarial Machine Learning}}.
\newblock
\urldef\tempurl%
\url{https://openreview.net/forum?id=7R62e4Wgim}
\showURL{%
\tempurl}


\bibitem[Ma et~al\mbox{.}(2023c)]%
        {ma2023generative}
\bibfield{author}{\bibinfo{person}{Yihan Ma}, \bibinfo{person}{Zhengyu Zhao},
  \bibinfo{person}{Xinlei He}, \bibinfo{person}{Zheng Li},
  \bibinfo{person}{Michael Backes}, {and} \bibinfo{person}{Yang Zhang}.}
  \bibinfo{year}{2023}\natexlab{c}.
\newblock \showarticletitle{Generative Watermarking Against Unauthorized
  Subject-Driven Image Synthesis}.
\newblock \bibinfo{journal}{\emph{arXiv preprint arXiv:2306.07754}}
  (\bibinfo{year}{2023}).
\newblock


\bibitem[Markov et~al\mbox{.}(2023)]%
        {markov2023holistic}
\bibfield{author}{\bibinfo{person}{Todor Markov}, \bibinfo{person}{Chong
  Zhang}, \bibinfo{person}{Sandhini Agarwal},
  \bibinfo{person}{Florentine~Eloundou Nekoul}, \bibinfo{person}{Theodore Lee},
  \bibinfo{person}{Steven Adler}, \bibinfo{person}{Angela Jiang}, {and}
  \bibinfo{person}{Lilian Weng}.} \bibinfo{year}{2023}\natexlab{}.
\newblock \showarticletitle{A holistic approach to undesired content detection
  in the real world}. In \bibinfo{booktitle}{\emph{Proceedings of the AAAI
  Conference on Artificial Intelligence}}, Vol.~\bibinfo{volume}{37}.
  \bibinfo{pages}{15009--15018}.
\newblock


\bibitem[Meehan et~al\mbox{.}(2020)]%
        {meehan2020non}
\bibfield{author}{\bibinfo{person}{Casey Meehan}, \bibinfo{person}{Kamalika
  Chaudhuri}, {and} \bibinfo{person}{Sanjoy Dasgupta}.}
  \bibinfo{year}{2020}\natexlab{}.
\newblock \showarticletitle{A non-parametric test to detect data-copying in
  generative models}. In \bibinfo{booktitle}{\emph{International Conference on
  Artificial Intelligence and Statistics}}.
\newblock


\bibitem[Melzi et~al\mbox{.}(2023)]%
        {melzi2023gandiffface}
\bibfield{author}{\bibinfo{person}{Pietro Melzi}, \bibinfo{person}{Christian
  Rathgeb}, \bibinfo{person}{Ruben Tolosana}, \bibinfo{person}{Ruben
  Vera-Rodriguez}, \bibinfo{person}{Dominik Lawatsch}, \bibinfo{person}{Florian
  Domin}, {and} \bibinfo{person}{Maxim Schaubert}.}
  \bibinfo{year}{2023}\natexlab{}.
\newblock \showarticletitle{Gandiffface: Controllable generation of synthetic
  datasets for face recognition with realistic variations}. In
  \bibinfo{booktitle}{\emph{Proceedings of the IEEE/CVF International
  Conference on Computer Vision}}. \bibinfo{pages}{3086--3095}.
\newblock


\bibitem[Mirsky and Lee(2021)]%
        {mirsky2021creation}
\bibfield{author}{\bibinfo{person}{Yisroel Mirsky} {and} \bibinfo{person}{Wenke
  Lee}.} \bibinfo{year}{2021}\natexlab{}.
\newblock \showarticletitle{The creation and detection of deepfakes: A survey}.
\newblock \bibinfo{journal}{\emph{ACM Computing Surveys (CSUR)}}
  \bibinfo{volume}{54}, \bibinfo{number}{1} (\bibinfo{year}{2021}),
  \bibinfo{pages}{1--41}.
\newblock


\bibitem[Mitchell et~al\mbox{.}(2023)]%
        {mitchell2023detectgpt}
\bibfield{author}{\bibinfo{person}{Eric Mitchell}, \bibinfo{person}{Yoonho
  Lee}, \bibinfo{person}{Alexander Khazatsky}, \bibinfo{person}{Christopher~D
  Manning}, {and} \bibinfo{person}{Chelsea Finn}.}
  \bibinfo{year}{2023}\natexlab{}.
\newblock \showarticletitle{Detectgpt: Zero-shot machine-generated text
  detection using probability curvature}. In
  \bibinfo{booktitle}{\emph{International Conference on Machine Learning}}.
  PMLR, \bibinfo{pages}{24950--24962}.
\newblock


\bibitem[Nichol et~al\mbox{.}(2021)]%
        {nichol2021glide}
\bibfield{author}{\bibinfo{person}{Alex Nichol}, \bibinfo{person}{Prafulla
  Dhariwal}, \bibinfo{person}{Aditya Ramesh}, \bibinfo{person}{Pranav Shyam},
  \bibinfo{person}{Pamela Mishkin}, \bibinfo{person}{Bob McGrew},
  \bibinfo{person}{Ilya Sutskever}, {and} \bibinfo{person}{Mark Chen}.}
  \bibinfo{year}{2021}\natexlab{}.
\newblock \showarticletitle{Glide: Towards photorealistic image generation and
  editing with text-guided diffusion models}.
\newblock \bibinfo{journal}{\emph{arXiv preprint arXiv:2112.10741}}
  (\bibinfo{year}{2021}).
\newblock


\bibitem[Pegoraro et~al\mbox{.}(2023)]%
        {pegoraro2023chatgpt}
\bibfield{author}{\bibinfo{person}{Alessandro Pegoraro},
  \bibinfo{person}{Kavita Kumari}, \bibinfo{person}{Hossein Fereidooni}, {and}
  \bibinfo{person}{Ahmad-Reza Sadeghi}.} \bibinfo{year}{2023}\natexlab{}.
\newblock \showarticletitle{To ChatGPT, or not to ChatGPT: That is the
  question!}
\newblock \bibinfo{journal}{\emph{arXiv preprint arXiv:2304.01487}}
  (\bibinfo{year}{2023}).
\newblock


\bibitem[Plant et~al\mbox{.}(2022)]%
        {plant2022you}
\bibfield{author}{\bibinfo{person}{Richard Plant}, \bibinfo{person}{Valerio
  Giuffrida}, {and} \bibinfo{person}{Dimitra Gkatzia}.}
  \bibinfo{year}{2022}\natexlab{}.
\newblock \showarticletitle{You Are What You Write: Preserving Privacy in the
  Era of Large Language Models}.
\newblock \bibinfo{journal}{\emph{arXiv preprint arXiv:2204.09391}}
  (\bibinfo{year}{2022}).
\newblock


\bibitem[Qi et~al\mbox{.}(2024)]%
        {qi2024hierarchical}
\bibfield{author}{\bibinfo{person}{Shuren Qi}, \bibinfo{person}{Yushu Zhang},
  \bibinfo{person}{Chao Wang}, \bibinfo{person}{Zhihua Xia},
  \bibinfo{person}{Jian Weng}, {and} \bibinfo{person}{Xiaochun Cao}.}
  \bibinfo{year}{2024}\natexlab{}.
\newblock \showarticletitle{Hierarchical Invariance for Robust and
  Interpretable Vision Tasks at Larger Scales}.
\newblock \bibinfo{journal}{\emph{arXiv preprint arXiv:2402.15430}}
  (\bibinfo{year}{2024}).
\newblock


\bibitem[Qu et~al\mbox{.}(2023)]%
        {qu2023unsafe}
\bibfield{author}{\bibinfo{person}{Yiting Qu}, \bibinfo{person}{Xinyue Shen},
  \bibinfo{person}{Xinlei He}, \bibinfo{person}{Michael Backes},
  \bibinfo{person}{Savvas Zannettou}, {and} \bibinfo{person}{Yang Zhang}.}
  \bibinfo{year}{2023}\natexlab{}.
\newblock \showarticletitle{Unsafe diffusion: On the generation of unsafe
  images and hateful memes from text-to-image models}. In
  \bibinfo{booktitle}{\emph{Proceedings of the 2023 ACM SIGSAC Conference on
  Computer and Communications Security}}. \bibinfo{pages}{3403--3417}.
\newblock


\bibitem[Rahman et~al\mbox{.}(2023)]%
        {rahman2023artifact}
\bibfield{author}{\bibinfo{person}{Md~Awsafur Rahman}, \bibinfo{person}{Bishmoy
  Paul}, \bibinfo{person}{Najibul~Haque Sarker}, \bibinfo{person}{Zaber
  Ibn~Abdul Hakim}, {and} \bibinfo{person}{Shaikh~Anowarul Fattah}.}
  \bibinfo{year}{2023}\natexlab{}.
\newblock \showarticletitle{Artifact: A large-scale dataset with artificial and
  factual images for generalizable and robust synthetic image detection}. In
  \bibinfo{booktitle}{\emph{2023 IEEE International Conference on Image
  Processing (ICIP)}}. IEEE, \bibinfo{pages}{2200--2204}.
\newblock


\bibitem[Rando et~al\mbox{.}(2022)]%
        {rando2022red}
\bibfield{author}{\bibinfo{person}{Javier Rando}, \bibinfo{person}{Daniel
  Paleka}, \bibinfo{person}{David Lindner}, \bibinfo{person}{Lennard Heim},
  {and} \bibinfo{person}{Florian Tram{\`e}r}.} \bibinfo{year}{2022}\natexlab{}.
\newblock \showarticletitle{Red-teaming the stable diffusion safety filter}.
\newblock \bibinfo{journal}{\emph{arXiv preprint arXiv:2210.04610}}
  (\bibinfo{year}{2022}).
\newblock


\bibitem[Rosati(2022)]%
        {rosati2022synscipass}
\bibfield{author}{\bibinfo{person}{Domenic Rosati}.}
  \bibinfo{year}{2022}\natexlab{}.
\newblock \showarticletitle{SynSciPass: detecting appropriate uses of
  scientific text generation}.
\newblock \bibinfo{journal}{\emph{arXiv preprint arXiv:2209.03742}}
  (\bibinfo{year}{2022}).
\newblock


\bibitem[Saleh et~al\mbox{.}(2024)]%
        {saleh2024messagebrokersgenerativeai}
\bibfield{author}{\bibinfo{person}{Alaa Saleh}, \bibinfo{person}{Roberto
  Morabito}, \bibinfo{person}{Sasu Tarkoma}, \bibinfo{person}{Susanna
  Pirttikangas}, {and} \bibinfo{person}{Lauri Lovén}.}
  \bibinfo{year}{2024}\natexlab{}.
\newblock \bibinfo{title}{Towards Message Brokers for Generative AI: Survey,
  Challenges, and Opportunities}.
\newblock
\newblock
\showeprint[arxiv]{2312.14647}~[cs.DC]
\urldef\tempurl%
\url{https://arxiv.org/abs/2312.14647}
\showURL{%
\tempurl}


\bibitem[Schramowski et~al\mbox{.}(2023)]%
        {schramowski2023safe}
\bibfield{author}{\bibinfo{person}{Patrick Schramowski},
  \bibinfo{person}{Manuel Brack}, \bibinfo{person}{Bj{\"o}rn Deiseroth}, {and}
  \bibinfo{person}{Kristian Kersting}.} \bibinfo{year}{2023}\natexlab{}.
\newblock \showarticletitle{Safe latent diffusion: Mitigating inappropriate
  degeneration in diffusion models}. In \bibinfo{booktitle}{\emph{Proceedings
  of the IEEE/CVF Conference on Computer Vision and Pattern Recognition}}.
  \bibinfo{pages}{22522--22531}.
\newblock


\bibitem[Schuhmann et~al\mbox{.}(2022)]%
        {schuhmann2022laion}
\bibfield{author}{\bibinfo{person}{Christoph Schuhmann},
  \bibinfo{person}{Romain Beaumont}, \bibinfo{person}{Richard Vencu},
  \bibinfo{person}{Cade Gordon}, \bibinfo{person}{Ross Wightman},
  \bibinfo{person}{Mehdi Cherti}, \bibinfo{person}{Theo Coombes},
  \bibinfo{person}{Aarush Katta}, \bibinfo{person}{Clayton Mullis},
  \bibinfo{person}{Mitchell Wortsman}, {et~al\mbox{.}}}
  \bibinfo{year}{2022}\natexlab{}.
\newblock \showarticletitle{Laion-5b: An open large-scale dataset for training
  next generation image-text models}.
\newblock \bibinfo{journal}{\emph{Advances in Neural Information Processing
  Systems}}  \bibinfo{volume}{35} (\bibinfo{year}{2022}),
  \bibinfo{pages}{25278--25294}.
\newblock


\bibitem[Sha et~al\mbox{.}(2023)]%
        {sha2023fake}
\bibfield{author}{\bibinfo{person}{Zeyang Sha}, \bibinfo{person}{Zheng Li},
  \bibinfo{person}{Ning Yu}, {and} \bibinfo{person}{Yang Zhang}.}
  \bibinfo{year}{2023}\natexlab{}.
\newblock \showarticletitle{De-fake: Detection and attribution of fake images
  generated by text-to-image generation models}. In
  \bibinfo{booktitle}{\emph{Proceedings of the 2023 ACM SIGSAC Conference on
  Computer and Communications Security}}. \bibinfo{pages}{3418--3432}.
\newblock


\bibitem[Shan et~al\mbox{.}(2023)]%
        {shan2023glaze}
\bibfield{author}{\bibinfo{person}{Shawn Shan}, \bibinfo{person}{Jenna Cryan},
  \bibinfo{person}{Emily Wenger}, \bibinfo{person}{Haitao Zheng},
  \bibinfo{person}{Rana Hanocka}, {and} \bibinfo{person}{Ben~Y Zhao}.}
  \bibinfo{year}{2023}\natexlab{}.
\newblock \showarticletitle{Glaze: Protecting Artists from Style Mimicry by
  $\{$Text-to-Image$\}$ Models}. In \bibinfo{booktitle}{\emph{32nd USENIX
  Security Symposium (USENIX Security 23)}}. \bibinfo{pages}{2187--2204}.
\newblock


\bibitem[Sheng et~al\mbox{.}(2019)]%
        {sheng2019woman}
\bibfield{author}{\bibinfo{person}{Emily Sheng}, \bibinfo{person}{Kai-Wei
  Chang}, \bibinfo{person}{Premkumar Natarajan}, {and} \bibinfo{person}{Nanyun
  Peng}.} \bibinfo{year}{2019}\natexlab{}.
\newblock \showarticletitle{The woman worked as a babysitter: On biases in
  language generation}.
\newblock \bibinfo{journal}{\emph{arXiv preprint arXiv:1909.01326}}
  (\bibinfo{year}{2019}).
\newblock


\bibitem[Sinitsa and Fried(2024)]%
        {sinitsa2024deep}
\bibfield{author}{\bibinfo{person}{Sergey Sinitsa} {and} \bibinfo{person}{Ohad
  Fried}.} \bibinfo{year}{2024}\natexlab{}.
\newblock \showarticletitle{Deep image fingerprint: Towards low budget
  synthetic image detection and model lineage analysis}. In
  \bibinfo{booktitle}{\emph{Proceedings of the IEEE/CVF Winter Conference on
  Applications of Computer Vision}}. \bibinfo{pages}{4067--4076}.
\newblock


\bibitem[Somepalli et~al\mbox{.}(2023)]%
        {somepalli2023diffusion}
\bibfield{author}{\bibinfo{person}{Gowthami Somepalli}, \bibinfo{person}{Vasu
  Singla}, \bibinfo{person}{Micah Goldblum}, \bibinfo{person}{Jonas Geiping},
  {and} \bibinfo{person}{Tom Goldstein}.} \bibinfo{year}{2023}\natexlab{}.
\newblock \showarticletitle{Diffusion art or digital forgery? investigating
  data replication in diffusion models}. In
  \bibinfo{booktitle}{\emph{Proceedings of the IEEE/CVF Conference on Computer
  Vision and Pattern Recognition}}. \bibinfo{pages}{6048--6058}.
\newblock


\bibitem[Tang et~al\mbox{.}(2024)]%
        {tang2024gendercare}
\bibfield{author}{\bibinfo{person}{Kunsheng Tang}, \bibinfo{person}{Wenbo
  Zhou}, \bibinfo{person}{Jie Zhang}, \bibinfo{person}{Aishan Liu},
  \bibinfo{person}{Gelei Deng}, \bibinfo{person}{Shuai Li},
  \bibinfo{person}{Peigui Qi}, \bibinfo{person}{Weiming Zhang},
  \bibinfo{person}{Tianwei Zhang}, {and} \bibinfo{person}{Nenghai Yu}.}
  \bibinfo{year}{2024}\natexlab{}.
\newblock \showarticletitle{GenderCARE: A Comprehensive Framework for Assessing
  and Reducing Gender Bias in Large Language Models}. In
  \bibinfo{booktitle}{\emph{ACM Conference on Computer and Communications
  Security(CCS 24)}}.
\newblock


\bibitem[Thambawita et~al\mbox{.}(2022)]%
        {thambawita2022singan}
\bibfield{author}{\bibinfo{person}{Vajira Thambawita}, \bibinfo{person}{Pegah
  Salehi}, \bibinfo{person}{Sajad~Amouei Sheshkal}, \bibinfo{person}{Steven~A
  Hicks}, \bibinfo{person}{Hugo~L Hammer}, \bibinfo{person}{Sravanthi Parasa},
  \bibinfo{person}{Thomas~de Lange}, \bibinfo{person}{P{\aa}l Halvorsen}, {and}
  \bibinfo{person}{Michael~A Riegler}.} \bibinfo{year}{2022}\natexlab{}.
\newblock \showarticletitle{SinGAN-Seg: Synthetic training data generation for
  medical image segmentation}.
\newblock \bibinfo{journal}{\emph{PloS one}} \bibinfo{volume}{17},
  \bibinfo{number}{5} (\bibinfo{year}{2022}), \bibinfo{pages}{e0267976}.
\newblock


\bibitem[Tirumala et~al\mbox{.}(2022)]%
        {tirumala2022memorization}
\bibfield{author}{\bibinfo{person}{Kushal Tirumala}, \bibinfo{person}{Aram
  Markosyan}, \bibinfo{person}{Luke Zettlemoyer}, {and} \bibinfo{person}{Armen
  Aghajanyan}.} \bibinfo{year}{2022}\natexlab{}.
\newblock \showarticletitle{Memorization without overfitting: Analyzing the
  training dynamics of large language models}.
\newblock \bibinfo{journal}{\emph{Advances in Neural Information Processing
  Systems}}  \bibinfo{volume}{35} (\bibinfo{year}{2022}),
  \bibinfo{pages}{38274--38290}.
\newblock


\bibitem[Tu et~al\mbox{.}(2023)]%
        {tu2023waterbench}
\bibfield{author}{\bibinfo{person}{Shangqing Tu}, \bibinfo{person}{Yuliang
  Sun}, \bibinfo{person}{Yushi Bai}, \bibinfo{person}{Jifan Yu},
  \bibinfo{person}{Lei Hou}, {and} \bibinfo{person}{Juanzi Li}.}
  \bibinfo{year}{2023}\natexlab{}.
\newblock \showarticletitle{Waterbench: Towards holistic evaluation of
  watermarks for large language models}.
\newblock \bibinfo{journal}{\emph{arXiv preprint arXiv:2311.07138}}
  (\bibinfo{year}{2023}).
\newblock


\bibitem[Tulchinskii et~al\mbox{.}(2024)]%
        {tulchinskii2024intrinsic}
\bibfield{author}{\bibinfo{person}{Eduard Tulchinskii},
  \bibinfo{person}{Kristian Kuznetsov}, \bibinfo{person}{Laida Kushnareva},
  \bibinfo{person}{Daniil Cherniavskii}, \bibinfo{person}{Sergey Nikolenko},
  \bibinfo{person}{Evgeny Burnaev}, \bibinfo{person}{Serguei Barannikov}, {and}
  \bibinfo{person}{Irina Piontkovskaya}.} \bibinfo{year}{2024}\natexlab{}.
\newblock \showarticletitle{Intrinsic dimension estimation for robust detection
  of ai-generated texts}.
\newblock \bibinfo{journal}{\emph{Advances in Neural Information Processing
  Systems}}  \bibinfo{volume}{36} (\bibinfo{year}{2024}).
\newblock


\bibitem[Uchendu et~al\mbox{.}(2021)]%
        {uchendu2021turingbench}
\bibfield{author}{\bibinfo{person}{Adaku Uchendu}, \bibinfo{person}{Zeyu Ma},
  \bibinfo{person}{Thai Le}, \bibinfo{person}{Rui Zhang}, {and}
  \bibinfo{person}{Dongwon Lee}.} \bibinfo{year}{2021}\natexlab{}.
\newblock \showarticletitle{TURINGBENCH: A Benchmark Environment for Turing
  Test in the Age of Neural Text Generation}. In
  \bibinfo{booktitle}{\emph{Findings of the Association for Computational
  Linguistics: EMNLP 2021}}. \bibinfo{pages}{2001--2016}.
\newblock


\bibitem[Van~Le et~al\mbox{.}(2023)]%
        {van2023anti}
\bibfield{author}{\bibinfo{person}{Thanh Van~Le}, \bibinfo{person}{Hao Phung},
  \bibinfo{person}{Thuan~Hoang Nguyen}, \bibinfo{person}{Quan Dao},
  \bibinfo{person}{Ngoc~N Tran}, {and} \bibinfo{person}{Anh Tran}.}
  \bibinfo{year}{2023}\natexlab{}.
\newblock \showarticletitle{Anti-DreamBooth: Protecting users from personalized
  text-to-image synthesis}. In \bibinfo{booktitle}{\emph{Proceedings of the
  IEEE/CVF International Conference on Computer Vision}}.
  \bibinfo{pages}{2116--2127}.
\newblock


\bibitem[Verdoliva(2020)]%
        {verdoliva2020media}
\bibfield{author}{\bibinfo{person}{Luisa Verdoliva}.}
  \bibinfo{year}{2020}\natexlab{}.
\newblock \showarticletitle{Media forensics and deepfakes: an overview}.
\newblock \bibinfo{journal}{\emph{IEEE Journal of Selected Topics in Signal
  Processing}} \bibinfo{volume}{14}, \bibinfo{number}{5}
  (\bibinfo{year}{2020}), \bibinfo{pages}{910--932}.
\newblock


\bibitem[Verma et~al\mbox{.}(2023)]%
        {verma2023ghostbuster}
\bibfield{author}{\bibinfo{person}{Vivek Verma}, \bibinfo{person}{Eve Fleisig},
  \bibinfo{person}{Nicholas Tomlin}, {and} \bibinfo{person}{Dan Klein}.}
  \bibinfo{year}{2023}\natexlab{}.
\newblock \showarticletitle{Ghostbuster: Detecting text ghostwritten by large
  language models}.
\newblock \bibinfo{journal}{\emph{arXiv preprint arXiv:2305.15047}}
  (\bibinfo{year}{2023}).
\newblock


\bibitem[Wang et~al\mbox{.}(2023c)]%
        {wang2023survey}
\bibfield{author}{\bibinfo{person}{Cunxiang Wang}, \bibinfo{person}{Xiaoze
  Liu}, \bibinfo{person}{Yuanhao Yue}, \bibinfo{person}{Xiangru Tang},
  \bibinfo{person}{Tianhang Zhang}, \bibinfo{person}{Cheng Jiayang},
  \bibinfo{person}{Yunzhi Yao}, \bibinfo{person}{Wenyang Gao},
  \bibinfo{person}{Xuming Hu}, \bibinfo{person}{Zehan Qi}, {et~al\mbox{.}}}
  \bibinfo{year}{2023}\natexlab{c}.
\newblock \showarticletitle{Survey on factuality in large language models:
  Knowledge, retrieval and domain-specificity}.
\newblock \bibinfo{journal}{\emph{arXiv preprint arXiv:2310.07521}}
  (\bibinfo{year}{2023}).
\newblock


\bibitem[Wang et~al\mbox{.}(2024a)]%
        {298240}
\bibfield{author}{\bibinfo{person}{Haichen Wang}, \bibinfo{person}{Shuchao
  Pang}, \bibinfo{person}{Zhigang Lu}, \bibinfo{person}{Yihang Rao},
  \bibinfo{person}{Yongbin Zhou}, {and} \bibinfo{person}{Minhui Xue}.}
  \bibinfo{year}{2024}\natexlab{a}.
\newblock \showarticletitle{dp-promise: Differentially Private Diffusion
  Probabilistic Models for Image Synthesis}. In \bibinfo{booktitle}{\emph{33rd
  USENIX Security Symposium (USENIX Security 24)}}.
  \bibinfo{pages}{1063--1080}.
\newblock
\showISBNx{978-1-939133-44-1}


\bibitem[Wang et~al\mbox{.}(2023b)]%
        {wang2023evaluating}
\bibfield{author}{\bibinfo{person}{Sheng-Yu Wang}, \bibinfo{person}{Alexei~A
  Efros}, \bibinfo{person}{Jun-Yan Zhu}, {and} \bibinfo{person}{Richard
  Zhang}.} \bibinfo{year}{2023}\natexlab{b}.
\newblock \showarticletitle{Evaluating data attribution for text-to-image
  models}. In \bibinfo{booktitle}{\emph{Proceedings of the IEEE/CVF
  International Conference on Computer Vision}}. \bibinfo{pages}{7192--7203}.
\newblock


\bibitem[Wang et~al\mbox{.}(2024b)]%
        {10646362}
\bibfield{author}{\bibinfo{person}{Tao Wang}, \bibinfo{person}{Yushu Zhang},
  \bibinfo{person}{Zixuan Yang}, \bibinfo{person}{Xiangli Xiao},
  \bibinfo{person}{Hua Zhang}, {and} \bibinfo{person}{Zhongyun Hua}.}
  \bibinfo{year}{2024}\natexlab{b}.
\newblock \showarticletitle{Seeing is not Believing: An Identity Hider for
  Human Vision Privacy Protection}.
\newblock \bibinfo{journal}{\emph{IEEE Transactions on Biometrics, Behavior,
  and Identity Science}} (\bibinfo{year}{2024}), \bibinfo{pages}{1--1}.
\newblock
\urldef\tempurl%
\url{https://doi.org/10.1109/TBIOM.2024.3449849}
\showDOI{\tempurl}


\bibitem[Wang et~al\mbox{.}(2023d)]%
        {wang2023identifiable}
\bibfield{author}{\bibinfo{person}{Tao Wang}, \bibinfo{person}{Yushu Zhang},
  \bibinfo{person}{Ruoyu Zhao}, \bibinfo{person}{Wenying Wen}, {and}
  \bibinfo{person}{Rushi Lan}.} \bibinfo{year}{2023}\natexlab{d}.
\newblock \showarticletitle{Identifiable Face Privacy Protection via Virtual
  Identity Transformation}.
\newblock \bibinfo{journal}{\emph{IEEE Signal Processing Letters}}
  (\bibinfo{year}{2023}).
\newblock


\bibitem[Wang et~al\mbox{.}(2022b)]%
        {wang2022survey}
\bibfield{author}{\bibinfo{person}{Yuntao Wang}, \bibinfo{person}{Zhou Su},
  \bibinfo{person}{Ning Zhang}, \bibinfo{person}{Rui Xing},
  \bibinfo{person}{Dongxiao Liu}, \bibinfo{person}{Tom~H Luan}, {and}
  \bibinfo{person}{Xuemin Shen}.} \bibinfo{year}{2022}\natexlab{b}.
\newblock \showarticletitle{A survey on metaverse: Fundamentals, security, and
  privacy}.
\newblock \bibinfo{journal}{\emph{IEEE Communications Surveys \& Tutorials}}
  (\bibinfo{year}{2022}).
\newblock


\bibitem[Wang et~al\mbox{.}(2023a)]%
        {wang2023dire}
\bibfield{author}{\bibinfo{person}{Zhendong Wang}, \bibinfo{person}{Jianmin
  Bao}, \bibinfo{person}{Wengang Zhou}, \bibinfo{person}{Weilun Wang},
  \bibinfo{person}{Hezhen Hu}, \bibinfo{person}{Hong Chen}, {and}
  \bibinfo{person}{Houqiang Li}.} \bibinfo{year}{2023}\natexlab{a}.
\newblock \showarticletitle{Dire for diffusion-generated image detection}. In
  \bibinfo{booktitle}{\emph{Proceedings of the IEEE/CVF International
  Conference on Computer Vision}}. \bibinfo{pages}{22445--22455}.
\newblock


\bibitem[Wang et~al\mbox{.}(2022a)]%
        {wang2022diffusiondb}
\bibfield{author}{\bibinfo{person}{Zijie~J Wang}, \bibinfo{person}{Evan
  Montoya}, \bibinfo{person}{David Munechika}, \bibinfo{person}{Haoyang Yang},
  \bibinfo{person}{Benjamin Hoover}, {and} \bibinfo{person}{Duen~Horng Chau}.}
  \bibinfo{year}{2022}\natexlab{a}.
\newblock \showarticletitle{Diffusiondb: A large-scale prompt gallery dataset
  for text-to-image generative models}.
\newblock \bibinfo{journal}{\emph{arXiv preprint arXiv:2210.14896}}
  (\bibinfo{year}{2022}).
\newblock


\bibitem[Webster(2023)]%
        {webster2023reproducible}
\bibfield{author}{\bibinfo{person}{Ryan Webster}.}
  \bibinfo{year}{2023}\natexlab{}.
\newblock \showarticletitle{A Reproducible Extraction of Training Images from
  Diffusion Models}.
\newblock \bibinfo{journal}{\emph{arXiv preprint arXiv:2305.08694}}
  (\bibinfo{year}{2023}).
\newblock


\bibitem[Webster et~al\mbox{.}(2019)]%
        {webster2019detecting}
\bibfield{author}{\bibinfo{person}{Ryan Webster}, \bibinfo{person}{Julien
  Rabin}, \bibinfo{person}{Loic Simon}, {and} \bibinfo{person}{Fr{\'e}d{\'e}ric
  Jurie}.} \bibinfo{year}{2019}\natexlab{}.
\newblock \showarticletitle{Detecting overfitting of deep generative networks
  via latent recovery}. In \bibinfo{booktitle}{\emph{Proceedings of the
  IEEE/CVF Conference on Computer Vision and Pattern Recognition}}.
  \bibinfo{pages}{11273--11282}.
\newblock


\bibitem[Wen et~al\mbox{.}(2023)]%
        {wen2023divide}
\bibfield{author}{\bibinfo{person}{Yunqian Wen}, \bibinfo{person}{Bo Liu},
  \bibinfo{person}{Jingyi Cao}, \bibinfo{person}{Rong Xie}, {and}
  \bibinfo{person}{Li Song}.} \bibinfo{year}{2023}\natexlab{}.
\newblock \showarticletitle{Divide and conquer: a two-step method for high
  quality face de-identification with model explainability}. In
  \bibinfo{booktitle}{\emph{Proceedings of the IEEE/CVF International
  Conference on Computer Vision}}. \bibinfo{pages}{5148--5157}.
\newblock


\bibitem[Wen et~al\mbox{.}(2022)]%
        {wen2022identitydp}
\bibfield{author}{\bibinfo{person}{Yunqian Wen}, \bibinfo{person}{Bo Liu},
  \bibinfo{person}{Ming Ding}, \bibinfo{person}{Rong Xie}, {and}
  \bibinfo{person}{Li Song}.} \bibinfo{year}{2022}\natexlab{}.
\newblock \showarticletitle{Identitydp: Differential private identification
  protection for face images}.
\newblock \bibinfo{journal}{\emph{Neurocomputing}}  \bibinfo{volume}{501}
  (\bibinfo{year}{2022}), \bibinfo{pages}{197--211}.
\newblock


\bibitem[Wu et~al\mbox{.}(2023)]%
        {wu2023promptrobust}
\bibfield{author}{\bibinfo{person}{Ruijia Wu}, \bibinfo{person}{Yuhang Wang},
  \bibinfo{person}{Huafeng Shi}, \bibinfo{person}{Zhipeng Yu},
  \bibinfo{person}{Yichao Wu}, {and} \bibinfo{person}{Ding Liang}.}
  \bibinfo{year}{2023}\natexlab{}.
\newblock \bibinfo{title}{Towards Prompt-robust Face Privacy Protection via
  Adversarial Decoupling Augmentation Framework}.
\newblock
\newblock
\showeprint[arxiv]{2305.03980}~[cs.CV]


\bibitem[Wu et~al\mbox{.}(2024)]%
        {294566}
\bibfield{author}{\bibinfo{person}{Yixin Wu}, \bibinfo{person}{Rui Wen},
  \bibinfo{person}{Michael Backes}, \bibinfo{person}{Pascal Berrang},
  \bibinfo{person}{Mathias Humbert}, \bibinfo{person}{Yun Shen}, {and}
  \bibinfo{person}{Yang Zhang}.} \bibinfo{year}{2024}\natexlab{}.
\newblock \showarticletitle{Quantifying Privacy Risks of Prompts in Visual
  Prompt Learning}. In \bibinfo{booktitle}{\emph{33rd USENIX Security Symposium
  (USENIX Security 24)}}. \bibinfo{address}{Philadelphia, PA},
  \bibinfo{pages}{5841--5858}.
\newblock


\bibitem[Xi et~al\mbox{.}(2023)]%
        {xi2023ai}
\bibfield{author}{\bibinfo{person}{Ziyi Xi}, \bibinfo{person}{Wenmin Huang},
  \bibinfo{person}{Kangkang Wei}, \bibinfo{person}{Weiqi Luo}, {and}
  \bibinfo{person}{Peijia Zheng}.} \bibinfo{year}{2023}\natexlab{}.
\newblock \showarticletitle{Ai-generated image detection using a
  cross-attention enhanced dual-stream network}. In
  \bibinfo{booktitle}{\emph{2023 Asia Pacific Signal and Information Processing
  Association Annual Summit and Conference (APSIPA ASC)}}. IEEE,
  \bibinfo{pages}{1463--1470}.
\newblock


\bibitem[Xiong et~al\mbox{.}(2023)]%
        {10.1145/3581783.3612448}
\bibfield{author}{\bibinfo{person}{Cheng Xiong}, \bibinfo{person}{Chuan Qin},
  \bibinfo{person}{Guorui Feng}, {and} \bibinfo{person}{Xinpeng Zhang}.}
  \bibinfo{year}{2023}\natexlab{}.
\newblock \showarticletitle{Flexible and Secure Watermarking for Latent
  Diffusion Model}. In \bibinfo{booktitle}{\emph{Proceedings of the 31st ACM
  International Conference on Multimedia}} (<conf-loc>, <city>Ottawa ON</city>,
  <country>Canada</country>, </conf-loc>) \emph{(\bibinfo{series}{MM '23})}.
  \bibinfo{publisher}{Association for Computing Machinery},
  \bibinfo{address}{New York, NY, USA}, \bibinfo{pages}{1668–1676}.
\newblock
\showISBNx{9798400701085}
\urldef\tempurl%
\url{https://doi.org/10.1145/3581783.3612448}
\showDOI{\tempurl}


\bibitem[Xu et~al\mbox{.}(2024a)]%
        {xu2024unleashing}
\bibfield{author}{\bibinfo{person}{Minrui Xu}, \bibinfo{person}{Hongyang Du},
  \bibinfo{person}{Dusit Niyato}, \bibinfo{person}{Jiawen Kang},
  \bibinfo{person}{Zehui Xiong}, \bibinfo{person}{Shiwen Mao},
  \bibinfo{person}{Zhu Han}, \bibinfo{person}{Abbas Jamalipour},
  \bibinfo{person}{Dong~In Kim}, \bibinfo{person}{Xuemin Shen},
  {et~al\mbox{.}}} \bibinfo{year}{2024}\natexlab{a}.
\newblock \showarticletitle{Unleashing the power of edge-cloud generative ai in
  mobile networks: A survey of aigc services}.
\newblock \bibinfo{journal}{\emph{IEEE Communications Surveys \& Tutorials}}
  (\bibinfo{year}{2024}).
\newblock


\bibitem[Xu et~al\mbox{.}(2024b)]%
        {xu2024walking}
\bibfield{author}{\bibinfo{person}{Rongwu Xu}, \bibinfo{person}{Zi'an Zhou},
  \bibinfo{person}{Tianwei Zhang}, \bibinfo{person}{Zehan Qi},
  \bibinfo{person}{Su Yao}, \bibinfo{person}{Ke Xu}, \bibinfo{person}{Wei Xu},
  {and} \bibinfo{person}{Han Qiu}.} \bibinfo{year}{2024}\natexlab{b}.
\newblock \showarticletitle{Walking in Others' Shoes: How Perspective-Taking
  Guides Large Language Models in Reducing Toxicity and Bias}.
\newblock \bibinfo{journal}{\emph{arXiv preprint arXiv:2407.15366}}
  (\bibinfo{year}{2024}).
\newblock


\bibitem[Yan et~al\mbox{.}(2023)]%
        {DeepfakeBench_YAN_NEURIPS2023}
\bibfield{author}{\bibinfo{person}{Zhiyuan Yan}, \bibinfo{person}{Yong Zhang},
  \bibinfo{person}{Xinhang Yuan}, \bibinfo{person}{Siwei Lyu}, {and}
  \bibinfo{person}{Baoyuan Wu}.} \bibinfo{year}{2023}\natexlab{}.
\newblock \showarticletitle{DeepfakeBench: A Comprehensive Benchmark of
  Deepfake Detection}. In \bibinfo{booktitle}{\emph{Advances in Neural
  Information Processing Systems}}, Vol.~\bibinfo{volume}{36}.
  \bibinfo{publisher}{Curran Associates, Inc.}, \bibinfo{pages}{4534--4565}.
\newblock


\bibitem[Yang et~al\mbox{.}(2024a)]%
        {yang2024leandojo}
\bibfield{author}{\bibinfo{person}{Kaiyu Yang}, \bibinfo{person}{Aidan Swope},
  \bibinfo{person}{Alex Gu}, \bibinfo{person}{Rahul Chalamala},
  \bibinfo{person}{Peiyang Song}, \bibinfo{person}{Shixing Yu},
  \bibinfo{person}{Saad Godil}, \bibinfo{person}{Ryan~J Prenger}, {and}
  \bibinfo{person}{Animashree Anandkumar}.} \bibinfo{year}{2024}\natexlab{a}.
\newblock \showarticletitle{Leandojo: Theorem proving with retrieval-augmented
  language models}.
\newblock \bibinfo{journal}{\emph{Advances in Neural Information Processing
  Systems}}  \bibinfo{volume}{36} (\bibinfo{year}{2024}).
\newblock


\bibitem[Yang et~al\mbox{.}(2022)]%
        {yang2022deepfake}
\bibfield{author}{\bibinfo{person}{Tianyun Yang}, \bibinfo{person}{Ziyao
  Huang}, \bibinfo{person}{Juan Cao}, \bibinfo{person}{Lei Li}, {and}
  \bibinfo{person}{Xirong Li}.} \bibinfo{year}{2022}\natexlab{}.
\newblock \showarticletitle{Deepfake network architecture attribution}. In
  \bibinfo{booktitle}{\emph{Proceedings of the AAAI Conference on Artificial
  Intelligence}}, Vol.~\bibinfo{volume}{36}. \bibinfo{pages}{4662--4670}.
\newblock


\bibitem[Yang et~al\mbox{.}(2023)]%
        {yang2023progressive}
\bibfield{author}{\bibinfo{person}{Tianyun Yang}, \bibinfo{person}{Danding
  Wang}, \bibinfo{person}{Fan Tang}, \bibinfo{person}{Xinying Zhao},
  \bibinfo{person}{Juan Cao}, {and} \bibinfo{person}{Sheng Tang}.}
  \bibinfo{year}{2023}\natexlab{}.
\newblock \showarticletitle{Progressive Open Space Expansion for Open-Set Model
  Attribution}. In \bibinfo{booktitle}{\emph{Proceedings of the IEEE/CVF
  Conference on Computer Vision and Pattern Recognition}}.
  \bibinfo{pages}{15856--15865}.
\newblock


\bibitem[Yang et~al\mbox{.}(2024b)]%
        {yang2024gaussian}
\bibfield{author}{\bibinfo{person}{Zijin Yang}, \bibinfo{person}{Kai Zeng},
  \bibinfo{person}{Kejiang Chen}, \bibinfo{person}{Han Fang},
  \bibinfo{person}{Weiming Zhang}, {and} \bibinfo{person}{Nenghai Yu}.}
  \bibinfo{year}{2024}\natexlab{b}.
\newblock \showarticletitle{Gaussian Shading: Provable Performance-Lossless
  Image Watermarking for Diffusion Models}. In
  \bibinfo{booktitle}{\emph{Proceedings of the IEEE/CVF Conference on Computer
  Vision and Pattern Recognition}}. \bibinfo{pages}{12162--12171}.
\newblock


\bibitem[Yao et~al\mbox{.}(2024)]%
        {yao2024PromptCARE}
\bibfield{author}{\bibinfo{person}{Hongwei Yao}, \bibinfo{person}{Jian Lou},
  \bibinfo{person}{Kui Ren}, {and} \bibinfo{person}{Zhan Qin}.}
  \bibinfo{year}{2024}\natexlab{}.
\newblock \showarticletitle{PromptCARE: Prompt Copyright Protection by
  Watermark Injection and Verification}. In \bibinfo{booktitle}{\emph{IEEE
  Symposium on Security and Privacy (S\&P)}}. \bibinfo{publisher}{IEEE}.
\newblock


\bibitem[Yao et~al\mbox{.}(2023)]%
        {yao2023ppup}
\bibfield{author}{\bibinfo{person}{Zhexin Yao}, \bibinfo{person}{Qiuming Liu},
  \bibinfo{person}{Jingkang Yang}, \bibinfo{person}{Yanan Chen}, {and}
  \bibinfo{person}{Zhen Wu}.} \bibinfo{year}{2023}\natexlab{}.
\newblock \showarticletitle{PPUP-GAN: A GAN-based privacy-protecting method for
  aerial photography}.
\newblock \bibinfo{journal}{\emph{Future Generation Computer Systems}}
  \bibinfo{volume}{145} (\bibinfo{year}{2023}), \bibinfo{pages}{284--292}.
\newblock


\bibitem[Yeh et~al\mbox{.}(2021)]%
        {yeh2021attack}
\bibfield{author}{\bibinfo{person}{Chin-Yuan Yeh}, \bibinfo{person}{Hsi-Wen
  Chen}, \bibinfo{person}{Hong-Han Shuai}, \bibinfo{person}{De-Nian Yang},
  {and} \bibinfo{person}{Ming-Syan Chen}.} \bibinfo{year}{2021}\natexlab{}.
\newblock \showarticletitle{Attack as the best defense: Nullifying
  image-to-image translation gans via limit-aware adversarial attack}. In
  \bibinfo{booktitle}{\emph{Proceedings of the IEEE/CVF International
  Conference on Computer Vision}}. \bibinfo{pages}{16188--16197}.
\newblock


\bibitem[Yu et~al\mbox{.}(2019)]%
        {yu2019attributing}
\bibfield{author}{\bibinfo{person}{Ning Yu}, \bibinfo{person}{Larry~S Davis},
  {and} \bibinfo{person}{Mario Fritz}.} \bibinfo{year}{2019}\natexlab{}.
\newblock \showarticletitle{Attributing fake images to gans: Learning and
  analyzing gan fingerprints}. In \bibinfo{booktitle}{\emph{Proceedings of the
  IEEE/CVF international conference on computer vision}}.
  \bibinfo{pages}{7556--7566}.
\newblock


\bibitem[Yu et~al\mbox{.}(2021)]%
        {yu2021artificial}
\bibfield{author}{\bibinfo{person}{Ning Yu}, \bibinfo{person}{Vladislav
  Skripniuk}, \bibinfo{person}{Sahar Abdelnabi}, {and} \bibinfo{person}{Mario
  Fritz}.} \bibinfo{year}{2021}\natexlab{}.
\newblock \showarticletitle{Artificial fingerprinting for generative models:
  Rooting deepfake attribution in training data}. In
  \bibinfo{booktitle}{\emph{Proceedings of the IEEE/CVF International
  conference on computer vision}}. \bibinfo{pages}{14448--14457}.
\newblock


\bibitem[Yuan et~al\mbox{.}(2022)]%
        {yuan2022generating}
\bibfield{author}{\bibinfo{person}{Zhuowen Yuan}, \bibinfo{person}{Zhengxin
  You}, \bibinfo{person}{Sheng Li}, \bibinfo{person}{Zhenxing Qian},
  \bibinfo{person}{Xinpeng Zhang}, {and} \bibinfo{person}{Alex Kot}.}
  \bibinfo{year}{2022}\natexlab{}.
\newblock \showarticletitle{On generating identifiable virtual faces}. In
  \bibinfo{booktitle}{\emph{Proceedings of the 30th ACM International
  Conference on Multimedia}}. \bibinfo{pages}{1465--1473}.
\newblock


\bibitem[Zeng et~al\mbox{.}(2023)]%
        {zeng2023securing}
\bibfield{author}{\bibinfo{person}{Yu Zeng}, \bibinfo{person}{Mo Zhou},
  \bibinfo{person}{Yuan Xue}, {and} \bibinfo{person}{Vishal~M Patel}.}
  \bibinfo{year}{2023}\natexlab{}.
\newblock \showarticletitle{Securing Deep Generative Models with Universal
  Adversarial Signature}.
\newblock \bibinfo{journal}{\emph{arXiv preprint arXiv:2305.16310}}
  (\bibinfo{year}{2023}).
\newblock


\bibitem[Zhang et~al\mbox{.}(2022)]%
        {zhang2022visual}
\bibfield{author}{\bibinfo{person}{Guangsheng Zhang}, \bibinfo{person}{Bo Liu},
  \bibinfo{person}{Tianqing Zhu}, \bibinfo{person}{Andi Zhou}, {and}
  \bibinfo{person}{Wanlei Zhou}.} \bibinfo{year}{2022}\natexlab{}.
\newblock \showarticletitle{Visual privacy attacks and defenses in deep
  learning: a survey}.
\newblock \bibinfo{journal}{\emph{Artificial Intelligence Review}}
  \bibinfo{volume}{55}, \bibinfo{number}{6} (\bibinfo{year}{2022}),
  \bibinfo{pages}{4347--4401}.
\newblock


\bibitem[Zhang et~al\mbox{.}(2024c)]%
        {zhang2024forget}
\bibfield{author}{\bibinfo{person}{Gong Zhang}, \bibinfo{person}{Kai Wang},
  \bibinfo{person}{Xingqian Xu}, \bibinfo{person}{Zhangyang Wang}, {and}
  \bibinfo{person}{Humphrey Shi}.} \bibinfo{year}{2024}\natexlab{c}.
\newblock \showarticletitle{Forget-me-not: Learning to forget in text-to-image
  diffusion models}. In \bibinfo{booktitle}{\emph{Proceedings of the IEEE/CVF
  Conference on Computer Vision and Pattern Recognition}}.
  \bibinfo{pages}{1755--1764}.
\newblock


\bibitem[Zhang et~al\mbox{.}(2024a)]%
        {zhang-etal-2024-r}
\bibfield{author}{\bibinfo{person}{Hanning Zhang}, \bibinfo{person}{Shizhe
  Diao}, \bibinfo{person}{Yong Lin}, \bibinfo{person}{Yi Fung},
  \bibinfo{person}{Qing Lian}, \bibinfo{person}{Xingyao Wang},
  \bibinfo{person}{Yangyi Chen}, \bibinfo{person}{Heng Ji}, {and}
  \bibinfo{person}{Tong Zhang}.} \bibinfo{year}{2024}\natexlab{a}.
\newblock \showarticletitle{{R}-Tuning: Instructing Large Language Models to
  Say {`}{I} Don{'}t Know{'}}. In \bibinfo{booktitle}{\emph{Proceedings of the
  2024 Conference of the North American Chapter of the Association for
  Computational Linguistics: Human Language Technologies (Volume 1: Long
  Papers)}}. \bibinfo{pages}{7113--7139}.
\newblock


\bibitem[Zhang et~al\mbox{.}(2024b)]%
        {zhang2024llm}
\bibfield{author}{\bibinfo{person}{Qihui Zhang}, \bibinfo{person}{Chujie Gao},
  \bibinfo{person}{Dongping Chen}, \bibinfo{person}{Yue Huang},
  \bibinfo{person}{Yixin Huang}, \bibinfo{person}{Zhenyang Sun},
  \bibinfo{person}{Shilin Zhang}, \bibinfo{person}{Weiye Li},
  \bibinfo{person}{Zhengyan Fu}, \bibinfo{person}{Yao Wan}, {et~al\mbox{.}}}
  \bibinfo{year}{2024}\natexlab{b}.
\newblock \showarticletitle{LLM-as-a-Coauthor: Can Mixed Human-Written and
  Machine-Generated Text Be Detected?}. In \bibinfo{booktitle}{\emph{Findings
  of the Association for Computational Linguistics: NAACL 2024}}.
  \bibinfo{pages}{409--436}.
\newblock


\bibitem[Zhao et~al\mbox{.}(2024)]%
        {zhao2024proactive}
\bibfield{author}{\bibinfo{person}{Yuan Zhao}, \bibinfo{person}{Bo Liu},
  \bibinfo{person}{Tianqing Zhu}, \bibinfo{person}{Ming Ding},
  \bibinfo{person}{Xin Yu}, {and} \bibinfo{person}{Wanlei Zhou}.}
  \bibinfo{year}{2024}\natexlab{}.
\newblock \showarticletitle{Proactive image manipulation detection via deep
  semi-fragile watermark}.
\newblock \bibinfo{journal}{\emph{Neurocomputing}}  \bibinfo{volume}{585}
  (\bibinfo{year}{2024}), \bibinfo{pages}{127593}.
\newblock


\bibitem[Zhao et~al\mbox{.}(2023)]%
        {zhao2023recipe}
\bibfield{author}{\bibinfo{person}{Yunqing Zhao}, \bibinfo{person}{Tianyu
  Pang}, \bibinfo{person}{Chao Du}, \bibinfo{person}{Xiao Yang},
  \bibinfo{person}{Ngai-Man Cheung}, {and} \bibinfo{person}{Min Lin}.}
  \bibinfo{year}{2023}\natexlab{}.
\newblock \showarticletitle{A recipe for watermarking diffusion models}.
\newblock \bibinfo{journal}{\emph{arXiv preprint arXiv:2303.10137}}
  (\bibinfo{year}{2023}).
\newblock


\bibitem[Zhong et~al\mbox{.}(2023)]%
        {zhong2023rich}
\bibfield{author}{\bibinfo{person}{Nan Zhong}, \bibinfo{person}{Yiran Xu},
  \bibinfo{person}{Zhenxing Qian}, {and} \bibinfo{person}{Xinpeng Zhang}.}
  \bibinfo{year}{2023}\natexlab{}.
\newblock \showarticletitle{Rich and Poor Texture Contrast: A Simple yet
  Effective Approach for AI-generated Image Detection}.
\newblock \bibinfo{journal}{\emph{arXiv preprint arXiv:2311.12397}}
  (\bibinfo{year}{2023}).
\newblock


\bibitem[Zhu et~al\mbox{.}(2024)]%
        {zhu2024genimage}
\bibfield{author}{\bibinfo{person}{Mingjian Zhu}, \bibinfo{person}{Hanting
  Chen}, \bibinfo{person}{Qiangyu Yan}, \bibinfo{person}{Xudong Huang},
  \bibinfo{person}{Guanyu Lin}, \bibinfo{person}{Wei Li},
  \bibinfo{person}{Zhijun Tu}, \bibinfo{person}{Hailin Hu},
  \bibinfo{person}{Jie Hu}, {and} \bibinfo{person}{Yunhe Wang}.}
  \bibinfo{year}{2024}\natexlab{}.
\newblock \showarticletitle{Genimage: A million-scale benchmark for detecting
  ai-generated image}.
\newblock \bibinfo{journal}{\emph{Advances in Neural Information Processing
  Systems}}  \bibinfo{volume}{36} (\bibinfo{year}{2024}).
\newblock


\bibitem[Zhu et~al\mbox{.}(2023)]%
        {10086559}
\bibfield{author}{\bibinfo{person}{Yao Zhu}, \bibinfo{person}{Yuefeng Chen},
  \bibinfo{person}{Xiaodan Li}, \bibinfo{person}{Rong Zhang},
  \bibinfo{person}{Xiang Tian}, \bibinfo{person}{Bolun Zheng}, {and}
  \bibinfo{person}{Yaowu Chen}.} \bibinfo{year}{2023}\natexlab{}.
\newblock \showarticletitle{Information-Containing Adversarial Perturbation for
  Combating Facial Manipulation Systems}.
\newblock \bibinfo{journal}{\emph{IEEE Transactions on Information Forensics
  and Security}}  \bibinfo{volume}{18} (\bibinfo{year}{2023}),
  \bibinfo{pages}{2046--2059}.
\newblock
\urldef\tempurl%
\url{https://doi.org/10.1109/TIFS.2023.3262156}
\showDOI{\tempurl}


\end{thebibliography}

\appendix

\end{document}